\documentclass[a4paper,11pt]{article}
\pdfoutput=1

\usepackage{jheppub} 

\usepackage[utf8]{inputenc}
\usepackage{float}
\usepackage{mathtools}
\usepackage{dsfont}
\usepackage{cancel}
\usepackage{amsmath}
\usepackage{amsfonts}
\usepackage{amssymb}
\usepackage{ytableau}
\usepackage{cellspace}
\usepackage{array}
\usepackage{makecell}
\usepackage{pgfplots}
\usepackage{tikz}
\usepackage{tikz-cd}
\usetikzlibrary{shapes.misc,shadows,fit,decorations.pathmorphing,positioning,trees,decorations.markings,decorations.pathreplacing,calc,shapes,patterns,arrows}
\usepackage{xcolor}
\usepackage{hyperref}
\usepackage{nowidow}
\usepackage{hhline}
\usepackage[capitalize]{cleveref}
\usepackage{comment}
\usepackage{graphicx}
\usepackage{subcaption}
\usepackage{environ}
\usepackage{enumitem}
\usepackage[section]{placeins}
\usepackage{tabularx}

\addtocounter{MaxMatrixCols}{1}
\renewcommand{\arraystretch}{1}

\hypersetup{
	pdftitle={2d $\mathcal{N}=(0,1)$ Gauge Theories and Spin(7) Orientifolds},   
	pdfauthor={\textcopyright\ Sebastian Franco, Alessandro Mininno, \'{A}ngel M. Uranga, Xingyang Yu},     
	pdfsubject={HEP},   
	pdfcreator={pdfLaTex},   
	pdfproducer={LaTex}, 
	pdfkeywords={},
		colorlinks=true,
}

\newcolumntype{M}[1]{>{\centering\arraybackslash}m{#1}}

\makeatletter
\newsavebox{\measure@tikzpicture}
\NewEnviron{scaletikzpicturetowidth}[1]{%
  \def\tikz@width{#1}%
  \begin{lrbox}{\measure@tikzpicture}%
  \BODY
  \end{lrbox}%
  \pgfmathparse{#1/\wd\measure@tikzpicture}%
  \BODY
}
\makeatother

\makeatletter
\def\squarecorner#1{
	\pgf@x=\the\wd\pgfnodeparttextbox%
	\pgfmathsetlength\pgf@xc{\pgfkeysvalueof{/pgf/inner xsep}}%
	\advance\pgf@x by 2\pgf@xc%
	\pgfmathsetlength\pgf@xb{\pgfkeysvalueof{/pgf/minimum width}}%
	\ifdim\pgf@x<\pgf@xb%
	\pgf@x=\pgf@xb%
	\fi%
	\pgf@y=\ht\pgfnodeparttextbox%
	\advance\pgf@y by\dp\pgfnodeparttextbox%
	\pgfmathsetlength\pgf@yc{\pgfkeysvalueof{/pgf/inner ysep}}%
	\advance\pgf@y by 2\pgf@yc%
	\pgfmathsetlength\pgf@yb{\pgfkeysvalueof{/pgf/minimum height}}%
	\ifdim\pgf@y<\pgf@yb%
	\pgf@y=\pgf@yb%
	\fi%
	\ifdim\pgf@x<\pgf@y%
	\pgf@x=\pgf@y%
	\else
	\pgf@y=\pgf@x%
	\fi
	\pgf@x=#1.5\pgf@x%
	\advance\pgf@x by.5\wd\pgfnodeparttextbox%
	\pgfmathsetlength\pgf@xa{\pgfkeysvalueof{/pgf/outer xsep}}%
	\advance\pgf@x by#1\pgf@xa%
	\pgf@y=#1.5\pgf@y%
	\advance\pgf@y by-.5\dp\pgfnodeparttextbox%
	\advance\pgf@y by.5\ht\pgfnodeparttextbox%
	\pgfmathsetlength\pgf@ya{\pgfkeysvalueof{/pgf/outer ysep}}%
	\advance\pgf@y by#1\pgf@ya%
}
\makeatother

\pgfdeclareshape{square}{
	\savedanchor\northeast{\squarecorner{}}
	\savedanchor\southwest{\squarecorner{-}}
	
	\foreach \x in {east,west} \foreach \y in {north,mid,base,south} {
		\inheritanchor[from=rectangle]{\y\space\x}
	}
	\foreach \x in {east,west,north,mid,base,south,center,text} {
		\inheritanchor[from=rectangle]{\x}
	}
	\inheritanchorborder[from=rectangle]
	\inheritbackgroundpath[from=rectangle]
}

\DeclareMathOperator{\Res}{Res}
\DeclareMathOperator{\U}{U}
\DeclareMathOperator{\SU}{SU}
\DeclareMathOperator{\SO}{SO}

\DeclareMathOperator{\USp}{USp}
\DeclareMathOperator{\Spin}{Spin}
\DeclareMathOperator{\SPP}{SPP}

\DeclareMathOperator{\PE}{PE}
\DeclareMathOperator{\PL}{PL}
\DeclareMathOperator{\HS}{HS}

\newcommand{\CC}{\mathbb{C}}

\newcommand{\ZZ}{\mathbb{Z}}
\newcommand{\ID}{\mathds{1}}
\newcommand{\IM}{\mathbf{M}}
\newcommand{\IX}{\mathbf{X}}
\newcommand{\IZ}{\mathbf{Z}}
\newcommand{\coma}{\, , \,}
\newcommand{\fstop}{\, .}
\newcommand{\hc}{\text{ h.c.}}

\def\quadro{\rotatebox[origin=c]{45}{$\blacksquare$}}

\newcommand{\drawsquare}[2]{\hbox{%
\rule{#2pt}{#1pt}\hskip-#2pt
\rule{#1pt}{#2pt}\hskip-#1pt
\rule[#1pt]{#1pt}{#2pt}}\rule[#1pt]{#2pt}{#2pt}\hskip-#2pt
\rule{#2pt}{#1pt}}

\newcommand{\fund}{~\raisebox{-.5pt}{\drawsquare{6.5}{0.4}}~}
\newcommand{\antifund}{~\overline{\raisebox{-.5pt}{\drawsquare{6.5}{0.4}}}~}

\definecolor{redX}{RGB}{218,59,38}
\definecolor{blueX}{RGB}{71,159,248}
\definecolor{yellowX}{RGB}{239,189,64}
\definecolor{greenX}{HTML}{54ae32}

\allowdisplaybreaks[1] 

\crefname{figure}{Figure}{Figures}
\crefname{section}{Section}{Sections}
\crefname{table}{Table}{Tables}

\ytableausetup{smalltableaux,centertableaux}

\preprint{IFT-UAM/CSIC-21-105, ZMP-HH/21-19}

\title{2d $\mathcal{N}=(0,1)$ Gauge Theories and Spin(7) Orientifolds}

\author[a,b,c]{Sebasti\'an Franco,}
\author[d,e]{Alessandro Mininno,}
\author[d]{\'{A}ngel M. Uranga,}
\author[f]{Xingyang Yu}

\affiliation[a]{Physics Department, The City College of the CUNY\\
	160 Convent Avenue, New York, NY 10031, USA}
\affiliation[b]{Physics Program and \textsuperscript{$c$}Initiative for the Theoretical Sciences\\
	The Graduate School and University Center, The City University of New York\\
	365 Fifth Avenue, New York NY 10016, USA}
\affiliation[d]{Instituto de F\'{\i}sica Te\'orica IFT-UAM/CSIC,\\
	C/ Nicol\'as Cabrera 13-15, 
	Campus de Cantoblanco, 28049 Madrid, Spain}
\affiliation[e]{II. Institut f\"ur Theoretische Physik, Universit\"at Hamburg,\\
Luruper Chaussee 149, 22607 Hamburg, Germany}
\affiliation[f]{Center for Cosmology and Particle Physics,\\
	Department of Physics, New York University,\\
	726 Broadway, New York, NY 10003, USA}

\emailAdd{sfranco@ccny.cuny.edu}
\emailAdd{alessandro.mininno@desy.de}
\emailAdd{angel.uranga@csic.es}
\emailAdd{xy1038@nyu.edu}

\abstract{We initiate the geometric engineering of 2d $\mathcal{N}=(0,1)$ gauge theories on D1-branes probing singularities. To do so, we introduce a new class of backgrounds obtained as quotients of Calabi-Yau 4-folds by a combination of an anti-holomorphic involution leading to a Spin(7) cone and worldsheet parity. We refer to such constructions as {\it Spin(7) orientifolds}. Spin(7) orientifolds explicitly realize the perspective on 2d $\mathcal{N}=(0,1)$  theories as real slices of $\mathcal{N}=(0,2)$ ones. Remarkably, this projection is geometrically realized as Joyce’s construction of Spin(7) manifolds via quotients of Calabi-Yau 4-folds by anti-holomorphic involutions. We illustrate this construction in numerous examples with both orbifold and non-orbifold parent singularities, discuss the role of the choice of vector structure in the orientifold quotient, and study partial resolutions.
}

\begin{document}

\maketitle

\flushbottom

\section{Introduction}

Engineering gauge theories in string or M-theory provides alternative perspectives, often geometric, on their dynamics. Such realizations typically lead to a deeper understanding of the theories at hand, suggest natural generalizations, and even contribute to the discovery of new results.

Our understanding of $2$d $\mathcal{N}=(0,2)$ gauge theories has significantly progressed in recent years. The new results include $c$-extremization \cite{Benini:2012cz,Benini:2013cda}, $\mathcal{N}=(0,2)$ triality \cite{Gadde:2013lxa} and connections to gauge theories in higher dimensions \cite{Benini:2013cda,Gadde:2013sca,Kutasov:2013ffl,Kutasov:2014hha,Benini:2015bwz}. These discoveries have fueled a renewed interest in the stringy engineering of such theories. A possible scenario involves realizing them on the world volume of D1-brane probing singular Calabi-Yau (CY) 4-folds.\footnote{For alternative setups leading to $2$d $\mathcal{N}=(0,2)$ gauge theories, see e.g. \cite{Benini:2013cda,Gadde:2013sca,Tatar:2015sga, Schafer-Nameki:2016cfr,Benini:2015bwz}.} Following the pioneering work of \cite{Garcia-Compean:1998sla}, a new class of brane configurations, denoted {\it  brane brick models}, was introduced in \cite{Franco:2015tya}. Brane brick models fully encode the $2$d $\mathcal{N}=(0,2)$ gauge theories probing toric CY 4-folds, to which they are connected by T-duality. Furthermore, they have significantly simplified the map between geometry and the corresponding gauge theories (see \cite{Franco:2016nwv,Franco:2016qxh,Franco:2016fxm,Franco:2017cjj,Franco:2018qsc,Franco:2020avj,Franco:2021elb} for further developments).

As usual, it is desirable to investigate theories with less supersymmetry. The next step corresponds to $2$d $\mathcal{N}=(0,1)$, namely minimally supersymmetric, theories. Such models are particularly interesting because while they are still supersymmetric, they no longer have holomorphy. While considerably less is known about them, new results about their dynamics have appeared in \cite{Gukov:2019lzi}, including the proposal of a new $2$d $\mathcal{N}=(0,1)$ triality. Once again, this raises the question of how to engineer these theories in string theory. In \cite{Gukov:2019lzi}, it was noted that the theories participating in $\mathcal{N}=(0,1)$ triality are, in a sense, ``real slices” of their ``complex” $\mathcal{N}=(0, 2)$ counterparts, both at the level of gauge theory description and effective non-linear sigma model. A more general formulation of such $\mathcal{N}=(0,2)/(0,1)$ correspondence was left as an open question. 

With these motivations in mind, in this paper we introduce {\it Spin(7) orientifolds}, a new class of backgrounds that combine Joyce’s construction of $\Spin(7)$ manifolds via the quotient of CY 4-folds by anti-holomorphic involutions with worldsheet parity, and construct $2$d $\mathcal{N}=(0,1)$ gauge theories on D1-branes probing them. Closely related ideas were presented in the insightful paper \cite{Forcella:2009jj,Amariti:2014ewa}, whose goal was to engineer 3d $\mathcal{N}=1$ theories on M2-branes.\footnote{For applications to F-theory of $\Spin(7)$ holonomy manifolds from CY 4-folds quotients see, e.g.,~\cite{Bonetti:2013fma,Bonetti:2013nka}.}

This paper is organized as follows. Section~\ref{sec:2dN01QFT} discusses the general structure and properties of $2$d $\mathcal{N}=(0,1)$ field theories. Section~\ref{sec:N02inN01Form} presents the decomposition of $\mathcal{N}=(0,2)$ supermultiplets in $\mathcal{N}=(0,1)$ language. Section~\ref{sec:2dN01orientifolds} explains the construction of $\Spin(7)$ cones and $\Spin(7)$ orientifolds starting from CY $4$-folds. Sections~\ref{sec:orientN01gaugeth} and \ref{sec:N01theorfromorienquot} discuss the field theory implementation of $\Spin(7)$ orientifolds. The connection between the anti-holomorphic involutions of the CY$_4$ and the gauge theory is studied in Section~\ref{sec:HSreview}. Section~\ref{sec:examplesenginN=01} considers $\Spin(7)$ orientifolds of $\CC^4$ and its orbifolds. In Section~\ref{sec:vectorstructure} we decribe how the choice of vector structure can lead to different  gauge theories associated to the same geometric involution. Section~\ref{sec:BeyondOrbif} presents $\Spin(7)$ orientifolds of generic, non-orbifold, parent CY$_4$’s. Finally, Section~\ref{sec:HiggsingPartResol} investigates the interplay between partial resolution and higgsing. Section~\ref{section_conclusions} collects our conclusions and outlook. Appendix~\ref{app:C4Z2Z2SPPC} contains additional examples that are used in Section~\ref{sec:HiggsingPartResol}.

\section{2d $\mathcal{N}=(0,1)$ Field Theories}
\label{sec:2dN01QFT}

In this section, we briefly review the general structure of $2$d $\mathcal{N}=(0,1)$ field theories. Instead of discussing all terms in the Lagrangian, we will focus on the main facts we will use in following sections. We refer the reader to \cite{Sakamoto:1984zk,Hull:1985jv,Brooks:1986uh,Brooks:1986gd,Gukov:2019lzi} for a more detailed presentation.

\subsection{Constructing 2d $\mathcal{N}=(0,1)$ gauge Theories}
\label{sec:2dN01construT}

We describe these theories in terms of $2$d $\mathcal{N}=(0,1)$ superspace $\left(x^0,x^1,\theta^+\right)$. 
There are three types of supermultiplets as elementary building blocks:
\begin{itemize}
\item Vector multiplet:
\begin{equation}\label{(0,1) vector}
	\begin{split}
		V_+&=\theta^+(A_0(x)+A_1(x))\coma\\
		V_-&=A_0(x)-A_1(x)+\theta^+\lambda_-(x)\fstop
	\end{split}
\end{equation}
It contains a gauge boson $A_{\pm}$ and a left-moving Majorana-Weyl fermion $\lambda_-$ in the adjoint representation. 

\item Scalar multiplet:
\begin{equation}
	\Phi(x,\theta)=\phi(x)+\theta^+\psi_+(x)\fstop
\end{equation}
It has a real scalar field $\phi$ and a right-moving Majorana-Weyl fermion $\psi_+$. 

\item Fermi multiplet:
\begin{equation}\label{(0,1) Fermi}
	\Lambda(x,\theta)=\psi_-(x)+\theta^+F(x)\fstop
\end{equation}
It has a left-moving Majorana-Weyl spinor as its only on-shell degree of freedom. Here $F$ is an auxiliary field. 
\end{itemize}

As usual, the kinetic terms for matter fields and their gauge couplings are given by
\begin{equation}
	\begin{split}
	\mathcal{L}_s+\mathcal{L}_F=\int d\theta^+~ \left(\frac{i}{2}\sum_i(\mathcal{D}_+\Phi_i\mathcal{D}_-\Phi_i)-\frac{1}{2}\sum_a(\Lambda_a\mathcal{D}_+\Lambda_a) \right)\coma
	\end{split}
	\label{eq:kingauge01mattfields}
\end{equation}
where $\mathcal{D}_{\pm}$ are super-covariant derivatives~\cite{Gukov:2019lzi}. 

These theories admit another interaction, which is an $\mathcal{N}=(0,1)$ analog of the $\mathcal{N}=(0,2)$ $J$-term interaction, or $\mathcal{N}=1$ superpotential:
\begin{equation}
	\mathcal{L}_J\equiv \int d \theta^+W^{(0,1)}=\int d\theta^+\sum_a (\Lambda_aJ^a(\Phi_i))\coma
	\label{eq:W01super}
\end{equation}
where $J^a(\Phi_i)$ are real functions of scalar fields. Both the quiver and $W^{(0,1)}$ are necessary for fully specifying any of the $\mathcal{N}=(0,1)$ gauge theories considered in this paper. From now on, we will refer to $W^{(0,1)}$ as the \emph{superpotential} for convenience.

After integrating out the auxiliary fields $F_a$, $\mathcal{L}_J$ produces various interactions, including Yukawa-like couplings 
\begin{equation}
	\sum_a \lambda_{-a}\frac{\partial J^a}{\partial \phi_i}\psi_{+i}\coma
\end{equation}
as well as a scalar potential 
\begin{equation}
	\frac{1}{2}\sum_a(J^a(\phi_i))^2\fstop
\end{equation}

\subsection{Anomalies}
\label{sec:Anomalies}

\begin{figure}[!htp]
	\centering
	\begin{tikzpicture}[scale=0.75]
	\draw[line width=1pt] (0,0) circle (1);
	\draw[line width=1pt,decoration=snake,decorate] (-3,0) -- (-1,0);
	\draw[line width=1pt,decoration=snake,decorate] (3,0) -- (1,0);
	\end{tikzpicture}
	\caption{Generic $1$-loop diagram associated with $2$d anomalies.}
	\label{fig:anom1loop}
\end{figure}
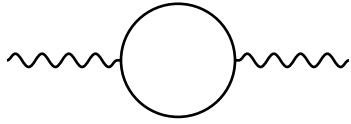

In $2$d, anomalies are given by $1$-loop diagrams of the generic form shown in Figure~\ref{fig:anom1loop}, where left- and right-moving fermions running in the loop contribute oppositely. 

Since $2$d $\mathcal{N}=(0,1)$ theories are chiral, left- and right-moving fermions are not necessarily paired up, and anomalies do not cancel automatically. For a given symmetry group, anomalies depend on the types and the representations of the fields transforming under it. Below, we focus on those groups and representations appearing in the $2$d $\mathcal{N}=(0,1)$ theories engineered in this paper.

\paragraph{Non-Abelian Anomalies}\mbox{}

\smallskip

Let us first consider pure non-Abelian $G^2$ gauge or global anomalies, where $G$ can be $\SU(N)$, $\SO(N)$ or $\USp(N)$ group.\footnote{In our convention $\USp(2)\simeq \SU(2)$, so $\USp(N)$ makes sense only if $N$ is even.} 
The corresponding anomaly is given by
\begin{equation}
	\text{Tr}[\gamma^3J_GJ_G]\coma
\end{equation}
where $\gamma^3$ is the chirality matrix in $2$d and $J_G$ is the current associated to $G$. The resulting anomaly from a field in representation $\rho$ of $G$ can be computed in terms of the Dynkin index $T(\rho)$: 
\begin{equation}
	T(\rho)=C_2(\rho)\frac{d(\rho)}{d(\text{adjoint})}\coma
\end{equation}
where $C_2(\rho)$ is the quadratic Casimir for representation $\rho$.

\begin{table}[!htp]
	\centering
\begin{tabular}{|Sc|Sc|Sc|Sc|Sc|}
	\hline
	$\SU(N)$	& fundamental & adjoint & antisymmetric & symmetric\\
	\hhline{|=|=|=|=|=|} 
	vector multiplet & $\times$ & $-N$ & $\times$ &$\times$ \\
	\hline
	Fermi multiplet & $-\dfrac{1}{2}$ & $-N$ & $\dfrac{-N+2}{2}$ & $\dfrac{-N-2}{2}$\\
	\hline
	scalar multiplet & $\dfrac{1}{2}$ & $N$ &$\dfrac{N-2}{2}$ & $\dfrac{N+2}{2}$\\
	\hline
\end{tabular}
\caption{Anomaly contributions of the $2$d $\mathcal{N}=(0,1)$ multiplets in various representations of $\SU(N)$. Since anomalies are quadratic in 2d, the same contributions apply for the conjugate representations.}
\label{tab:SUanomaly}
\end{table}

\begin{table}[!htp]
	\centering
\begin{tabular}{|Sc|Sc|Sc|Sc|}
	\hline
	$\SO(N)$	& fundamental & antisymmetric (adjoint) & symmetric\\
	\hhline{|=|=|=|=|} 
	vector multiplet & $\times$ & $-N+2$ & $\times$ \\
	\hline
	Fermi multiplet & $-1$ & $-N+2$ & $-N-2$\\
	\hline
	scalar multiplet & $1$ & $N-2$ &$N+2$\\
	\hhline{|=|=|=|=|} 
	$\USp(N)$	& fundamental & antisymmetric  & symmetric (adjoint)\\
	\hhline{|=|=|=|=|} 
	vector multiplet & $\times$ & $\times$ & $-N-2$ \\
	\hline
	Fermi multiplet & $-1$ & $-N+2$ & $-N-2$\\
	\hline
	scalar multiplet & $1$ & $N-2$ &$N+2$\\
	\hline
\end{tabular}
\caption{Anomaly contributions of the $2$d $\mathcal{N}=(0,1)$ multiplets in various representations of $\SO(N)$ and $\USp(N)$.}
\label{tab:SOUSpanomaly}
\end{table}

In Table~\ref{tab:SUanomaly} we present anomaly contributions for superfields in the most common representations of $\SU(N)$. In Table~\ref{tab:SOUSpanomaly}, we present anomaly contributions of different types of superfields carrying various representations of $\SO(N)$ and $\USp(N)$ groups, computed using Dynkin indices listed in \cite{Yamatsu:2015npn}. 

In the case of gauge groups, anomalies must vanish for consistency of the theory at the quantum level. This leads to important constraints in our construction of $2$d $\mathcal{N}=(0,1)$ theories, that may require the introduction of extra flavors to cancel  anomalies. We will illustrate this with concrete examples in following sections.

Unlike gauge symmetries, global symmetries may indeed be anomalous. One important property of global anomalies is that they are preserved along the Renormalization Group (RG) flow. Therefore, they can be used to check dualities between two or more theories, namely whether these UV-different theories are IR-equivalent. Examples of using global anomalies to check dualities in $2$d $\mathcal{N}=(0,1)$ theories can be found in \cite{Gukov:2019lzi} and also in our upcoming work \cite{Franco:2021branetriality}.

\paragraph{Abelian Anomalies}\mbox{}

\smallskip

For $\U(N)$ groups of the worldvolume theories on D-brane probes, in addition to non-Abelian anomalies, the $\U(1)$ factors can generically have $\U(1)_i^2$ and mixed $\U(1)_i\U(1)_j$ Abelian anomalies. As before, the $\U(1)$ groups can be either gauged or global.

The theories studied in this paper generically have non-vanishing Abelian gauge anomalies. Similarly to the discussion in \cite{Franco:2015tna, Franco:2017cjj}, we expect that such anomalies are canceled by the bulk fields in the closed string sector via a generalized Green-Schwarz (GS) mechanism (see \cite{Ibanez:1998qp,Mohri:1997ef} for derivations in 4d ${\cal N}=1$ and $2$d ${\cal N}=(0,2)$ theories realized at orbifolds/orientifold singularities).

\subsection{Triality}
\label{sec:triality}

Recently, an IR triality between 2d $\mathcal{N}=(0,1)$ theories with $\SO$ and $\USp$ gauge groups was proposed in \cite{Gukov:2019lzi}. Evidence for the proposal includes matching of anomalies and elliptic genera. This new triality can be regarded as a cousin of the 2d $\mathcal{N}=(0,2)$ triality introduced in \cite{Gadde:2013lxa}. Interestingly, 2d $\mathcal{N}=(0,2)$ triality, together with Seiberg duality for 4d gauge theories \cite{Seiberg:1994pq}, extend to an infinite family of order $(m+1)$ dualities of $m$-graded quiver theories \cite{Franco:2016tcm,Franco:2017lpa,Closset:2018axq}.

It is natural to ask whether, within the context of gauge theories on the worldvolume of D-branes probing singularities, the $\mathcal{N}=(0,1)$ triality admits a geometric explanation. The similarity between the theories in \cite{Gukov:2019lzi} and the ones constructed in this paper hints that this is the case. This question will be addressed in \cite{Franco:2021branetriality}, where we will show that $\mathcal{N}=(0,1)$ triality follows from the non-uniqueness of the map between $\Spin(7)$ cones and 2d $\mathcal{N}=(0,1)$ gauge theories.

\section{$\mathcal{N}=(0,2)$ Field Theories in $\mathcal{N}=(0,1)$ Formalism}
\label{sec:N02inN01Form}

In this paper, we will construct $2$d $\mathcal{N}=(0,1)$ theories from $2$d $\mathcal{N}=(0,2)$ theories via orientifold quotients. Therefore, it is useful to decompose $\mathcal{N}=(0,2)$ theories in terms of the $\mathcal{N}=(0,1)$ formalism.

$\mathcal{N}=(0,2)$ theories can be expressed in superspace $\left(x^0,x^1, \theta^+, \bar{\theta}^+\right)$ and have three types of multiplets: vector, chiral and Fermi. These multiplets and the Lagrangian can be further expressed in $\mathcal{N}=(0,1)$ language using the superspace $\left(x^0, x^1, \theta^+\right)$.

\newpage

\paragraph{$\mathcal{N}=(0,2)$ vector multiplet}\mbox{}

\smallskip

The $\mathcal{N}=(0,2)$ vector multiplet $V^{(0,2)}$ contains a gauge boson, a left-moving chiral fermion and an auxiliary field. It decomposes into $\mathcal{N}=(0,1)$ multiplets as follows:
\begin{equation}
\renewcommand{\arraystretch}{1.1}
    \begin{array}{c}
           \mathcal{N}=(0,2) \text{ vector multiplet } V_i^{(0,2)}  \\
         \swarrow \qquad \searrow  \\
    \mathcal{N}=(0,1) \text{ vector multiplet } V_i  \quad \oplus\quad  \mathcal{N}=(0,1) \text{ Fermi multiplet } \Lambda_{i}^R\fstop
    \end{array}
\end{equation}
The chiral fermion in $V^{(0,2)}_i$ is separated into two Majorana-Weyl fermions, one of which is included in $V_i$ and the other is in $\Lambda_{i}^R$. The auxiliary field in $V^{(0,2)}_i$ becomes the one in $\Lambda_{i}^R$.

The kinetic term of $V_i^{(0,2)}$ in the Lagrangian can be expressed in $\mathcal{N}=(0,1)$ superspace as kinetic terms of $V_i$ and $\Lambda_{i}^R$:
\begin{equation}
	\mathcal{L}_{\text{gauge}}^{(0,2)}\rightarrow \mathcal{L}_{\text{gauge}}-\frac{1}{2}\int d \theta^+\sum_a(\Lambda_{i}^R\mathcal{D}_+\Lambda_{i}^R)\coma
\end{equation}
where $\mathcal{L}_{\text{gauge}}$ is the kinetic term of an $\mathcal{N}=(0,1)$ vector multiplet.

\paragraph{$\mathcal{N}={(0,2)}$ chiral multiplet}\mbox{}

\smallskip

The $\mathcal{N}=(0,2)$ chiral superfield contains a complex scalar $\phi^c$ and a right-moving chiral fermion $\psi^c_+$. Its expansion is
\begin{equation}
	\Phi^{(0,2)}_m=\phi^c_m +\theta^+\psi_{+m}^c-i\theta^+ \bar{\theta}^+D_+\phi^c_m \fstop
\end{equation}
It decomposes into $\mathcal{N}=(0,1)$ multiplets as follows:
\begin{equation}
\renewcommand{\arraystretch}{1.1}
    \begin{array}{c}
           \mathcal{N}=(0,2) \text{ chiral multiplet } \Phi^{(0,2)}_m  \\
         \swarrow \qquad \searrow  \\
    \mathcal{N}=(0,1) \text{ scalar multiplet } \Phi^{1}_m \quad \oplus \quad \mathcal{N}=(0,1) \text{ scalar multiplet } \Phi_m^2\fstop
    \end{array}
\end{equation}
The two $\mathcal{N}=(0,1)$ scalar multiplets $\Phi_m^{1,2}$ can be further combined into an $\mathcal{N}=(0,1)$ \emph{complex scalar multiplet}, so that the above decomposition is rewritten as \begin{equation}\label{(0,2) chiral in (0,1) complex scalar}
\renewcommand{\arraystretch}{1.1}
    \begin{array}{c}
           \mathcal{N}=(0,2) \text{ chiral multiplet } \Phi^{(0,2)}_m  \\
         \downarrow  \\
     \mathcal{N}=(0,1) \text{ complex scalar multiplet } \Phi_m\fstop
    \end{array}
\end{equation}
The kinetic terms of the matter fields in $\Phi^{(0,2)}_m$ and their gauge couplings are included in the term $\mathcal{L}^{(0,2)}_{\text{chiral}}$ in $\mathcal{N}=(0,2)$ superspace. As an example, let us consider a chiral multiplet $\Phi^{(0,2)}_m$ transforming under a $\U(1)$ gauge group. In this case, $\mathcal{L}_{\text{chiral}}^{(0,2)}$ reads: 
\begin{equation}
\begin{split}
		\mathcal{L}^{(0,2)}_{\text{chiral}}&=-\frac{i}{2}\int d\theta^+ d\bar{\theta}^+ (\Phi^{(0,2)}_m)^\dagger\mathcal{D}_-^{(0,2)}\Phi^{(0,2)}_m\coma
\end{split}
\end{equation}
where with $\dagger$ we mean the Hermitian conjugate\footnote{I.e. complex conjugate and transposition,  $\left(\bar{\Phi}^{(0,2)}\right)^T$.} of $\Phi^{(0,2)}_m$. The above Lagrangian can be regarded as a combination of two parts:
\begin{equation}\label{two parts of (0,2) chiral kinetic terms}
	\mathcal{L}^{(0,2)}_{\text{chiral}}=\text{Kinetic terms of $\Phi^{(0,2)}_m$}+\text{Interaction terms between $V^{(0,2)}$ and $\Phi^{(0,2)}_m$,}
\end{equation}
which can be further expressed in terms of $\mathcal{N}=(0,1)$ multiplets as 
\begin{equation}\label{(0,2) chiral lagrangian in (0,1) multiplets}
\begin{split}
	\mathcal{L}^{(0,2)}_{\text{chiral}}\rightarrow &~~\text{Kinetic terms of $\mathcal{N}=(0,1)$  complex scalar $\Phi_m$}\\
	&+\text{Interaction terms between $V$ and $\Phi_m$}\\
	&+\text{Interaction terms between $\Lambda^R$ and $\Phi_m$ .}
\end{split}
\end{equation}
$V$ and $\Lambda^R$ here are $\mathcal{N}=(0,1)$ vector and adjoint Fermi multiplets coming from the decomposition of $V^{(0,2)}$. From now on, the superscript $R$ is used to emphasize that a superfield is real.

$\mathcal{L}^{(0,2)}_{\text{chiral}}$ can be expressed in $\mathcal{N}=(0,1)$ superspace using~\cref{eq:kingauge01mattfields,eq:W01super}. It becomes
\begin{equation}
\begin{split}
	\mathcal{L}^{(0,2)}_{\text{chiral}}&\rightarrow \mathcal{L}_{s}+\int d \theta^+ W^{(0,1)}\\
	&=-\frac{i}{4}\int d \theta^+[\mathcal{D}_+\Phi^\dagger_m\mathcal{D}_-\Phi_m+\mathcal{D}_+\Phi_m\mathcal{D}_-\Phi^\dagger_m]+\int d\theta^+\Lambda^R \Phi_m^{\dagger}\Phi_m\fstop
\end{split}
\end{equation}

\paragraph{$\mathcal{N}=(0,2)$ Fermi Multiplet}\mbox{}

\smallskip

The $\mathcal{N}=(0,2)$ Fermi multiplet contains a left-moving chiral fermion $\lambda_{-a}^c$ and an auxiliary field $G_a$. It can be expanded as
\begin{equation}\label{(0,2) Fermi}
	\Lambda^{(0,2)}_a=\lambda^c_{-a}-\theta^+G_a-i\theta^+\bar{\theta}^+D_+\lambda^c_{-a}-\bar{\theta}^+E_a^{(0,2)}(\Phi^{(0,2)}_m)\coma
\end{equation}
where $E_a^{(0,2)}\left(\Phi^{(0,2)}_m\right)$ is a holomorphic function of chiral multiplets, called $E$-term. The decomposition of an $\mathcal{N}=(0,2)$ Fermi multiplet into $\mathcal{N}=(0,1)$ multiplets is 
\begin{equation}
\renewcommand{\arraystretch}{1.1}
    \begin{array}{c}
           \mathcal{N}=(0,2) \text{ Fermi multiplet } \Lambda^{(0,2)}_a   \\
         \swarrow \qquad \searrow  \\
    \mathcal{N}=(0,1) \text{ Fermi multiplet } \Lambda_a^1 \quad  \oplus\quad  \mathcal{N}=(0,1) \text{ Fermi multiplet } \Lambda_a^2\fstop
    \end{array}
\end{equation}
The two $\mathcal{N}=(0,1)$ Fermi multiplets can be further combined into an $\mathcal{N}=(0,1)$ \emph{complex Fermi multiplet}. The decomposition of $\mathcal{N}=(0,2)$ Fermi multiplet is then 
\begin{equation}
\renewcommand{\arraystretch}{1.1}
    \begin{array}{c}
           \mathcal{N}=(0,2) \text{ Fermi multiplet } \Lambda^{(0,2)}_a   \\
         \downarrow  \\
     \mathcal{N}=(0,1) \text{ complex Fermi multiplet } \Lambda_a\fstop
    \end{array}
\end{equation}

In $\mathcal{N}=(0,2)$ theories, in addition to the $E$-term, there is another holomorphic function $J^{(0,2)a}(\Phi_m)$ of chiral fields associated to the Fermi multiplet $\Lambda^{(0,2)}_a$. The kinetic terms for the Fermi multiplet and its couplings to chiral multiplets are 
\begin{equation}\label{kinetic terms of (0,2) Fermi}
\begin{split}
\mathcal{L}_{\text{Fermi}}^{(0,2)}+\mathcal{L}^{(0,2)}_J&=-\frac{1}{2}\int d\theta^+d\bar{\theta}^+(\Lambda^{(0,2)a})^{\dagger }\Lambda^{(0,2)}_a-\frac{1}{\sqrt{2}}\int d\theta^+\Lambda^{(0,2)}_aJ^{(0,2)a}|_{\bar{\theta}^+=0}-\hc
\end{split}
\end{equation}
There is a symmetry under exchanging $J^{(0,2)a}\leftrightarrow E^{(0,2)}_a$, which corresponds to exchanging $\Lambda_a^{(0,2)}\leftrightarrow (\Lambda^{(0,2)})^{\dagger a}$.

In order to express the above Lagrangian terms for $\mathcal{N}=(0,2)$ Fermi multiplets in $\mathcal{N}=(0,1)$ superspace, we first decompose $\Phi^{(0,2)}_m$ chiral fields into $\mathcal{N}=(0,1)$ complex scalar multiplets $\Phi_m$, as in~\eqref{(0,2) chiral in (0,1) complex scalar}. Then, we introduce $\mathcal{N}=(0,1)$ complex scalar multiplets $E_a(\Phi_m)$ and $J^a(\Phi_m)$ as functions of $\Phi_m$. The field components of $E_a(\Phi_m)$ and $J^a(\Phi_m)$ are given by
\begin{equation}
\begin{split}
	E_a(\Phi_m)=E_a(\phi_m)-\theta^+\frac{\partial E_a}{\partial\phi_m}\psi_+^m,\\
	J^a(\Phi_m)=J^a(\phi_m)-\theta^+\frac{\partial J^a}{\partial \phi_m}\psi_+^m.
\end{split}
\end{equation}
where $\phi_m$ and $\psi^m_+$ are component fields of the $\mathcal{N}=(0,1)$ complex scalar multiplet $\Phi_m$. 
The terms for an $\mathcal{N}=(0,2)$ Fermi multiplet $\Lambda_a^{(0,2)}$ in the Lagrangian can then be expressed in terms of $\mathcal{N}=(0,1)$ superspace and multiplets as 
 \begin{equation}
\begin{split}
 	\mathcal{L}_{\text{Fermi}}^{(0,2)}+\mathcal{L}^{(0,2)}_J\rightarrow&\, \mathcal{L}_{F}+\int d\theta^+W^{(0,1)}\\
 	\rightarrow&\,-\frac{1}{2}\int d \theta^+(\Lambda_a\mathcal{D}_+\Lambda_a)+\\
 	&+\int d\theta^+[\Lambda_a(J^a(\Phi_m)+E^{\dagger a}(\Phi^\dagger_m))+\Lambda^{\dagger a}(E_a(\Phi_m)+J^\dagger_a(\Phi_m^\dagger))]\fstop
 \end{split}
 \end{equation}

\paragraph{$\mathcal{N}=(0,1)$ superpotential of $\mathcal{N}=(0,2)$ gauge theories}\mbox{}

\smallskip

To conclude this section, for an $\mathcal{N}=(0,2)$ field theory with vector multiplets $V^{(0,2)}_i$, chiral multiplets $\Phi^{(0,2)}_m$ and Fermi multiplets $\Lambda^{(0,2)}_a$, the generic $\mathcal{N}=(0,2)$ Lagrangian can be expressed in terms of $\mathcal{N}=(0,1)$ multiplets and superspace as 
\begin{equation}
	\mathcal{L}=\mathcal{L}_{\text{gauge}}+\mathcal{L}_s+\mathcal{L}_F+\int d \theta^+ W^{(0,1)}\coma
\end{equation}
where $\mathcal{L}_{\text{gauge}}$, $\mathcal{L}_s$ and $\mathcal{L}_{F}$ are the usual kinetic terms for vector, scalar and Fermi superfields. The $\mathcal{N}=(0,1)$ superpotential $W^{(0,1)}$ reads
\begin{equation}\label{(0,1) superpotential of (0,2) theory}
	W^{(0,1)}=\sum_{i}\sum_{n}\Lambda_{i}^R\Phi^\dagger_n\Phi_n+\sum_a\int d\theta^+[\Lambda_a(J^a(\Phi_m)+E^{\dagger a}(\Phi^\dagger_m))+\Lambda^{\dagger a}(E_a(\Phi_m)+J^\dagger_a(\Phi_m^\dagger))]\coma
\end{equation}
where the sum over $n$ in the first term means the sum over all complex scalar multiplets transforming under a given gauge group $i$.

\section{2d $\mathcal{N}=(0,1)$ Theories and Orientifolds}
\label{sec:2dN01orientifolds}

In this section, we discuss the construction of $\Spin(7)$ and $\Spin(7)$ orientifolds starting from 
CY 4-folds. We also explain the general structure of the $\mathcal{N}=(0,1)$ theories on D$1$-branes probing $\Spin(7)$ orientifolds, which are obtained from the 
$\mathcal{N}=(0,2)$ gauge theories associated to the parent CY$_4$ via a $\ZZ_2$ orientifold quotient. While we will focus on the case in which the CY$_4$ is toric, our construction applies in general. Concrete examples will be covered in Sections~\ref{sec:examplesenginN=01} to \ref{sec:HiggsingPartResol}.

\subsection{Spin(7) Cones and Spin(7) Orientifolds from CY$_4$}
\label{section_Spin(7)_from_CY4}

Our aim in this section is to set the stage for $\Spin(7)$ orientifolds probed by D$1$-branes. The construction of the corresponding gauge theories on D$1$-branes will be introduced in~\cref{sec:orientN01gaugeth,sec:HSreview,sec:N01theorfromorienquot}.

We start discussing $\Spin(7)$ manifolds, which are eight dimensional Riemannian manifolds of special holonomy group $\Spin(7)$. Every $\Spin(7)$ manifold is equipped with a globally well-defined $4$-form $\Omega^{(4)}$, called Cayley $4$-form. 

$\Spin(7)$ manifolds are interesting because they lead to minimally supersymmetric theories. For instance, consider Type IIB string theory on a $\IM_2\times \IX_8$, where $\IM_2$ is $2$d Minkowski space and $\IX_8$ is a $\Spin(7)$ manifold. The number of supercharges is broken from $32$ real supercharges to $2$, since $\Spin(7)$ manifolds preserves $1/16$ of the original supersymmetry.\footnote{For more details of why $\Spin(7)$ preserves $1/16$ SUSY, we refer the reader to \cite{Becker:2000jc} and \cite{Heckman:2018mxl}.} 

Probing the singularity of such $\Spin(7)$ manifold with a stack of $N$ D$1$-branes breaks SUSY even further. We would be left with only $1$ real supercharge on the $2$d worldvolume, hence engineering $2$d $\mathcal{N}=(0,1)$ theories.

However, in this paper we focus on an alternative, yet related, way to achieve $2$d $\mathcal{N}=(0,1)$ theories, using an orientifold construction based on the following observation. An explicit construction of $\Spin(7)$ manifolds was introduced by Joyce in \cite{Joyce:1999nk}. Start with a Calabi-Yau $4$-fold $\IM_8$ equipped with the holomorphic (4,0)-form $\Omega^{(4,0)}$ and K\"ahler form $J^{(1,1)}$. One can always define a $4$-form 
\begin{equation}\label{Cayley-form from CY4}
\Omega^{(4)}=\text{Re}\left(
\Omega^{(4,0)}\right)+\frac{1}{2}J^{(1,1)}\wedge J^{(1,1)}\coma
\end{equation}
which is stabilized by a $\Spin(7)$ subgroup of the general $\SO(8)$ holonomy of a $8$d Riemannian manifold.

We can now consider the parent CY$_4$ geometry, and perform an orientifold by $\Omega \sigma$, where $\Omega$ denotes worldsheet parity\footnote{We hope the context suffices for the reader not to confuse it with the holomorphic $4$-form.} and $\sigma$ is an anti-holomorphic involution keeping the real 4-form (\ref{Cayley-form from CY4}) invariant. It is easy to check, in analogy with the above arguments, that the supersymmetry preserved by D1-brane probing this orientifold singularity is $2$d $\mathcal{N}=(0,1)$. Hence, we refer to this construction as \textit{Spin(7) orientifolds}. One motivation for considering these orientifolds is that they naturally realize the ``real projection” of the ``complex” $\mathcal{N}=(0,2)$ theories mentioned in \cite{Gukov:2019lzi}. The theories on D1-branes probing $\Spin(7)$ cones, without the orientifold projection, are also interesting and we plan to investigate them in future work.

\subsection{Spin(7) Orientifolds in the Field Theory}
\label{sec:orientN01gaugeth}

We now discuss the field theory implementation of the $\Spin(7)$ orientifold construction. The field theory involution must act anti-holomorphically on the chiral fields of the parent gauge theory. Its connection to $\sigma$ will be addressed in Section~\ref{sec:HSreview}. Further details on the theory obtained via the orientifold projection will be given in Section~\ref{sec:N01theorfromorienquot}. The construction follows the standard orientifolding procedure. Anti-holomorphic orientifolds have appeared in the literature in other contexts, see, e.g., \cite{Blumenhagen:2000wh, Aldazabal:2000dg}.

Such involution must be a $\mathbb{Z}_2$ symmetry of the parent gauge theory, namely a symmetry of both its quiver and superpotential. Given the anti-holomorphicity of the transformation, it is convenient to write the superpotential in $\mathcal{N}=(0,1)$ language, as in \eqref{(0,1) superpotential of (0,2) theory}.

We will use indices $i,j=1,\ldots, g$, to label gauge groups in the parent theory. We will also use $\alpha_i,\beta_j=1,\ldots,N_i$ for Chan-Paton indices, equivalently (anti) fundamental color indices of $\U(N_i)$ in the gauge theory. Every bifundamental field $\Phi_{ij}$ in the gauge theory (adjoint if $i=j$) should be regarded as an $N_i \times N_j$ matrix to be contracted with the corresponding Chan-Paton factors, namely open string states are of the form $\Phi_{ij, \alpha_i \beta_j} |\alpha_i,\beta_j\rangle$. In what follows, we will keep the color/Chan-Paton indices implicit.

Below, we present the transformation properties of each type of field under the generator of the orientifold group.

\paragraph{Vector multiplets}\mbox{}

\smallskip

Gauge fields transform as follows
\begin{equation}
A^i_{\mu} \to -\gamma_{\Omega_{i'}} A^{i'T}_{\mu} \gamma_{\Omega_{i'}}^{-1} \, ,
\label{A_involution}
\end{equation}
where the transposition acts on color indices and $\gamma_{\Omega_{i}}$ is a matrix encoding the action of worldsheet parity on the Chan-Paton degrees of freedom at the node $i$. All matrices in this expression are $N_i \times N_i$ dimensional, with $N_i=N_{i'}$.

For gauge groups that are mapped to themselves, i.e., when $i=i'$, the fact that the involution squares to the identity gives rise to the standard constraint
\begin{equation}
\gamma_{\Omega_i}^T \gamma_{\Omega_i}^{-1} = \pm \ID_{N_i} \, .
\end{equation}
The two canonical solutions to this equation are the identity matrix $\gamma_{\Omega_i}=\ID_{N_i}$, for the positive sign, and the symplectic matrix $\gamma_{\Omega_i}=J=i\epsilon_{N_i/2}$, for the negative sign. Plugging each of them back into \eqref{A_involution}, they respectively lead to gauge fields in the antisymmetric or symmetric representation, namely in the adjoint representations of the resulting $\SO(N_i)$ or $\USp(N_i)$ gauge groups. The corresponding gaugino is projected accordingly, completing an $\mathcal{N}=(0,1)$ vector multiplet. Our general discussion allows for independent ranks for different gauge groups. That said, in the explicit examples considered later, we will assume that the ranks in the parent theory are such that all the ranks in the orientifolded theory are equal.

\paragraph{Scalar multiplets}\mbox{}

\smallskip

Let us consider complex $\mathcal{N}=(0,1)$ scalar fields or, equivalently, the $\mathcal{N}=(0,2)$ chiral fields in the parent theory. The anti-holomorphicity of the geometric involution implies that we have to take their Hermitian conjugate and their transformation becomes
\begin{equation}
X^m_{ij} \to \eta_{mn} \gamma_{\Omega_{i'}} \bar{X}^{n}_{i'j'} \gamma_{\Omega_{j'}}^{-1} \, ,
\label{scalar_involution}
\end{equation}
where the bar indicates conjugation. We can understand the conjugation as the net result of two operations. First, we have the transposition of the matrix $X^{n}_{i'j'}$, which effectively exchanges its two endpoints. This corresponds to the usual orientation reversal between fields and their images, which is characteristic of orientifolds and is also present in holomorphic orientifolds. In addition, we take the Hermitian conjugate, which is the matrix counterpart of the conjugation involved in the anti-holomorphic involution. This leads to an additional orientation flip.

While expressions like \eqref{scalar_involution} are rather standard, this is a good point to carefully state the meaning of each of the matrices in it. Color indices are implicit. As mentioned earlier, $\gamma_{\Omega_{i'}}$ and $\gamma_{\Omega_{j'}}$ encode the action of worldsheet parity on the color indices at nodes $i'$ and $j'$, and they are $N_{i'}\times N_{i'}$ and $N_{j'}\times N_{j'}$ matrices, respectively. $X^{n}_{i'j'}$ is an $N_{i'}\times N_{j'}$ matrix, for which the transposition and Hermitian conjugation, independently, transpose the color indices. We also include the indices $m,n=1,\ldots,n^\chi_{ij}$, with $n^\chi_{ij}$ the number of $\mathcal{N}=(0,2)$ chiral fields between nodes $i$ and $j$. $\eta$ is an $n^\chi_{ij} \times n^\chi_{ij}$ matrix corresponding to the representation of the $\mathbb{Z}_2$ group generated by the field theory involution under which the $X^{m}_{ij}$ fields transform.\footnote{In principle, this representation might be reducible. The irreducible representations of $\mathbb{Z}_2$ are either 1- or 2-dimensional.} We sum over the repeated index $n$. $i'$ and $j'$ indicate the nodes connected by the field, and are clearly not summed over. Eq.~\eqref{scalar_involution} also applies to fields that are mapped to themselves.

The condition that the orientifold action is an involution implies that $\eta \cdot \eta^T = \ID$. In the explicit examples presented later, we will mostly use $\eta=\pm 1$ (in the 1-dimensional representation case) or $\eta=\pm \left(\begin{smallmatrix} 0 & 1 \\ 1 & 0 \end{smallmatrix}\right)$, which implements a non-trivial exchange between two pairs of fields. In view of this, from now on we will reduce $\eta_{mn} \bar{X}^{n}_{i'j'}$ to  $\pm \bar{X}^{m'}_{i'j'}$, in order to simplify expressions.

The transformation \eqref{scalar_involution} and the ones for Fermi superfields that we present below, simplify considerably in the case of Abelian parents. As usual, this is sufficient for connecting the gauge theories to the probed geometries, along the lines that will be discussed in Section~\ref{sec:N01theorfromorienquot}.

\paragraph{Fermi multiplets}\mbox{}

\smallskip

Contrary to scalar fields, whose transformation always involves conjugation in order to account for the anti-holomorphicity of the geometric involution, Fermi fields may or may not be conjugated. 

Let us first consider the $\mathcal{N}=(0,1)$ complex Fermi multiplets in the parent, i.e. the $\mathcal{N}=(0,2)$ Fermi multiplets in the original theory. Their transformation is either\footnote{Here we use the simplified notation introduced earlier in the case of scalar multiplets, instead of including an $\eta$ matrix as in \eqref{scalar_involution}.}
\begin{equation}
\Lambda^m_{ij} \to \pm \gamma_{\Omega_{i'}} \bar{\Lambda}^{m'}_{i'j'} \gamma_{\Omega_{j'}}^{-1}\coma
\label{Lambda_involution-antih}
\end{equation}
or
\begin{equation}
\Lambda^m_{ij} \to \pm \gamma_{\Omega_{i'}} \Lambda^{m' \, T}_{j'i'} \gamma_{\Omega_{j'}}^{-1}\fstop
\label{Lambda_involution-h}
\end{equation}
Notice that the second transformation only involves transposition, without complex conjugation.

The signs and the presence or absence of complex conjugation in the transformations of each Fermi in \cref{Lambda_involution-antih,Lambda_involution-h} are determined by imposing the transformation of the chirals and requiring the invariance of the superpotential $W^{(0,1)}$ of the parent theory. As mentioned earlier, focusing on the Abelian theory is sufficient for this. 

The decomposition of $\mathcal{N}=(0,2)$ vector multiplets gives rise to additional $\mathcal{N}=(0,1)$ adjoint Fermi fields $\Lambda_i^R$, as explained in Section~\ref{sec:N02inN01Form}. Invariance of $W^{(0,1)}$ in the parent fully determines the transformation of the $\Lambda^R_{i}$, which is given by
\begin{equation}
\Lambda_i^R \to \gamma_{\Omega_{i'}} \Lambda_{i'}^{R \, \, T} \gamma_{\Omega_{i'}}^{-1} \, .
\label{Lambda^R_projection}
\end{equation}
The relative sign between \eqref{A_involution} and \eqref{Lambda^R_projection} implies that for $i=i'$, an $\SO$ or $\USp$ projection of the gauge group is correlated with a projection of $\Lambda_i^R$ into a symmetric or antisymmetric representation, respectively.

The construction of the $\Spin(7)$ orientifolds we have just presented exclusively uses information from the gauge theory. In coming sections, we will explain how it can be connected to the geometry.\footnote{In the case of toric CY$_4$, perfect matchings of the corresponding brane brick models are powerful tools in connecting gauge theory and geometry \cite{Franco:2015tya}. It is therefore natural to ask whether and, if so, how the anti-holomorphic involution translates into perfect matchings. Preliminary investigations suggest that, at least, the involution of chiral fields maps to an anti-holomorphic involution of the perfect matchings. It would be interesting to study this question in the future.} The anti-holomorphic involution of the generators of the parent CY$_4$ geometry can be mapped to an action on scalars. This, combined with the invariance of the parent superpotential, determines the transformation of the Fermi superfields.

\subsection{Orientifold Projection of the Quiver}
\label{sec:N01theorfromorienquot}

\subsubsection*{Quiver}

In this section we explicitly discuss all possible orientifold projections of the quiver following from the rules in Section~\ref{sec:orientN01gaugeth}. The different types of $\mathcal{N}=(0,1)$ superfields, combined with their various transformations, lead to several possibilities.

\subsubsection*{Gauge groups}

\begin{figure}[!htp]
    \centering
    \begin{subfigure}[t]{0.49\textwidth}
    \centering
    \begin{tikzpicture}[scale=2.1]
    \node (L) at (0,0) {
    \begin{tikzpicture}[scale=2]
    	\node[draw=black,line width=1pt,circle,fill=yellowX,minimum width=0.75cm,inner sep=1pt,label={[yshift=-1.75cm]:$\U\left(N_i\right)$}] (A) at (0,0) {$i$};
    	\node[draw=black,line width=1pt,circle,fill=yellowX,minimum width=0.75cm,inner sep=1pt,label={[yshift=-1.75cm]:$\U\left(N_{i'}\right)$}] (B) at (1.25,0) {$i'$};
    	\draw[line width=1pt,postaction={draw,redX,dash pattern= on 3pt off 5pt,dash phase=4pt}] [line width=1pt,black,dash pattern= on 3pt off 5pt] (A) -- (0.35,0.35);
    	\draw[line width=1pt,postaction={draw,redX,dash pattern= on 3pt off 5pt,dash phase=4pt}] [line width=1pt,black,dash pattern= on 3pt off 5pt] (A) -- (0.35,-0.35);
    	\draw[line width=1pt,postaction={draw,redX,dash pattern= on 3pt off 5pt,dash phase=4pt}] [line width=1pt,black,dash pattern= on 3pt off 5pt] (B) -- (0.95,0.35);
    	\draw[line width=1pt,postaction={draw,redX,dash pattern= on 3pt off 5pt,dash phase=4pt}] [line width=1pt,black,dash pattern= on 3pt off 5pt] (B) -- (0.95,-0.35);
    	\node at (0.625,0) {$\cdots$};
    \end{tikzpicture}
    };
    \node (R) at (2,0) {
    \begin{tikzpicture}[scale=2]
    	\node[draw=black,line width=1pt,circle,fill=yellowX,minimum width=0.75cm,inner sep=1pt,label={[yshift=-1.75cm]:$\U\left(N_{i}\right)$}] (C) at (0,0) {$i$};
    	\draw[line width=1pt,postaction={draw,redX,dash pattern= on 3pt off 5pt,dash phase=4pt}] [line width=1pt,black,dash pattern= on 3pt off 5pt] (C) -- (0.35,0.35);
    	\draw[line width=1pt,postaction={draw,redX,dash pattern= on 3pt off 5pt,dash phase=4pt}] [line width=1pt,black,dash pattern= on 3pt off 5pt] (C) -- (0.35,-0.35);
    \end{tikzpicture}
    };
    \draw[-Triangle,blueX,line width=1mm] (1,0.17) -- (1.5,0.17);
    	\end{tikzpicture}    
    \caption{Two nodes mapped  
    according to Rule~\ref{rule:1a}.}
    	\label{fig:rule1a}
    \end{subfigure}\hfill
    \begin{subfigure}[t]{0.49\textwidth}
    	\centering 
    	\begin{tikzpicture}[scale=2.1]
    	\node (L) at (0,0) {
    	\begin{tikzpicture}[scale=2]
    		\node[draw=black,line width=1pt,circle,fill=yellowX,minimum width=0.75cm,inner sep=1pt,label={[yshift=-1.75cm]:$\U\left(N_i\right)$}] (B) at (0,0) {$i$};
    	\draw[line width=1pt,postaction={draw,redX,dash pattern= on 3pt off 5pt,dash phase=4pt}] [line width=1pt,black,dash pattern= on 3pt off 5pt] (B) -- (0.35,0.35);
    	\draw[line width=1pt,postaction={draw,redX,dash pattern= on 3pt off 5pt,dash phase=4pt}] [line width=1pt,black,dash pattern= on 3pt off 5pt] (B) -- (0.35,-0.35);
    	\end{tikzpicture}
    	};
    	\node (R) at (2,0) {
    		\begin{tikzpicture}[scale=2]
    	\node[draw=black,line width=1pt,circle,fill=yellowX,minimum width=0.75cm,inner sep=1pt,label={[yshift=-1.75cm]:$G_i(N_i)$}] (C) at (0,0) {$i$};
    	\draw[line width=1pt,postaction={draw,redX,dash pattern= on 3pt off 5pt,dash phase=4pt}] [line width=1pt,black,dash pattern= on 3pt off 5pt] (C) -- (0.35,0.35);
    	\draw[line width=1pt,postaction={draw,redX,dash pattern= on 3pt off 5pt,dash phase=4pt}] [line width=1pt,black,dash pattern= on 3pt off 5pt] (C) -- (0.35,-0.35);
    	\end{tikzpicture}
    	};
    	\draw[-Triangle,blueX,line width=1mm] (0.75,0.17) -- (1.25,0.17);
    	\end{tikzpicture}
    	\caption{A node mapped according to Rule~\ref{rule:1b}.}
    	\label{fig:rule1b}
    \end{subfigure}
    \caption{The two possible identifications of gauge groups. The group $G_i(N_i)$ can be either $\SO(N_i)$ or $\USp(N_i)$. Dashed black and red lines represent fields that can be either scalar or Fermi fields.}
    \label{fig:gaugegroupident}
\end{figure}
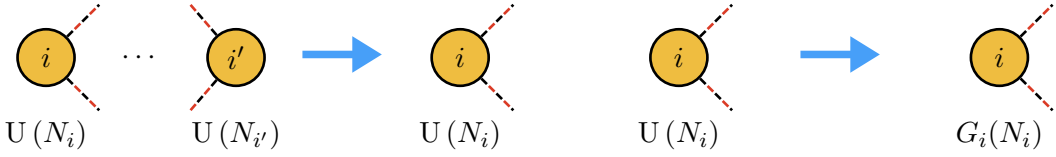

The orientifold projections for gauge groups can be one of the following two possibilities:

\begin{enumerate}[label=1\alph*.,ref=1\alph*]
\item\label{rule:1a} Every node $i\neq i'$ gives rise to a gauge factor $\U(N_i)$, as shown in Figure~\ref{fig:rule1a}. 
\item\label{rule:1b} Every node $i=i'$ gives rise to a gauge factor $\SO(N_i)$ or $\USp(N_i)$, for $\gamma_{\Omega_i}=\ID$ or $J$, respectively, as schematically shown in Figure~\ref{fig:rule1b}. 
\end{enumerate}

\subsubsection*{Matter fields}

\bigskip

\noindent\underline{$\mathcal{N}=(0,2)$ Chiral and Fermi fields}

\medskip

We start with the projection of $\mathcal{N}=(0,2)$ chiral and Fermi multiplets, equivalently $\mathcal{N}=(0,1)$ complex scalar and Fermi multiplets. Unless explicitly mentioned, the rules below apply to both scalar and Fermi fields. In figures, we will use dashed black and red lines to indicate fields that can be of the two types and we use $\mathcal{N}=(0,1)$ language. To organize the presentation, we will distinguish between the case in which a field is mapped to a different image and when it is mapped to itself.

\bigskip 

\begin{center}
{\it Fields mapped to other fields}
\end{center}

\smallskip

The two rules that follow apply to both to fields transforming anti-holomorphically, as in \eqref{scalar_involution} and \eqref{Lambda_involution-antih}, or holomorphically, as in \eqref{Lambda_involution-h}. While the resulting quiver does not depend on the presence of conjugation, such details do affect how the final fields precisely emerge from the original theory and, therefore, the projection of the superpotential. 

\begin{enumerate}[label=2\alph*.,ref=2\alph*]
\item\label{rule:2a} Consider a bifundamental or adjoint field $X_{ij}$ of the parent theory, for $j \neq i'$, which transforms into (the conjugate of) a different image field $X_{i'j'}$. The two fields, $X_{ij}$ and $X_{i'j'}$, are projected down to a single complex bifundamental (or adjoint) $X_{ij}$.\footnote{There is no distinction between $\fund$ and $\antifund$ whenever the resulting gauge group is $\SO$ or $\USp$.} Various possibilities are shown in Figure~\ref{fig:rule2a}.
\end{enumerate} 

\begin{figure}[H]
\centering
    \begin{subfigure}[t]{\textwidth}
    \vspace*{-2.7cm} 
    \centering
    \begin{tikzpicture}[scale=3]
    \draw[help lines,white] (-1,0) grid (2,0);
    \node (L) at (0,0) {
    \begin{tikzpicture}[scale=2]
    	\node[draw=black,line width=1pt,circle,fill=yellowX,minimum width=0.75cm,inner sep=1pt,label={[xshift=-1cm,yshift=-0.5cm]:$\U\left(N_i\right)$}] (A) at (0,0.5) {$i$};
    	\node[draw=black,line width=1pt,circle,fill=yellowX,minimum width=0.75cm,inner sep=1pt,label={[xshift=-1cm,yshift=-1cm]:$\U\left(N_j\right)$}] (B) at (0,-0.5) {$j$};
    	\node[draw=black,line width=1pt,circle,fill=yellowX,minimum width=0.75cm,inner sep=1pt,label={[xshift=1cm,yshift=-0.5cm]:$\U\left(N_{i'}\right)$}] (C) at (1,0.5) {$i'$};
    	\node[draw=black,line width=1pt,circle,fill=yellowX,minimum width=0.75cm,inner sep=1pt,label={[xshift=1.1cm,yshift=-1cm]:$\U\left(N_{j'}\right)$}] (D) at (1,-0.5) {$j'$};
    	\draw[line width=1pt,postaction={draw,redX,dash pattern= on 3pt off 5pt,dash phase=4pt}] [line width=1pt,black,dash pattern= on 3pt off 5pt] (A) -- (0.25,0.75);
    	\draw[line width=1pt,postaction={draw,redX,dash pattern= on 3pt off 5pt,dash phase=4pt}] [line width=1pt,black,dash pattern= on 3pt off 5pt] (B) -- (0.25,-0.75);
    	\draw[line width=1pt,postaction={draw,redX,dash pattern= on 3pt off 5pt,dash phase=4pt}] [line width=1pt,black,dash pattern= on 3pt off 5pt] (C) -- (0.75,0.75);
    	\draw[line width=1pt,postaction={draw,redX,dash pattern= on 3pt off 5pt,dash phase=4pt}] [line width=1pt,black,dash pattern= on 3pt off 5pt] (D) -- (0.75,-0.75);
    	\node at (0.5,0.5) {$\ldots$};
    	\node at (0.5,-0.5) {$\ldots$};
    	\draw[line width=1pt,postaction={draw,redX,dash pattern= on 3pt off 5pt,dash phase=4pt}] [line width=1pt,black,dash pattern= on 3pt off 5pt] (A)  --node[left,midway,xshift=-0.5cm]{$X_{ij}$} node[fill=white,text opacity=1,fill opacity=1,draw=black,rectangle,thin,dash pattern= on 0pt off 0pt] {$2$} (B);
    	\draw[line width=1pt,postaction={draw,redX,dash pattern= on 3pt off 5pt,dash phase=4pt}] [line width=1pt,black,dash pattern= on 3pt off 5pt] (C) -- node[fill=white,text opacity=1,fill opacity=1,draw=black,rectangle,thin,dash pattern= on 0pt off 0pt] {$2$} node[right,midway,xshift=0.5cm]{$X_{i'j'}$} (D);
    \end{tikzpicture}
    };
    \node (R) at (2,0) {
    \begin{tikzpicture}[scale=2]
    	\node[draw=black,line width=1pt,circle,fill=yellowX,minimum width=0.75cm,inner sep=1pt,label={[xshift=1cm,yshift=-0.4cm]:$\U(N_{i})$}] (E) at (0,0.5) {$i$};
    	\node[draw=black,line width=1pt,circle,fill=yellowX,minimum width=0.75cm,inner sep=1pt,label={[xshift=1.1cm,yshift=-1.2cm]:$\U\left(N_{j}\right)$}] (F) at (0,-0.5) {$j$};
    	\draw[line width=1pt,postaction={draw,redX,dash pattern= on 3pt off 5pt,dash phase=4pt}] [line width=1pt,black,dash pattern= on 3pt off 5pt] (E) --node[right,midway,xshift=0.3cm]{$X_{ij}$} node[fill=white,text opacity=1,fill opacity=1,draw=black,rectangle,thin,dash pattern= on 0pt off 0pt] {$2$} (F);
    \end{tikzpicture}
    };
    \draw[-Triangle,blueX,line width=1mm] (1,0) -- (1.35,0);
    	\end{tikzpicture}
    	\caption{Pairs of bifundamentals that do not share any node.}
    	\label{fig:two_bifundamentals_four_groups}
    \end{subfigure}
    \begin{subfigure}[t]{\textwidth}
    \vspace*{0.7cm} 
    \centering
    	\begin{tikzpicture}[scale=3]
    	\draw[help lines,white] (-1,0) grid (2,0);
    	\node (L) at (0,0) {
    \begin{tikzpicture}[scale=2]
     	\node[draw=black,line width=1pt,circle,fill=yellowX,minimum width=0.75cm,inner sep=1pt,label={[xshift=-1cm,yshift=-0.5cm]:$\U\left(N_i\right)$}] (A) at (0.5,0.5) {$i$};
    	\node[draw=black,line width=1pt,circle,fill=yellowX,minimum width=0.75cm,inner sep=1pt,label={[xshift=-1cm,yshift=-1cm]:$\U\left(N_j\right)$}] (B) at (0,-0.5) {$j$};
    	\node[draw=black,line width=1pt,circle,fill=yellowX,minimum width=0.75cm,inner sep=1pt,label={[xshift=1.1cm,yshift=-1cm]:$\U\left(N_{j'}\right)$}] (C) at (1,-0.5) {$j'$};
    	\draw[line width=1pt,postaction={draw,redX,dash pattern= on 3pt off 5pt,dash phase=4pt}] [line width=1pt,black,dash pattern= on 3pt off 5pt] (B) -- (0.25,-0.75);
    	\draw[line width=1pt,postaction={draw,redX,dash pattern= on 3pt off 5pt,dash phase=4pt}] [line width=1pt,black,dash pattern= on 3pt off 5pt] (C) -- (0.75,-0.75);
    	\node at (0.5,-0.5) {$\ldots$};
    	\draw[line width=1pt,postaction={draw,redX,dash pattern= on 3pt off 5pt,dash phase=4pt}] [line width=1pt,black,dash pattern= on 3pt off 5pt] (A) --node[left,midway,xshift=-0.25cm,yshift=0.25cm]{$X_{ij}$} node[fill=white,text opacity=1,fill opacity=1,draw=black,rectangle,thin,dash pattern= on 0pt off 0pt] {$2$} (B);
    	\draw[line width=1pt,postaction={draw,redX,dash pattern= on 3pt off 5pt,dash phase=4pt}] [line width=1pt,black,dash pattern= on 3pt off 5pt] (A) --node[right,midway,xshift=0.25cm,yshift=0.25cm]{$X_{ij'}$} node[fill=white,text opacity=1,fill opacity=1,draw=black,rectangle,thin,dash pattern= on 0pt off 0pt] {$2$} (C);
    \end{tikzpicture}
    };
    \node (R) at (2,0) {
    \begin{tikzpicture}[scale=2]
    	\node[draw=black,line width=1pt,circle,fill=yellowX,minimum width=0.75cm,inner sep=1pt,label={[xshift=1.1cm,yshift=-0.4cm]:$G_i(N_i)$}] (E) at (0,0.5) {$i$};
    	\node[draw=black,line width=1pt,circle,fill=yellowX,minimum width=0.75cm,inner sep=1pt,label={[xshift=1.1cm,yshift=-1.2cm]:$\U\left(N_{j}\right)$}] (F) at (0,-0.5) {$j$};
    	\draw[line width=1pt,postaction={draw,redX,dash pattern= on 3pt off 5pt,dash phase=4pt}] [line width=1pt,black,dash pattern= on 3pt off 5pt] (E) --node[right,midway,xshift=0.3cm]{$X_{ij}$} node[fill=white,text opacity=1,fill opacity=1,draw=black,rectangle,thin,dash pattern= on 0pt off 0pt] {$2$} (F);
    \end{tikzpicture}
    };
    \draw[-Triangle,blueX,line width=1mm] (1,0) -- (1.35,0);
    	\end{tikzpicture}
    	\caption{Pairs of bifundamentals with a common node.}
    	\label{fig:two_bifundamentals_three_groups}
    \end{subfigure}
    \begin{subfigure}[t]{\textwidth}
    \vspace*{0.7cm} 
    \centering
    	\begin{tikzpicture}[scale=3]
    	\draw[help lines,white] (-1,0) grid (2,0);
    	\node (L) at (-0.05,0) {
    \begin{tikzpicture}[scale=2]
    	\node[draw=black,line width=1pt,circle,fill=yellowX,minimum width=0.75cm,inner sep=1pt,label={[xshift=-1cm,yshift=-0.5cm]:$\U\left(N_i\right)$}] (A) at (0,0.75) {$i$};
    	\node[draw=black,line width=1pt,circle,fill=yellowX,minimum width=0.75cm,inner sep=1pt,label={[xshift=-1cm,yshift=-1cm]:$\U\left(N_j\right)$}] (B) at (0,-0.75) {$j$};
    	\draw[line width=1pt,postaction={draw,redX,dash pattern= on 3pt off 5pt,dash phase=4pt}] [line width=1pt,black,dash pattern= on 3pt off 5pt] (A) to[bend left=20] node[fill=white,text opacity=1,fill opacity=1,draw=black,rectangle,thin,dash pattern= on 0pt off 0pt] {$2$} node[right,midway,xshift=0.3cm] {$Y_{ij}$} (B);
    	\draw[line width=1pt,postaction={draw,redX,dash pattern= on 3pt off 5pt,dash phase=4pt}] [line width=1pt,black,dash pattern= on 3pt off 5pt] (A) to[bend right=20] node[fill=white,text opacity=1,fill opacity=1,draw=black,rectangle,thin,dash pattern= on 0pt off 0pt] {$2$} node[left,midway,xshift=-0.3cm] {$X_{ij}$} (B);
    \end{tikzpicture}
    };
    \node (R) at (2,0) {
    \begin{tikzpicture}[scale=2]
    	\node[draw=black,line width=1pt,circle,fill=yellowX,minimum width=0.75cm,inner sep=1pt,label={[xshift=1.1cm,yshift=-0.4cm]:$G_i\left(N_i\right)$}] (C) at (0,0.75) {$i$};
    	\node[draw=black,line width=1pt,circle,fill=yellowX,minimum width=0.75cm,inner sep=1pt,label={[xshift=1.1cm,yshift=-1.2cm]:$G_j\left(N_j\right)$}] (D) at (0,-0.75) {$j$};
    	\draw[line width=1pt,postaction={draw,redX,dash pattern= on 3pt off 5pt,dash phase=4pt}] [line width=1pt,black,dash pattern= on 3pt off 5pt] (C) -- node[fill=white,text opacity=1,fill opacity=1,draw=black,rectangle,thin,dash pattern= on 0pt off 0pt] {$2$} node[right,midway,xshift=0.3cm] {$X_{ij}$} (D);
    \end{tikzpicture}
    };
    \draw[-Triangle,blueX,line width=1mm] (1,0) -- (1.35,0);
    	\end{tikzpicture}
    	\caption{Pairs of bifundamentals sharing both nodes.}
    	\label{fig:two-bifundamentals_two_groups}
    \end{subfigure}
    \end{figure}
\begin{figure}[H]\ContinuedFloat
    \begin{subfigure}[t]{\textwidth}
    \centering
    \begin{tikzpicture}[scale=3]
    \draw[help lines,white] (-1,0) grid (2,0);
    \node (L) at (-0.3,0) {
    \begin{tikzpicture}[scale=2]
    \draw[line width=1pt,postaction={draw,redX,dash pattern= on 3pt off 5pt,dash phase=4pt}] [line width=1pt,black,dash pattern= on 3pt off 5pt] (0,0.75) circle (0.25) node[xshift=0.85cm] {$X_{ii}$} node[fill=white,text opacity=1,fill opacity=1,draw=black,rectangle,yshift=0.5cm,thin,dash pattern= on 0pt off 0pt] {$2$};
    \draw[line width=1pt,postaction={draw,redX,dash pattern= on 3pt off 5pt,dash phase=4pt}] [line width=1pt,black,dash pattern= on 3pt off 5pt] (0,-0.75) circle (0.25) node[xshift=0.9cm] {$X_{i'i'}$} node[fill=white,text opacity=1,fill opacity=1,draw=black,rectangle,yshift=-0.5cm,thin,dash pattern= on 0pt off 0pt] {$2$};
    \node[draw=black,line width=1pt,circle,fill=yellowX,minimum width=0.75cm,inner sep=1pt,label={[xshift=-1cm,yshift=-0.7cm]:$\U\left(N_i\right)$}] (A) at (0,0.5) {$i$};
    	\node[draw=black,line width=1pt,circle,fill=yellowX,minimum width=0.75cm,inner sep=1pt,label={[xshift=-1cm,yshift=-0.7cm]:$\U\left(N_{i'}\right)$}] (B) at (0,-0.5) {$i'$};
    	\draw[line width=1pt,postaction={draw,redX,dash pattern= on 3pt off 5pt,dash phase=4pt}] [line width=1pt,black,dash pattern= on 3pt off 5pt] (A) -- (-0.25,0.25);
    	\draw[line width=1pt,postaction={draw,redX,dash pattern= on 3pt off 5pt,dash phase=4pt}] [line width=1pt,black,dash pattern= on 3pt off 5pt] (A) -- (0.25,0.25);
    	\draw[line width=1pt,postaction={draw,redX,dash pattern= on 3pt off 5pt,dash phase=4pt}] [line width=1pt,black,dash pattern= on 3pt off 5pt] (B) -- (-0.25,-0.25);
    	\draw[line width=1pt,postaction={draw,redX,dash pattern= on 3pt off 5pt,dash phase=4pt}] [line width=1pt,black,dash pattern= on 3pt off 5pt] (B) -- (0.25,-0.25);
    	\node at (0,0) {$\vdots$};
    \end{tikzpicture}
    };
    \node (R) at (1.6,0) {
    \begin{tikzpicture}[scale=2]
	\draw[line width=1pt,postaction={draw,redX,dash pattern= on 3pt off 5pt,dash phase=4pt}] [line width=1pt,black,dash pattern= on 3pt off 5pt] (0.25,0) circle (0.25) node[fill=white,text opacity=1,fill opacity=1,draw=black,rectangle,xshift=0.5cm,thin,dash pattern= on 0pt off 0pt] {$2$} node[yshift=0.75cm] {$X_{ii}$};  
	\node[draw=black,line width=1pt,circle,fill=yellowX,minimum width=0.75cm,inner sep=1pt,label={[xshift=-0.25cm,yshift=-1.5cm]:$\U\left(N_{i}\right)$}] (A) at (0,0) {$i$};
    \end{tikzpicture}
    };
    \draw[-Triangle,blueX,line width=1mm] (0.82,0) -- (1.17,0);
    \end{tikzpicture}
    	\caption{A pair of adjoint fields whose nodes are mapped to each other.}
    	\label{fig:two_adjoint_two_groups}
    \end{subfigure}
    \caption{Various instances of Rule~\ref{rule:2a}. 
    These pictures apply to both fields that are mapped anti-holomorphically (via \eqref{scalar_involution} or \eqref{Lambda_involution-antih}) or holomorphically (via \eqref{Lambda_involution-h}). The group $G_i(N_i)$ can be either $\SO(N_i)$ or $\USp(N_i)$.}
    \label{fig:rule2a}
\end{figure}
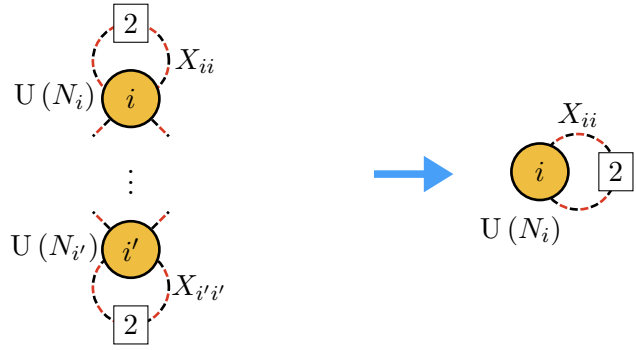 
 
 \begin{enumerate}[label=2\alph*.,ref=2\alph*,resume]
\item\label{rule:2b} Consider two bifundamental or adjoint fields $X_{ii'}$ and $Y_{i'i}$, which transform into (the conjugate of) each other. They give rise to two complex fields, one in the symmetric representation and the other one in the antisymmetric representation of the resulting unitary (for $i\neq i'$) or $\SO/\USp$ (for $i=i'$) node.\footnote{We thank Massimo Porrati for discussions on this point.} From now on, we indicate symmetric and antisymmetric representations with star and diamond symbols, respectively. This rule is illustrated in Figure~\ref{fig:rule2b}.
\end{enumerate} 

\begin{figure}[H]
    \centering
    \begin{subfigure}[t]{\textwidth}
    \vspace*{-0.6cm} 
    \centering
    \begin{tikzpicture}[scale=3]
    \draw[help lines,white] (-1,0) grid (2,0);
    \node (L) at (0,0) {
    \begin{tikzpicture}[scale=2]
    	\node[draw=black,line width=1pt,circle,fill=yellowX,minimum width=0.75cm,inner sep=1pt,label={[yshift=-1.75cm]:$\U\left(N_i\right)$}] (A) at (0,0) {$i$};
    	\node[draw=black,line width=1pt,circle,fill=yellowX,minimum width=0.75cm,inner sep=1pt,label={[yshift=-1.75cm]:$\U\left(N_{i'}\right)$}] (B) at (1.5,0) {$i'$};
    	\draw[line width=1pt,postaction={draw,redX,dash pattern= on 3pt off 5pt,dash phase=4pt}] [line width=1pt,black,dash pattern= on 3pt off 5pt] (A) to[bend left=20] node[fill=white,text opacity=1,fill opacity=1,draw=black,rectangle,thin,dash pattern= on 0pt off 0pt] {$2$} node[above,midway,yshift=0.3cm] {$X_{ii'}$} (B);
    	\draw[line width=1pt,postaction={draw,redX,dash pattern= on 3pt off 5pt,dash phase=4pt}] [line width=1pt,black,dash pattern= on 3pt off 5pt] (A) to[bend right=20] node[fill=white,text opacity=1,fill opacity=1,draw=black,rectangle,thin,dash pattern= on 0pt off 0pt] {$2$} node[below,midway,yshift=-0.3cm] {$X_{i'i}$} (B);
    	\end{tikzpicture}
    };
    \node (R) at (2,0.05) {
    \begin{tikzpicture}[scale=2]
	\draw[line width=1pt,postaction={draw,redX,dash pattern= on 3pt off 5pt,dash phase=4pt}] [line width=1pt,black,dash pattern= on 3pt off 5pt] (4,0) circle (0.45) node[yshift=0.90cm,xshift=0cm,star,star points=5, star point ratio=2.25, inner sep=1pt, fill=black, draw,dash pattern= on 0pt off 0pt] {}  node[fill=white,text opacity=1,fill opacity=1,draw=black,rectangle,xshift=0.85cm,thin,dash pattern= on 0pt off 0pt] {$2$} node[yshift=1.2cm,xshift=0.5cm] {$X_{iiS}$};
	\draw[line width=1pt,postaction={draw,redX,dash pattern= on 3pt off 5pt,dash phase=4pt}] [line width=1pt,black,dash pattern= on 3pt off 5pt] (3.75,0) circle (0.25)  node[yshift=0.5cm] {\scriptsize{$\quadro$}} node[fill=white,text opacity=1,fill opacity=1,draw=black,rectangle,xshift=0.45cm,thin,dash pattern= on 0pt off 0pt] {$2$} node[yshift=1.2cm,xshift=0cm] {$X_{iiA}$};
	\node[draw=black,line width=1pt,circle,fill=yellowX,minimum width=0.75cm,inner sep=1pt,label={[xshift=-0.5cm,yshift=-1.5cm]:$\U\left(N_{i}\right)$}] (A) at (3.5,0) {$i$};
    	\end{tikzpicture}
    };
    \draw[-Triangle,blueX,line width=1mm] (0.82,0) -- (1.17,0);
    \end{tikzpicture}
	\caption{Two bifundamental fields connecting a node and its image.}
        	\label{fig:two_bifundamentals_one_group}
    \end{subfigure}  \end{figure}%
\begin{figure}[H]\ContinuedFloat
    \begin{subfigure}[t]{\textwidth}
    \vspace*{-0.4cm} 
    \centering
    \begin{tikzpicture}[scale=3]
    \draw[help lines,white] (-1,0) grid (2,0);
    \node (L) at (-0.05,0.05) {\begin{tikzpicture}[scale=2]
    \draw[line width=1pt,postaction={draw,redX,dash pattern= on 3pt off 5pt,dash phase=4pt}] [line width=1pt,black,dash pattern= on 3pt off 5pt] (1.05,0) circle (0.45)  node[fill=white,text opacity=1,fill opacity=1,draw=black,rectangle,xshift=-0.85cm,thin,dash pattern= on 0pt off 0pt] {$2$} node[yshift=1.2cm,xshift=-0.5cm] {$X_{ii}$};
	\draw[line width=1pt,postaction={draw,redX,dash pattern= on 3pt off 5pt,dash phase=4pt}] [line width=1pt,black,dash pattern= on 3pt off 5pt] (1.25,0) circle (0.25)  node[fill=white,text opacity=1,fill opacity=1,draw=black,rectangle,xshift=-0.5cm,thin,dash pattern= on 0pt off 0pt] {$2$} node[yshift=1.2cm] {$Y_{ii}$};
	\node[draw=black,line width=1pt,circle,fill=yellowX,minimum width=0.75cm,inner sep=1pt,label={[xshift=0.5cm,yshift=-1.5cm]:$\U\left(N_i\right)$}] (A) at (1.5,0) {$i$};
    \end{tikzpicture}
    };
    \node (R) at (2,0.05) {
    \begin{tikzpicture}[scale=2]
	\draw[line width=1pt,postaction={draw,redX,dash pattern= on 3pt off 5pt,dash phase=4pt}] [line width=1pt,black,dash pattern= on 3pt off 5pt] (4,0) circle (0.45) node[yshift=0.90cm,xshift=0cm,star,star points=5, star point ratio=2.25, inner sep=1pt, fill=black, draw,dash pattern= on 0pt off 0pt] {}  node[fill=white,text opacity=1,fill opacity=1,draw=black,rectangle,xshift=0.85cm,thin,dash pattern= on 0pt off 0pt] {$2$} node[yshift=1.2cm,xshift=0.5cm] {$X_{iiS}$};
	\draw[line width=1pt,postaction={draw,redX,dash pattern= on 3pt off 5pt,dash phase=4pt}] [line width=1pt,black,dash pattern= on 3pt off 5pt] (3.75,0) circle (0.25)  node[yshift=0.5cm] {\scriptsize{$\quadro$}} node[fill=white,text opacity=1,fill opacity=1,draw=black,rectangle,xshift=0.45cm,thin,dash pattern= on 0pt off 0pt] {$2$} node[yshift=1.2cm,xshift=0cm] {$X_{iiA}$};
	\node[draw=black,line width=1pt,circle,fill=yellowX,minimum width=0.75cm,inner sep=1pt,label={[xshift=-0.5cm,yshift=-1.5cm]:$G_i\left(N_{i}\right)$}] (A) at (3.5,0) {$i$};
    \end{tikzpicture}
    };
    \draw[-Triangle,blueX,line width=1mm] (0.82,0) -- (1.17,0);
    \end{tikzpicture}
    	\caption{Two adjoint fields sharing on a node that is mapped to itself.}
    	\label{fig:two_adjoint_one_group}
    \end{subfigure}
     \caption{The two instances of Rule~\ref{rule:2b}, depending on whether the original fields are bifundamental ($i\neq i'$) or adjoint ($i=i'$). This picture applies to both fields that are mapped anti-holomorphically (via \eqref{scalar_involution} or \eqref{Lambda_involution-antih}) or holomorphically (via \eqref{Lambda_involution-h}). The group $G_i(N_i)$ can be either $\SO(N_i)$ or $\USp(N_i)$.}
    \label{fig:rule2b}
\end{figure}

\bigskip 

\begin{center}
{\it Fields mapped to themselves}
\end{center}

\smallskip

In this case, the transformation of the quiver depends crucially on whether the map is anti-holomorphic or holomorphic. Therefore, in the figures we indicate it over the arrow connecting the parent to the orientifolded theory.

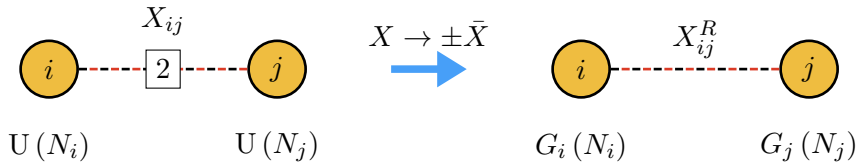
\begin{figure}[H]
\centering
    	\begin{tikzpicture}[scale=2]
    	\begin{scope}
    	\node[draw=black,line width=1pt,circle,fill=yellowX,minimum width=0.75cm,inner sep=1pt,label={[yshift=-1.75cm]:$\U\left(N_i\right)$}] (A) at (0,0) {$i$};
    	\node[draw=black,line width=1pt,circle,fill=yellowX,minimum width=0.75cm,inner sep=1pt,label={[yshift=-1.75cm]:$\U\left(N_j\right)$}] (B) at (1.5,0) {$j$};
    	\draw[line width=1pt,postaction={draw,redX,dash pattern= on 3pt off 5pt,dash phase=4pt}] [line width=1pt,black,dash pattern= on 3pt off 5pt] (A) -- node[above,midway,yshift=0.3cm] {$X_{ij}$} node[fill=white,text opacity=1,fill opacity=1,draw=black,rectangle,thin,dash pattern= on 0pt off 0pt] {$2$} (B);
    	\node[draw=black,line width=1pt,circle,fill=yellowX,minimum width=0.75cm,inner sep=1pt,label={[yshift=-1.75cm]:$G_i\left(N_i\right)$}] (C) at (3.5,0) {$i$};
    	\node[draw=black,line width=1pt,circle,fill=yellowX,minimum width=0.75cm,inner sep=1pt,label={[yshift=-1.75cm]:$G_j\left(N_j\right)$}] (D) at (5,0) {$j$};
    	\draw[line width=1pt,postaction={draw,redX,dash pattern= on 3pt off 5pt,dash phase=4pt}] [line width=1pt,black,dash pattern= on 3pt off 5pt] (C) -- node[above,midway] {$X_{ij}^R$} (D);
    	\draw[-Triangle,blueX,line width=1mm] (2.25,0) -- node[above,midway,yshift=0.1cm] {\color{black}{$X\rightarrow \pm \bar{X}$}} (2.75,0);
    	\end{scope}
    	\end{tikzpicture}
        \caption{Rule~\ref{rule:3a}, in which a complex bifundamental scalar or Fermi is mapped to itself anti-holomorphically via \eqref{scalar_involution} or \eqref{Lambda_involution-antih}. $G_i$ and $G_j$ are the same type of $\SO$ or $\USp$ gauge group.}
  	\label{fig:rule3a}
\end{figure}

\begin{enumerate}[label=3\alph*.,ref=3\alph*]
\item\label{rule:3a} A bifundamental field $X_{ij}$ that is mapped to itself anti-holomorphically via \eqref{scalar_involution} or \eqref{Lambda_involution-antih}, with the nodes $i$ and $j$ also being their own images, gives rise to a real $\mathcal{N}=(0,1)$ field  transforming under the bifundamental of $G_i\left(N_i\right)\times G_j(N_{j})$, where $G_i$ and $G_j$ are the same type of $\SO$ or $\USp$ gauge group.\footnote{We will later elaborate on why these two gauge groups should be of the same type.} Figure~\ref{fig:rule3a} illustrates this rule.
\end{enumerate}

\begin{figure}[H]
    \centering
    	\begin{tikzpicture}[scale=2]
    	\node[draw=black,line width=1pt,circle,fill=yellowX,minimum width=0.75cm,inner sep=1pt,label={[yshift=-1.75cm]:$\U\left(N_i\right)$}] (A) at (0,0) {$i$};
    	\node[draw=black,line width=1pt,circle,fill=yellowX,minimum width=0.75cm,inner sep=1pt,label={[yshift=-1.75cm]:$\U\left(N_{i'}\right)$}] (B) at (1.5,0) {$i'$};
    	\draw[line width=1pt,redX] (A) -- node[above,midway,yshift=0.3cm] {$\Lambda_{ii'}$} node[fill=white,text opacity=1,fill opacity=1,draw=black,rectangle,thin] {$2$} (B);
	\draw[line width=1pt,redX] (3.75,0) circle (0.25)  node[yshift=0.5cm] {\scriptsize{$\quadro$}} node[fill=white,text opacity=1,fill opacity=1,draw=black,rectangle,thin,xshift=0.5cm] {$2$} node[yshift=0.9cm] {$\Lambda_{iiA}$};
	\node[draw=black,line width=1pt,circle,fill=yellowX,minimum width=0.75cm,inner sep=1pt,label={[xshift=-0.5cm,yshift=-1.5cm]:$\U\left(N_{i}\right)$}] (A) at (3.5,0) {$i$};
	\node at (4.4,0) {or};
	\draw[line width=1pt,redX] (5.25,0) circle (0.25)  node[yshift=0.5cm,star,star points=5, star point ratio=2.25, inner sep=1pt, fill=redX, draw] {} node[fill=white,text opacity=1,fill opacity=1,draw=black,rectangle,thin,xshift=0.5cm] {$2$} node[yshift=0.9cm] {$\Lambda_{iiS}$};
	\node[draw=black,line width=1pt,circle,fill=yellowX,minimum width=0.75cm,inner sep=1pt,label={[xshift=-0.5cm,yshift=-1.5cm]:$\U\left(N_{i}\right)$}] (A) at (5,0) {$i$};
    	\draw[-Triangle,blueX,line width=1mm] (2.25,0) -- node[above,midway,yshift=0.1cm] {\color{black}{$\Lambda\rightarrow \pm \Lambda^T$}} (2.75,0);
    	\end{tikzpicture}
        \caption{Rule~\ref{rule:3b}, in which a Fermi connecting a node to its image is mapped to itself holomorphically via \eqref{Lambda_involution-h}.}
        \label{fig:rule3b}
\end{figure}
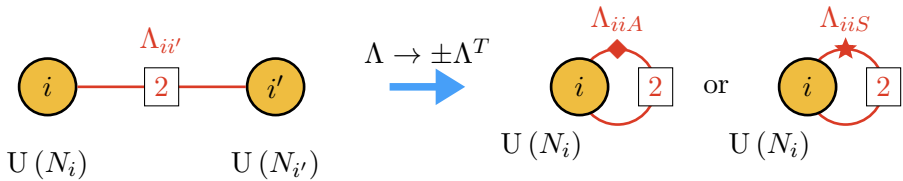

\begin{enumerate}[label=3\alph*.,ref=3\alph*,resume]
\item\label{rule:3b} There is another possibility for a bifundamental Fermi field $\Lambda_{ii'}$ stretching between a node and its image. Such a field can only be mapped to itself in the case of a holomorphic transformation \eqref{Lambda_involution-h}.\footnote{For this reason, there is no analogue of this rule for chiral or Fermi fields transforming anti-holomorphically.} This gives rise to a complex Fermi superfield in the symmetric/antisymmetric representation of the resulting $\U\left(N_{i}\right)$ group for a $+/-$ sign, respectively, as shown in Figure~\ref{fig:rule3b}.
\end{enumerate}

\begin{figure}[H]
    \centering
    \begin{tikzpicture}[scale=2]
   \begin{scope}[yshift=-1.25cm]
   \draw[line width=1pt,redX] (1.25,0) circle (0.25)  node[yshift=0.75cm] {$\Lambda_{ii}$} node[fill=white,text opacity=1,fill opacity=1,draw=black,rectangle,thin,xshift=-0.5cm] {$2$};
	\node[draw=black,line width=1pt,circle,fill=yellowX,minimum width=0.75cm,inner sep=1pt] (A) at (1.5,0) {$i$};
	\draw[line width=1pt,redX] (3.75,0) circle (0.25)  node[yshift=0.5cm] {\color{redX}{\scriptsize{$\quadro$}}} node[fill=white,text opacity=1,fill opacity=1,draw=black,rectangle,thin,xshift=0.5cm] {$2$} node[yshift=0.9cm] {$\Lambda_{iiA}$};;
	\node[draw=black,line width=1pt,circle,fill=yellowX,minimum width=0.75cm,inner sep=1pt,label={[xshift=-0.5cm,yshift=-1.5cm]:$G_i(N_{i})$}] (A) at (3.5,0) {$i$};
	\node at (4.4,0) {or};
	\draw[line width=1pt,redX] (5.25,0) circle (0.25)  node[yshift=0.5cm,star,star points=5, star point ratio=2.25, inner sep=1pt, fill=redX, draw=redX] {} node[fill=white,text opacity=1,fill opacity=1,draw=black,rectangle,thin,xshift=0.5cm] {$2$} node[yshift=0.9cm] {$\Lambda_{iiS}$};;
	\node[draw=black,line width=1pt,circle,fill=yellowX,minimum width=0.75cm,inner sep=1pt,label={[xshift=-0.5cm,yshift=-1.5cm]:$G_i(N_{i})$}] (A) at (5,0) {$i$};
   \draw[-Triangle,blueX,line width=1mm] (2.25,0) -- node[above,midway,yshift=0.1cm] {\color{black}{$\Lambda\rightarrow \pm \Lambda^T$}} (2.75,0);
   \end{scope}
    \end{tikzpicture}
       \caption{Rule~\ref{rule:3c}, in which a complex adjoint Fermi is mapped to itself holomorphically via \eqref{Lambda_involution-h}. The group $G_i(N_i)$ can be either $\SO(N_i)$ or $\USp(N_i)$.}
    \label{fig:rule3c}
\end{figure}
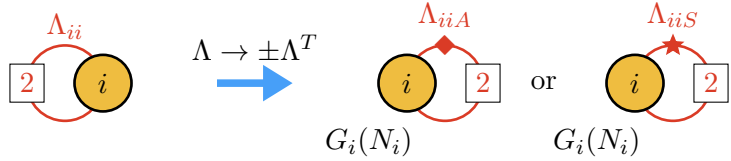

\begin{enumerate}[label=3\alph*.,ref=3\alph*,resume]
\item\label{rule:3c} Closely related to Rule~\ref{rule:3b}, consider an adjoint complex Fermi field $\Lambda_{ii}$ that is mapped to itself via the holomorphic transformation \eqref{Lambda_involution-h}. As shown in Figure~\ref{fig:rule3c}, this gives rise to a complex Fermi field in the symmetric or antisymmetric representation of the resulting gauge group for a $+/-$ sign, respectively. In this case, the $\pm$ sign in \eqref{Lambda_involution-h} correlates the projection of such Fermi with the one of the corresponding vector multiplet, which is controlled by \eqref{A_involution}. In particular, a $+$ sign implies the opposite projection, and hence we obtain symmetric/antisymmetric for $\SO/\USp$. Similarly, a $-$ sign implies the same projection, and we obtain antisymmetric/symmetric for $\SO/\USp$. 
\end{enumerate}

\begin{figure}[H]
    \centering
    \begin{tikzpicture}[scale=2]
    \begin{scope}
	\draw[line width=1pt,postaction={draw,redX,dash pattern= on 3pt off 5pt,dash phase=4pt}] [line width=1pt,black,dash pattern= on 3pt off 5pt] (1.25,0) circle (0.25)  node[yshift=0.75cm] {$X_{ii}$} node[fill=white,text opacity=1,fill opacity=1,draw=black,rectangle,thin,xshift=-0.5cm,dash pattern= on 0pt off 0pt,thin] {$2$};
	\node[draw=black,line width=1pt,circle,fill=yellowX,minimum width=0.75cm,inner sep=1pt] (A) at (1.5,0) {$i$};
	\draw[line width=1pt,postaction={draw,redX,dash pattern= on 3pt off 5pt,dash phase=4pt}] [line width=1pt,black,dash pattern= on 3pt off 5pt] (4,0) circle (0.45) node[xshift=0.90cm,xshift=0cm,star,star points=5, star point ratio=2.25, inner sep=1pt, fill=black, draw,dash pattern= on 0pt off 0pt] {}   node[yshift=1.2cm,xshift=0.5cm] {$X_{iiS}^R$};
	\draw[line width=1pt,postaction={draw,redX,dash pattern= on 3pt off 5pt,dash phase=4pt}] [line width=1pt,black,dash pattern= on 3pt off 5pt] (3.75,0) circle (0.25)  node[xshift=0.5cm] {\scriptsize{$\quadro$}} node[yshift=1.2cm,xshift=0cm] {$X_{iiS}^R$};
	\node[draw=black,line width=1pt,circle,fill=yellowX,minimum width=0.75cm,inner sep=1pt,label={[xshift=-0.5cm,yshift=-1.5cm]:$G_i(N_{i})$}] (A) at (3.5,0) {$i$};
   \draw[-Triangle,blueX,line width=1mm] (2.25,0) -- node[above,midway,yshift=0.1cm] {\color{black}{$X\rightarrow \pm \bar{X}$}} (2.75,0);
   \end{scope}
    \end{tikzpicture}
       \caption{Rule~\ref{rule:3d}, in which a complex adjoint scalar or Fermi is mapped to itself anti-holomorphically via \eqref{scalar_involution} or \eqref{Lambda_involution-antih}. The group $G_i(N_i)$ can be either $\SO(N_i)$ or $\USp(N_i)$.}
    	\label{fig:rule3d}
\end{figure}
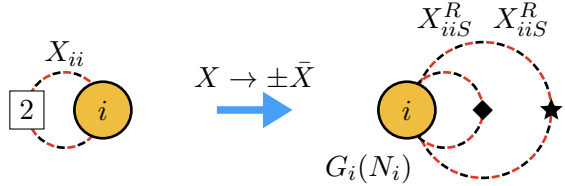

\begin{enumerate}[label=3\alph*.,ref=3\alph*,resume]
\item\label{rule:3d} Consider an adjoint complex scalar or Fermi field that is mapped to itself via the anti-holomorphic transformation in \eqref{scalar_involution} or \eqref{Lambda_involution-antih}. This gives rise to two real scalar or Fermi fields, one symmetric and one antisymmetric of node $i$. This can be understood as projecting the real and imaginary parts of the parent field with opposite signs. The sign in \eqref{A_involution} determines the projection of the real part relative to the $\SO$ or $\USp$ projection of the gauge group as in Rule~\ref{rule:3c}. This case is shown in Figure~\ref{fig:rule3d}.
\end{enumerate}

\medskip

\noindent\underline{$\mathcal{N}=(0,1)$ real Fermi fields from $\mathcal{N}=(0,2)$ vector multiplets}

\medskip

Finally, let us consider the projection of the $\mathcal{N}=(0,1)$ adjoint real Fermi fields $\Lambda_{ii}^R$ coming from the $\mathcal{N}=(0,2)$ vector multiplets. Such fields always transform according to \eqref{Lambda^R_projection}. Therefore, there are only two possibilities, depending on whether the corresponding node is mapped to a different node or to itself.

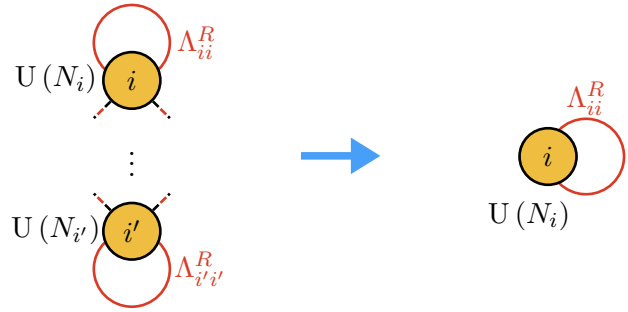
\begin{figure}[H]
\centering
    \begin{tikzpicture}[scale=3]
    \draw[help lines,white] (-1,0) grid (2,0);
    \node (L) at (-0.3,0) {
    \begin{tikzpicture}[scale=2]
    \draw[line width=1pt,redX]  (0,0.75) circle (0.25) node[xshift=0.85cm] {$\Lambda^R_{ii}$};
    \draw[line width=1pt,redX] (0,-0.75) circle (0.25) node[xshift=0.9cm] {$\Lambda^R_{i'i'}$};
    \node[draw=black,line width=1pt,circle,fill=yellowX,minimum width=0.75cm,inner sep=1pt,label={[xshift=-1cm,yshift=-0.7cm]:$\U\left(N_i\right)$}] (A) at (0,0.5) {$i$};
    	\node[draw=black,line width=1pt,circle,fill=yellowX,minimum width=0.75cm,inner sep=1pt,label={[xshift=-1cm,yshift=-0.7cm]:$\U\left(N_{i'}\right)$}] (B) at (0,-0.5) {$i'$};
    	\draw[line width=1pt,postaction={draw,redX,dash pattern= on 3pt off 5pt,dash phase=4pt}] [line width=1pt,black,dash pattern= on 3pt off 5pt] (A) -- (-0.25,0.25);
    	\draw[line width=1pt,postaction={draw,redX,dash pattern= on 3pt off 5pt,dash phase=4pt}] [line width=1pt,black,dash pattern= on 3pt off 5pt] (A) -- (0.25,0.25);
    	\draw[line width=1pt,postaction={draw,redX,dash pattern= on 3pt off 5pt,dash phase=4pt}] [line width=1pt,black,dash pattern= on 3pt off 5pt] (B) -- (-0.25,-0.25);
    	\draw[line width=1pt,postaction={draw,redX,dash pattern= on 3pt off 5pt,dash phase=4pt}] [line width=1pt,black,dash pattern= on 3pt off 5pt] (B) -- (0.25,-0.25);
    	\node at (0,0) {$\vdots$};
    \end{tikzpicture}
    };
    \node (R) at (1.6,0) {
    \begin{tikzpicture}[scale=2]
	\draw[line width=1pt,redX]  (0.25,0) circle (0.25) node[yshift=0.77cm] {$\Lambda^R_{ii}$};  
	\node[draw=black,line width=1pt,circle,fill=yellowX,minimum width=0.75cm,inner sep=1pt,label={[xshift=-0.25cm,yshift=-1.5cm]:$\U\left(N_{i}\right)$}] (A) at (0,0) {$i$};
    \end{tikzpicture}
    };
    \draw[-Triangle,blueX,line width=1mm] (0.5,0) -- (0.85,0);
    \end{tikzpicture}
    \caption{Rule~\ref{rule:4a}, in which two real Fermi fields are mapped to each other into a single real Fermi field.}
    \label{fig:rule4a}
\end{figure}

\begin{enumerate}[label=4\alph*.,ref=4\alph*]
\item\label{rule:4a} Consider a real Fermi $\Lambda^R_{ii}$ which transforms via \eqref{Lambda^R_projection} into $\Lambda^R_{i'i'}$, with $i'\neq i$. The two fields are projected down to a single real Fermi $\Lambda^R_{ii}$, as in Figure~\ref{fig:rule4a}.
\end{enumerate}

\begin{figure}[H]
\centering
    \begin{tikzpicture}[scale=2]
   \begin{scope}[yshift=-1.25cm]
   \draw[line width=1pt,redX] (1.25,0) circle (0.25)  node[yshift=0.75cm] {$\Lambda^R_{ii}$};
	\node[draw=black,line width=1pt,circle,fill=yellowX,minimum width=0.75cm,inner sep=1pt] (A) at (1.5,0) {$i$};
	\draw[line width=1pt,redX] (3.75,0) circle (0.25)  node[yshift=0.5cm] {\color{redX}{\scriptsize{$\quadro$}}}  node[yshift=0.9cm] {$\Lambda^R_{iiA}$};;
	\node[draw=black,line width=1pt,circle,fill=yellowX,minimum width=0.75cm,inner sep=1pt,label={[xshift=-0.5cm,yshift=-1.5cm]:$\USp(N_{i})$}] (A) at (3.5,0) {$i$};
	\node at (4.4,0) {or};
	\draw[line width=1pt,redX] (5.25,0) circle (0.25)  node[yshift=0.5cm,star,star points=5, star point ratio=2.25, inner sep=1pt, fill=redX, draw=redX] {} node[yshift=0.9cm] {$\Lambda^R_{iiS}$};;
	\node[draw=black,line width=1pt,circle,fill=yellowX,minimum width=0.75cm,inner sep=1pt,label={[xshift=-0.5cm,yshift=-1.5cm]:$\SO(N_{i})$}] (A) at (5,0) {$i$};
   \draw[-Triangle,blueX,line width=1mm] (2.25,0) -- 
   (2.75,0);
   \end{scope}
    \end{tikzpicture}
    \caption{Rule~\ref{rule:4b}, in which one real Fermi field is mapped to itself.}
    \label{fig:rule4b}
\end{figure}
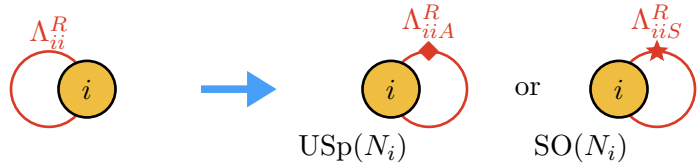

\begin{enumerate}[label=4\alph*.,ref=4\alph*,resume]
\item\label{rule:4b} Consider a real Fermi $\Lambda^R_{ii}$ which is mapped to itself, with $i'\neq i$. Due to the relative sign between \eqref{A_involution} and \eqref{Lambda^R_projection}, this gives rise to a symmetric or antisymmetric real Fermi for an $\SO$ or $\USp$ projection of the node $i$, respectively. We show the result in Figure~\ref{fig:rule4b}.
\end{enumerate}

In general, it is possible for the theories constructed with the orientifolding procedure described above to suffer from gauge anomalies. Such anomalies can be canceled by the addition of appropriate scalar or Fermi flavors. In string theory, this corresponds to introducing flavor D$5$/D$9$-branes to cancel the local RR tadpole arising when orientifold planes are present. In Section~\ref{sec:exampC4Z2beyonduni}, we present an example in which flavor fields are needed in order to cancel the gauge anomalies.

\subsubsection*{Superpotential}

The superpotential of the orientifold theory is obtained from the parent superpotential by keeping the invariant terms and projecting out half of the other terms, which are identified in pairs. In the surviving terms, the parent fields must be replaced by their images under the orientifold projection.

\subsubsection*{A constraint on the relative projections of nodes connected by matter}

Requiring that the orientifold group acts on the gauge theory as an involution, leads to interesting relations between the transformation of matter fields and gauge groups. In particular, focusing on bifundamental fields, applying the transformations \eqref{scalar_involution}, \eqref{Lambda_involution-antih} or \eqref{Lambda_involution-h} twice and demanding that they amount to the identity, leads to correlations between the $\eta$ and $\gamma_{\Omega}$ matrices. For example, for a pair of nodes $i$ and $j$ connected by a single field or by a pair of fields with $\eta=\pm \left(\begin{smallmatrix} 0 & 1 \\ 1 & 0 \end{smallmatrix}\right)$ which transform anti-holomorphically, we must have $\gamma_{\Omega_i}=\gamma_{\Omega_j}$. Most of the examples we will consider later are of these two types. On the other hand, $\eta=\left(\begin{smallmatrix} 0 & 1 \\ -1 & 0 \end{smallmatrix}\right)$ implies that $\gamma_{\Omega_i}$ and $\gamma_{\Omega_j}$ are of opposite types.

\subsection{Anti-Holomorphic Involutions from the Mesonic Moduli Space}
\label{sec:HSreview}

The anti-holomorphic involution $\sigma$ of a CY$_4$ underlying Joyce's construction can be beautifully connected to the anti-holomorphic involution of the associated $\mathcal{N}=(0,2)$ gauge theory. The CY$_4$ arises as the mesonic moduli space of the parent gauge theory. Consequently, the complex coordinates parameterizing the CY$_4$ correspond to mesonic operators. Below, we present an algorithmic procedure for identifying anti-holomorphic involutions of CY$_4$ cones leading to $\Spin(7)$ manifolds. Combined with the map of generators to the gauge theory, this provides an alternative method for constructing $\Spin(7)$ orientifolds. This approach is analogous to the one introduced in \cite{Franco:2007ii} for 4d orientifolds. As usual, the construction focuses on the Abelian case of the gauge theories, but the results extend to general ranks.

In general, we can define the moduli space as the polynomial ring of the chiral fields modded by the ideal generated by the $J$- and $E$-terms, i.e.
\begin{equation}
    \mathcal{M}=\left(\CC[X_1,\ldots, X_n]/\left\langle J_{ij},E_{ij}\right\rangle\right)//\U(1)^G\coma 
\end{equation}
where $G$ is the number of $\U(1)$ gauge groups in the theory, and $n$ is the number of chiral fields. It is then possible to associate a GLSM to such a moduli space, given by a set of fields $p_a$ such that
\begin{equation}
    \mathcal{M}=\left(\CC[p_1,\ldots, p_m]//Q_{EJ}\right)//Q_{D}\coma
\end{equation}
where $Q_{EJ}$ and $Q_D$ are matrices containing $\U(1)$ charges of the $p_a$ that implement the $J$-, $E$- and $D$-terms. The mesonic moduli space is obtained by considering combinations of fields $p_a$ that are invariant under the action of these $\U(1)$'s. For details on this construction we refer to~\cite{Franco:2015tya}.

A tool that has proven to be powerful to compute such gauge invariant operators is the Hilbert series (HS)~\cite{Benvenuti:2006qr,Feng:2007ur}. The explicit expression of the HS is
\begin{equation}
    \HS(\mathbf{x},\mathbf{p})= \PE\left[\sum_{a=1}^m \mathbf{x}^{Q^a}p_a\right]\coma
\end{equation}
where $Q^a=(Q^a_{EJ},Q^a_{D})$ are the charges of the field $p_a$ represented by the collective fugacity $\mathbf{x}$. The function $\PE$ is called the Plethystic Exponential (PE) and is defined as 
\begin{equation}
    \PE\left[f(t)\right] = \PE\left[\sum_{k=0}^\infty c_k t^k\right]=\exp\left[\sum_{k=1}^\infty \frac{1}{k}\left(f\left(t^k\right)-f(0)\right)\right]=\prod_{k=1}^\infty\frac{1}{(1-t^k)^{c_k}}\fstop
\end{equation}
Performing the Molien integral over the fugacities $\mathbf{x}$, we obtain the HS of the mesonic moduli space $\mathcal{M}$:
\begin{equation}
    \HS(\mathbf{p};\mathcal{M})=\oint_{|\mathbf{x}|=1}\frac{d\mathbf{x}}{2\pi i \mathbf{x}} \HS(\mathbf{x},\mathbf{p})\fstop
\end{equation}
Such HS contains the generators of the mesonic moduli space and their relations. This information can be extracted using the Plethystic Logarithm (PL):
\begin{equation}
    \PL[\HS(\mathbf{p};\mathcal{M})]=\sum_{k=1}^\infty \frac{\mu(k)}{k}\ln\left[\HS\left(\mathbf{p}^k;\mathcal{M}\right)\right]\coma
\end{equation}
where $\mu$ is the M\"obius function. The resulting series can be finite, and in that case, the mesonic moduli space is said to be a \textit{complete intersection}, or it can be an infinite sum of positive and negative monomials in $\mathbf{p}$. The generators are identified with the positive monomials, while the relations are associated with the negative monomials. The generators for all examples in the paper have been computed using such HS techniques.

The generators, subject to their relations, are the coordinates that parameterize the toric CY$_4$ under consideration. From the point of view of the gauge theory, these coordinates are mesons and we call them $M_a$, with $a$ running from $1$ to the number of mesons. The anti-holomorphic involution $\sigma$ acts on these coordinates by mapping each $M_a$ to a possibly different $\bar{M}_b$, with $\bar{M}_b$ being the complex conjugate of $M_b$, i.e.
\begin{equation}
    M_a \rightarrow \pm \bar{M}_b\fstop
\end{equation}
This transformation must be consistent with the relations among the generators. 
 
As explained in Section~\ref{section_Spin(7)_from_CY4}, in order to obtain a $\Spin(7)$ structure, $\sigma$ must preserve the Cayley $4$-form. A sufficient condition for this to happen is that $\Omega^{(4,0)}\rightarrow \bar{\Omega}^{(0,4)}$ \cite{Joyce:1999nk}. Consider a CY$_4$ with $n$ generators $M_a$, $a=1, \cdots, n$ and $k$ relations among them $F_\alpha(M_a)=0$, $\alpha=1,\cdots, k$. The holomorphic 4-form is computed in terms of the Poincar\'{e} residue
\begin{equation}
	\Omega^{(4,0)}=\text{Res}\frac{d M_1\wedge \cdots \wedge d M_n}{\prod_{\alpha =1}^{k}F_\alpha(M_a)} \fstop
\end{equation}
With this formula, it is straightforward to verify that all the involutions considered in this paper satisfy $\Omega^{(4,0)}\rightarrow \bar{\Omega}^{(0,4)}$. In the following sections, we will show this explicitly in some examples.\footnote{When the HS is not a complete intersection, the number of relations is redundant. It is then possible to reduce them to their effective number, and $F_\alpha(M_a)$ represents the minimal number of relations that are necessary in order to get a $4$-form, given $n$ generators, i.e. $\alpha=1,\ldots,n-4$. Moreover, given an $\Omega^{(4,0)}$, after applying the anti-holomorphic involution, it might be necessary to use such relations to obtain the corresponding $\bar{\Omega}^{(0,4)}$. Generically, the resulting $(0,4)$-form that is obtained by the involution, is not simply the complex conjugate of $\Omega^{(4,0)}$. An explicit example of this is given in Section~\ref{sec:D3examp}.}

The procedure outlined above provides a geometric criterion for identifying an anti-holomorphic involution $\sigma$ leading to a $\Spin(7)$ orientifold. Using the definition of the generators as gauge invariant chiral operators in the field theory, we can translate $\sigma$ into the anti-holomorphic involution that acts on the chiral fields. Finally, we can complete such involution with the transformations of Fermi fields in the form of \cref{Lambda_involution-antih,Lambda_involution-h,Lambda^R_projection} such that it corresponds to a $\mathbb{Z}_2$ symmetry of the $\mathcal{N}=(0,2)$ gauge theory, as discussed in Section~\ref{sec:orientN01gaugeth}.

An important observation is that the relation between the geometric anti-holomorphic involution $\sigma$ that accompanies the orientifold action, and the action on the $\mathcal{N}=(0,2)$ theory, is not one-to-one. In particular, this non-uniqueness goes beyond the obvious one due to choices of signs and $\gamma_\Omega$'s in \cref{scalar_involution,Lambda_involution-antih,Lambda_involution-h,Lambda^R_projection}. Indeed, certain orientifolded geometries defined by an involution $\sigma$ of CY$_4$ correspond to a unique action on the $\mathcal{N}=(0,2)$ quiver (up to those obvious choices), but others can admit several genuinely different possible actions from the field theory point of view. These are distinguished by the action of the orientifold on the gauge factor, in particular by the presence or absence of groups mapped to themselves. In more mathematical terms, this is related  to the presence or the absence of vector structure in type IIB singularities with orientifolds. We will discuss this in more detail and present illustrative examples in Section~\ref{sec:vectorstructure}. 

\section{$\mathbb{C}^4$ and its Orbifolds}
\label{sec:examplesenginN=01}

In this section, we construct the $2$d gauge theories on D1-branes over $\Spin(7)$ orientifolds of $\mathbb{C}^4$ and its Abelian orbifold $\mathbb{C}^4/\mathbb{Z}_2$.

\subsection{$\mathbb{C}^4$}
\label{sec:C4example}

Let first consider the simplest CY$_4$, i.e. $\CC^4$, and construct its $\Spin(7)$ orientifold. Its toric diagram is shown in Figure~\ref{fig:C4toricdiagram}.

\begin{figure}[!htp]
    \centering
	\begin{tikzpicture}[scale=1.5]
	\draw[thick,gray,-Triangle] (0,0,0) -- node[above,pos=1] {$x$} (1.5,0,0);
	\draw[thick,gray,-Triangle] (0,0,0) -- node[left,pos=1] {$y$} (0,1.5,0);
	\draw[thick,gray,-Triangle] (0,0,0) -- node[below,pos=1] {$z$} (0,0,1.5);
	\node[draw=black,line width=1pt,circle,fill=black,minimum width=0.2cm,inner sep=1pt] (p1) at (0,0,0) {};
	\node[draw=black,line width=1pt,circle,fill=black,minimum width=0.2cm,inner sep=1pt] (p2) at (1,0,0) {};
	\node[draw=black,line width=1pt,circle,fill=black,minimum width=0.2cm,inner sep=1pt] (p3) at (0,1,0) {};
	\node[draw=black,line width=1pt,circle,fill=black,minimum width=0.2cm,inner sep=1pt] (p4) at (0,0,1) {};
	\draw[line width=1pt] (p1)--(p2)--(p4)--(p3)--(p2);
	\draw[line width=1pt] (p1)--(p3);
	\draw[line width=1pt] (p1)--(p4);
	\end{tikzpicture}
	\caption{Toric diagram for $\CC^4$.}
	\label{fig:C4toricdiagram}
\end{figure}

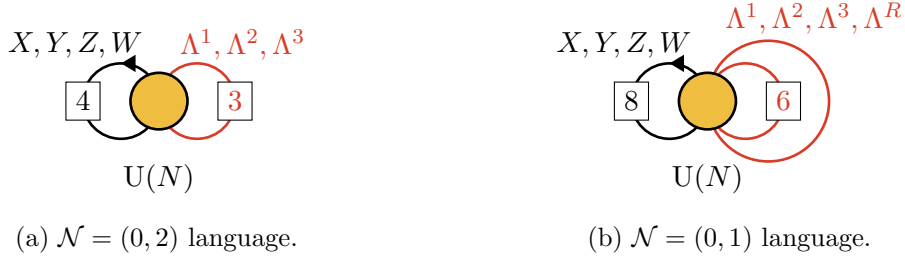
\begin{figure}[!htp]
	\centering
	\begin{subfigure}[t]{0.49\textwidth}
	\centering
	\begin{tikzpicture}[scale=2]
	\draw[line width=1pt,redX] (0.25,0) circle (0.25) node[fill=white,text opacity=1,fill opacity=1,draw=black,rectangle,xshift=0.5cm,thin] {\color{redX}{$3$}} node[xshift=0.6cm,yshift=0.75cm] {\color{redX}{$\Lambda^1,\Lambda^2,\Lambda^3$}};
	\draw[line width=1pt,decoration={markings, mark=at position 0.25 with{\arrow{Triangle}}}, postaction={decorate}] (-0.25,0) circle (0.25) node[fill=white,text opacity=1,fill opacity=1,draw=black,rectangle,xshift=-0.5cm,thin] {$4$} node[xshift=-0.6cm,yshift=0.75cm] {$X,Y,Z,W$};
	\node[draw=black,line width=1pt,circle,fill=yellowX,minimum width=0.75cm,inner sep=1pt,label={[yshift=-1.75cm]:$\U(N)$}] (A) at (0,0) {};
	\end{tikzpicture}
	\caption{$\mathcal{N}=(0,2)$ language.}
	\label{fig:C4quivN02}
	\end{subfigure}
	\begin{subfigure}[t]{0.49\textwidth}
	\centering
	\begin{tikzpicture}[scale=2]
	\draw[line width=1pt,redX] (0.25,0) circle (0.25) node[fill=white,text opacity=1,fill opacity=1,draw=black,rectangle,xshift=0.5cm,thin] {\color{redX}{$6$}};
	\draw[line width=1pt,redX] (0.4,0) circle (0.4) node[xshift=0.6cm,yshift=1.1cm] {\color{redX}{$\Lambda^1,\Lambda^2,\Lambda^3,\Lambda^R$}};
	\draw[line width=1pt,decoration={markings, mark=at position 0.25 with{\arrow{Triangle}}}, postaction={decorate}] (-0.25,0) circle (0.25) node[fill=white,text opacity=1,fill opacity=1,draw=black,rectangle,xshift=-0.5cm,thin] {$8$} node[xshift=-0.6cm,yshift=0.75cm] {$X,Y,Z,W$};
	\node[draw=black,line width=1pt,circle,fill=yellowX,minimum width=0.75cm,inner sep=1pt,label={[yshift=-1.75cm]:$\U(N)$}] (A) at (0,0) {};
	\end{tikzpicture}
	\caption{$\mathcal{N}=(0,1)$ language.}
	\label{fig:C4quivN01}
	\end{subfigure}
	\caption{Quiver diagrams for $\CC^4$ in $\mathcal{N}=(0,2)$ and $\mathcal{N}=(0,1)$ language. $\Lambda^R$ is the real Fermi coming from the $\mathcal{N}=(0,2)$ vector multiplet.}
	\label{fig:C4quiv}
\end{figure}

The parent $2$d worldvolume theory on D1-branes over $\mathbb{C}^4$ is the dimensional reduction of $4$d $\mathcal{N}=4$ super Yang-Mills (SYM) and has $\mathcal{N}=(8,8)$ SUSY. In $\mathcal{N}=(0,2)$ language, this theory is given by the quiver shown in Figure~\ref{fig:C4quivN02}, and the following $J$- and $E$-terms for the Fermi fields:
\begin{equation}
\renewcommand{\arraystretch}{1.1}
\begin{array}{lclc}
& J        &\text{\hspace{.5cm}} & E \\
\Lambda^{1}     \,:\, & YZ-ZY           & \quad               &  WX-XW   \\
\Lambda^{2}     \,:\, & ZX-XZ           & \quad               &  WY-YW   \\
\Lambda^{3}     \,:\, & XY-YX           & \quad               &  WZ-ZW   
\end{array}
\label{J_E_C4}
\end{equation}

Before performing the orientifold quotient, it is useful to rewrite this theory in $\mathcal{N}=(0,1)$ superspace. In $\mathcal{N}=(0,1)$ language, this theory has a vector multiplet associated with the $\U(N)$ gauge group, four complex scalar multiplets $\left(X,Y,Z\right.$ and $\left.W\right)$, three complex Fermi multiplets $\left(\Lambda^{i}\coma i=1,2,3\right)$ and one real Fermi multiplet $\left(\Lambda^R\right)$ from the $\mathcal{N}=(0,2)$ vector multiplet. The quiver is shown in Figure~\ref{fig:C4quivN01}. 
The corresponding $\mathcal{N}=(0,1)$ superpotential is given by 
\begin{equation}
\begin{split}
    W^{(0,1)}= &\,W^{(0,2)}+\Lambda^{4R}(X^\dagger X+Y^\dagger Y+Z^\dagger Z+W^\dagger W) \\
    =& \,\Lambda^1(YZ-ZY)+ \Lambda^{1\dagger}(WX-XW)+\hc\\
    +& \,\Lambda^2(ZX-XZ)+\Lambda^{2\dagger}(WY-YW)+\hc\\
    +& \,\Lambda^3(XY-YX)+\Lambda^{3\dagger}(WZ-ZW)+\hc\\
    +& \,\Lambda^{4R}(X^\dagger X+Y^\dagger Y+Z^\dagger Z+W^\dagger W) \coma
\end{split}
\label{eq:W01C4}
\end{equation}
where $W^{(0,2)}$ indicates the superpotential obtained from the $J$- and $E$-terms in \eqref{J_E_C4}.

For this theory, computing the HS for identifying the generators parameterizing the moduli space is not necessary, since these mesons are in one-to-one correspondence with the chiral superfields. The four complex coordinates $(x,y,z,w)$ of $\CC^4$ map to the four $\mathcal{N}=(0,1)$ complex scalar fields 
\begin{equation}
    (x,y,z,w)\Leftrightarrow (X,Y,Z,W)\fstop
    \label{eq:coordC4}
\end{equation}
In the Abelian case, the space is freely generated, i.e. there are no relations among the generators. This can be easily understood in $\mathcal{N}=(0,2)$ language, where the $J$- and $E$-terms in~\eqref{J_E_C4} are automatically vanishing.

Now we are ready to find an anti-holomorphic involution $\sigma$ of $\mathbb{C}^4$ and construct the gauge theory for the corresponding $\Spin(7)$ orientifold. We will choose a specific form of $\sigma$. All other possible $\sigma$'s are in fact equivalent to it via the $\SO(8)$ global symmetry of $\mathbb{C}^4$.

\subsubsection{The Orientifold Theory}

Let us consider the anti-holomorphic involution under which the $\U(N)$ gauge group is mapped to itself and the chiral fields transform as 
\begin{equation}\label{involution of c4 chiral}
    X\rightarrow \gamma_{\Omega} \bar{X} \gamma_{\Omega}^{-1}\coma  Y\rightarrow \gamma_{\Omega} \bar{Y}\gamma_{\Omega}^{-1}\coma  Z\rightarrow \gamma_{\Omega} \bar{Z}\gamma_{\Omega}^{-1}\coma W\rightarrow \gamma_{\Omega} \bar{W}\gamma_{\Omega}^{-1}\fstop
\end{equation}

Requiring the invariance of the superpotential $W^{(0,1)}$ in \eqref{eq:W01C4}, we obtain the action on the Fermi multiplets 
\begin{equation}\label{involution of c4 Fermi}
    \Lambda^1\rightarrow \gamma_{\Omega} \bar{\Lambda}^1\gamma_{\Omega}^{-1} \coma \Lambda^2\rightarrow \gamma_{\Omega} \bar{\Lambda}^2\gamma_{\Omega}^{-1} \coma \Lambda^3 \rightarrow \gamma_{\Omega} \bar{\Lambda}^3\gamma_{\Omega}^{-1}\coma \Lambda^{4R}\rightarrow\gamma_{\Omega} \Lambda^{4R\,\,T}\gamma_{\Omega}^{-1} \fstop
\end{equation}

From a geometric point of view, the anti-holomorphic involution $\sigma$ is simply given by  
\begin{equation}
     (x,y,z,w)\mapsto (\bar{x},\bar{y},\bar{z},\bar{w})\fstop
    \label{eq:univ_C4sigma}
\end{equation}
The holomorphic $4$-form $\Omega^{(4,0)}$ and K\"ahler form $J^{(1,1)}$ of $\mathbb{C}^4$ are given by
\begin{equation}
    \Omega^{(4,0)}=dx\wedge dy\wedge dz\wedge dw\coma J^{(1,1)}=\sum_{x_i\in \{ x,y,z,w \}} dx_i\wedge d\bar{x}_i
    \label{eq:OmegaJdef}
\end{equation}
They transform under $\sigma$ as 
\begin{equation}
    \Omega^{(4,0)}\rightarrow \bar{\Omega}^{(0,4)}\coma J^{(1,1)}\rightarrow -J^{(1,1)}.
    \label{eq:sigmaonOmegaJ}
\end{equation}
One can then easily check that the Cayley $4$-form defined in~\eqref{Cayley-form from CY4} is indeed invariant under this involution $\sigma$.

The orientifold theory can be derived by projecting over the involution. As discussed in Section~\ref{sec:orientN01gaugeth}, $\gamma_{\Omega}$ equal to $\ID_{N_a}$ or $J$ corresponds to the $\SO(N)$ or $\USp(N)$ gauge group after projection. We will construct the $\SO(N)$ theory in detail below. The $\USp(N)$ theory can be derived following the same procedure. 

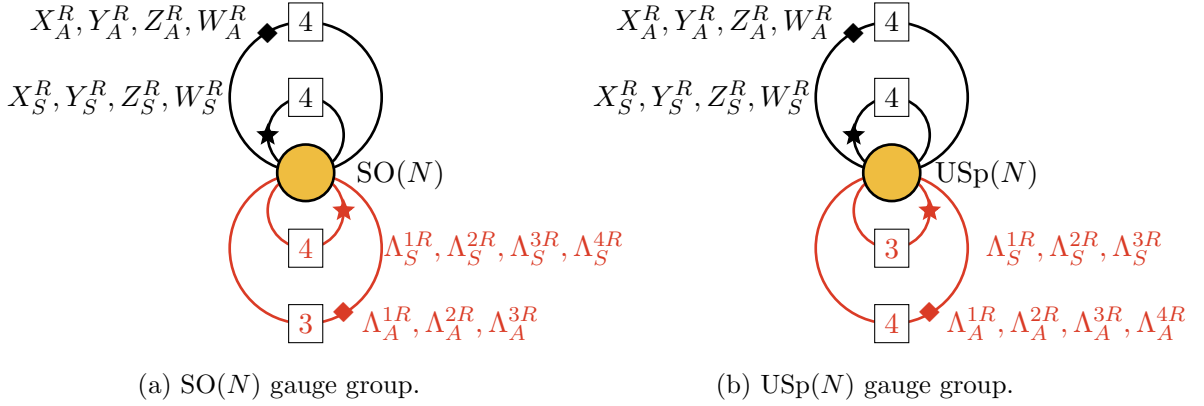
\begin{figure}[H]
	\centering
	\begin{subfigure}[t]{0.49\textwidth}
	\centering
	\begin{tikzpicture}[scale=2]
	\draw[line width=1pt] (0,0.5) circle (0.5) node[yshift=0.85cm,xshift=-0.5cm] {\scriptsize{$\quadro$}} node[fill=white,text opacity=1,fill opacity=1,draw=black,rectangle,yshift=1cm,thin] {$4$} node[yshift=1cm,xshift=-2.2cm] {$X_A^R,Y_A^R,Z_A^R,W_A^R$};
	\draw[line width=1pt] (0,0.25) circle (0.25) node[xshift=-0.5cm,star,star points=5, star point ratio=2.25, inner sep=1pt, fill=black, draw] {} node[fill=white,text opacity=1,fill opacity=1,draw=black,rectangle,yshift=0.5cm,thin] {$4$} node[yshift=0.5cm,xshift=-2.5cm] {$X_S^R,Y_S^R,Z_S^R,W_S^R$};
	\draw[line width=1pt,redX] (0,-0.5) circle (0.5) node[yshift=-0.85cm,xshift=0.5cm] {\scriptsize{\color{redX}{$\quadro$}}} node[yshift=-1cm,xshift=1.9cm] {\color{redX}{$\Lambda^{1R}_A,\Lambda^{2R}_A,\Lambda^{3R}_A$}} node[fill=white,text opacity=1,fill opacity=1,draw=black,rectangle,yshift=-1cm,thin] {\color{redX}{$3$}};
	\draw[line width=1pt,redX] (0,-0.25) circle (0.25) node[xshift=0.5cm,star,star points=5, star point ratio=2.25, inner sep=1pt, fill=redX, draw=redX] {} node[fill=white,text opacity=1,fill opacity=1,draw=black,rectangle,yshift=-0.5cm,thin] {\color{redX}{$4$}} node[yshift=-0.5cm,xshift=2.6cm] {\color{redX}{$\Lambda^{1R}_S,\Lambda^{2R}_S,\Lambda^{3R}_S,\Lambda^{4R}_S$}};
	\node[draw=black,line width=1pt,circle,fill=yellowX,minimum width=0.75cm,inner sep=1pt,label={[xshift=1.25cm,yshift=-0.75cm]:$\SO(N)$}] (A) at (0,0) {};
	\end{tikzpicture}
	\caption{$\SO(N)$ gauge group.}
	\label{fig:univ_o_theory_c4_SO}
	\end{subfigure}\hfill
	\begin{subfigure}[t]{0.49\textwidth}
	\centering
	\begin{tikzpicture}[scale=2]
	\draw[line width=1pt] (0,0.5) circle (0.5) node[yshift=0.85cm,xshift=-0.5cm] {\scriptsize{$\quadro$}} node[fill=white,text opacity=1,fill opacity=1,draw=black,rectangle,yshift=1cm,thin] {$4$} node[yshift=1cm,xshift=-2.2cm] {$X_A^R,Y_A^R,Z_A^R,W_A^R$};
	\draw[line width=1pt] (0,0.25) circle (0.25) node[xshift=-0.5cm,star,star points=5, star point ratio=2.25, inner sep=1pt, fill=black, draw] {} node[fill=white,text opacity=1,fill opacity=1,draw=black,rectangle,yshift=0.5cm,thin] {$4$} node[yshift=0.5cm,xshift=-2.5cm] {$X_S^R,Y_S^R,Z_S^R,W_S^R$};
	\draw[line width=1pt,redX] (0,-0.5) circle (0.5) node[yshift=-0.85cm,xshift=0.5cm] {\scriptsize{\color{redX}{$\quadro$}}} 	node[yshift=-1cm,xshift=2.3cm] {\color{redX}{$\Lambda^{1R}_A,\Lambda^{2R}_A,\Lambda^{3R}_A,\Lambda^{4R}_A$}} node[fill=white,text opacity=1,fill opacity=1,draw=black,rectangle,yshift=-1cm,thin] {\color{redX}{$4$}};
	\draw[line width=1pt,redX] (0,-0.25) circle (0.25) node[xshift=0.5cm,star,star points=5, star point ratio=2.25, inner sep=1pt, fill=redX, draw=redX] {} node[fill=white,text opacity=1,fill opacity=1,draw=black,rectangle,yshift=-0.5cm,thin] {\color{redX}{$3$}} node[yshift=-0.5cm,xshift=2.4cm] {\color{redX}{$\Lambda^{1R}_S,\Lambda^{2R}_S,\Lambda^{3R}_S$}};
	\node[draw=black,line width=1pt,circle,fill=yellowX,minimum width=0.75cm,inner sep=1pt,label={[xshift=1.25cm,yshift=-0.75cm]:$\USp(N)$}] (A) at (0,0) {};
	\end{tikzpicture}
	\caption{$\USp(N)$ gauge group.}
	\label{fig:univ_o_theory_c4_USp}
	\end{subfigure}
	\caption{Quiver diagrams for the orientifold theories associated with the anti-holomorphic involution of $\mathbb{C}^4$ in  \eqref{eq:univ_C4sigma}.}
	\label{fig:univ_o_theory_c4}
\end{figure}

The $\SO(N)$ gauge theory contains four real scalar superfields in the symmetric representation and four real scalar superfields in the antisymmetric representation. We will use subscripts $S$ and $A$ to keep track of representations. There are also four real Fermi superfields $\left(\Lambda^a_S\right.$ with $a=1,2, 3$ and $\Lambda^{4R}$) in the symmetric representation, and three real Fermi superfields $\left(\Lambda^a_A\right.$ with $a=1,2, 3$) in the antisymmetric representation. The origin of these matter multiplets from the parent theory is as follows
\begin{equation}
\begin{array}{ccccccc}
    X&\Rightarrow & X_S^R,X_A^R\coma & \ \ \ \ \ \ & \Lambda^1&\Rightarrow & \Lambda^{1R}_S,\Lambda^{1R}_A\coma \\[.1cm]
    Y&\Rightarrow & Y_S^R,Y_A^R\coma & & \Lambda^2&\Rightarrow &  \Lambda^{2R}_S,\Lambda^{2R}_A\coma \\[.1cm]
    Z&\Rightarrow & Z_S^R,Z_A^R\coma & & \Lambda^3&\Rightarrow &  \Lambda^{3R}_S,\Lambda^{3R}_A\coma \\[.1cm]
    W&\Rightarrow & W_S^R,W_A^R\coma & & \Lambda^{4R}&\Rightarrow & \Lambda^{4R}_S\fstop\\
\end{array}
\label{eq:C4fieldredefinuniv}
\end{equation}
The field content of the resulting $\SO(N)$ gauge theory is summarized by the quiver in Figure~\ref{fig:univ_o_theory_c4_SO}. The quiver for the $\USp(N)$ theory is shown in Figure~\ref{fig:univ_o_theory_c4_USp}. Redefining the fields according to Eq.~\eqref{eq:C4fieldredefinuniv}, it is possible to derive the $W^{(0,1)}$ after the involution from Eq.~\eqref{eq:W01C4}.

Finally, computing the $\SO(N)^2$ anomaly contributions from different $\mathcal{N}=(0,1)$ fields using Table \ref{tab:SOUSpanomaly}, we obtain
\begin{equation}
    \underbrace{-(N-2)}_{\shortstack{Vector}}\underbrace{-4(N+2)-3(N-2)}_{\shortstack{Fermi}}\underbrace{+4(N+2)+4(N-2)}_{\shortstack{Scalar}}=0\fstop
\end{equation}
Therefore, this theory is free of gauge anomalies.

While the Spin(7) orientifold construction generically produces 2d $\mathcal{N}=(0,1)$ theories, special cases such as this one can have enhanced SUSY. This theory in fact enjoys $\mathcal{N}=(4,4)$ SUSY. To see this more explicitly, let us define the four complex coordinates of $\mathbb{C}^4$ in terms of the $8$d space transverse to the D1-branes as 
\begin{equation}
	(x,y,z,w)\equiv (x_2+ix_6, x_3+ix_7, x_4+ix_8, x_5+ix_9)\coma
\end{equation}
where $x_i$, $i=0,1,\cdots, 9$ are real spacetime coordinates. In terms of them, the geometric involution \eqref{eq:univ_C4sigma} becomes 
\begin{equation}
	(x_2,x_3,x_4,x_5,x_6,x_7,x_8,x_9)\rightarrow (x_2,x_3,x_4,x_5,-x_6,-x_7,-x_8,-x_9)\coma
\end{equation}
giving rise to a codimension-4 fixed locus, i.e., an O5-plane. The brane setup is therefore,
\begin{equation}\label{D1 on O5}
	\begin{array}{c|cccccccccc}
	 		& \, 0 \, & \, 1 \, & \, 2 \, & \, 3 \, & \, 4 \, & \, 5 \, & \, 6 \, & \, 7 \, & \, 8 \, & \, 9 \, \\
	 		\hline
		\text{D1} & \bullet & \bullet & \times & \times & \times & \times & \times & \times & \times & \times  \\
		\text{O5}& \bullet & \bullet & \bullet & \bullet & \bullet & \bullet & \times & \times & \times & \times 
	\end{array}
\end{equation}
where $\bullet$ and $\times$ indicate directions in which an object extends or does not extend, respectively. The configuration preserves $\mathcal{N}=(4,4)$ SUSY in the $2$d spacetime of the gauge theory, given by $(x_0, x_1)$. The field theory has $\SO(N)$ or $\USp(N)$ gauge symmetry, depending on the charge of the O5-plane. 

The extended SUSY can also be seen at the level of the gauge theory. The field content can be organized into $\mathcal{N}=(4,4)$ multiplets. For example, in the $\SO(N)$ case, we have
\begin{equation}
\begin{array}{rcc}
	V\oplus \Lambda^{(1,2,3)R}_A \oplus X^R_A,Y^R_A,Z^R_A,W^R_A & \rightarrow & \mathcal{N}=(4,4) ~ \text{vector multiplet} \\ & & \text{(adjoint=antisymmetric)} \\[.4cm]
	\Lambda^{4R}\oplus \Lambda_{S}^{(1,2,3)R}\oplus X^R_S,Y^R_S,Z^R_S,W^R_S &\rightarrow & \mathcal{N}=(4,4) ~ \text{hypermultiplet} \\ & & \text{(symmetric)}
\end{array}
\end{equation}
where $V$ is the $\mathcal{N}=(0,1)$ vector multiplet of the $\SO(N)$ gauge group. 

Note also that the $\SO(4)\times \SO(4)$ R-symmetry group of $\mathcal{N}=(4,4)$ supersymmetry is completely manifest in our realization. An $\SO(4)$ factor corresponds to geometric rotations in the directions transverse to the D1-branes and along the O5-plane, i.e. 2345 in \eqref{D1 on O5}.  On the other hand, the second $\SO(4)$ corresponds to rotations in the directions transverse to the O5-plane, i.e. 6789 in \eqref{D1 on O5}. The above multiplets fill out representations of $\SO(4)^2$ (noticing that the representation including the 3 Fermi multiplets must be completed by including the gauginos in the $\mathcal{N}=(0,1)$ vector multiplet, as befits an R-symmetry). It is easy to check that the interactions are also compatible with this symmetry.

Naively, one can consider seemingly different involutions $\sigma$ preserving the Cayley 4-form and construct the corresponding orientifold theories. However, the resulting theories will always be the same $2$d $\mathcal{N}=(4,4)$ $\SO(N)/\USp(N)$ gauge theory worked out above. All such anti-holomorphic involutions are equivalent, since they are connected by $\SO(8)$ rotations of the eight real coordinates of $\mathbb{C}^4$ and lead to the same brane configuration with D1-branes on top of an O5-plane. 

For example, consider the anti-holomorphic involution $(x,y,z,w)\rightarrow(\bar{y},\bar{x},\bar{z},-\bar{w})$, under which the Cayley 4-form is also invariant. Using the $\SO(8)$ global symmetry, we can redefine the eight real coordinates of $\mathbb{C}^4$ as 
\begin{equation}
\left(x_2^\prime,x_3^\prime,x_4^\prime,x_5^\prime,x_6^\prime,x_7^\prime,x_8^\prime,x_9^\prime\right)\equiv\left(\frac{x_2+x_3}{2},\frac{x_6-x_7}{2},x_4,x_9,\frac{x_2-x_3}{2},\frac{x_6+x_7}{2},x_8,x_5\right)\fstop    
\end{equation}
Then, the fixed locus of the involution corresponds to an O5-plane extended along $x_i^\prime$, $i=2,\ldots,5$. This is exactly the same orientifold configuration in \eqref{D1 on O5}. Therefore, despite the seemingly different involution, the $2$d gauge theory on D1-branes is the same up to field redefinitions.

\subsection{A Universal Involution}
\label{sec:universal involution}

Interestingly, the anti-holomorphic involution of $\mathbb{C}^4$ can be generalized to any CY$_4$. Consider the gauge theory associated to a generic toric CY$_4$. From the field theory perspective, it is always possible to define an involution as follows. First, all gauge groups are mapped to themselves. In addition, all chiral fields transform as 
\begin{equation}\label{eq:universal involution of chiral}
	X_{ij}\rightarrow \gamma_{\Omega_i}\bar{X}_{ij}\gamma_{\Omega_j}^{-1}\coma
\end{equation}
i.e. every chiral field is mapped to itself anti-holomorphically. This in turn implies that the $J$- and $E$-terms for every Fermi $\Lambda_{ij}$ transform as
\begin{equation}
	J_{ji}\rightarrow \gamma_{\Omega_j}\bar{J}_{ji}\gamma_{\Omega_i}^{-1}, E_{ij}\rightarrow \gamma_{\Omega_i}\bar{E}_{ij}\gamma_{\Omega_j}^{-1}\fstop
\end{equation}
Invariance of the superpotential $W^{(0,1)}$ implies that the action on the Fermi fields must be 
\begin{equation}\label{eq:universal involution of Fermi}
	\Lambda_{ij}\rightarrow \gamma_{\Omega_i}\bar{\Lambda}_{ij} \gamma_{\Omega_j}^{-1}\fstop
\end{equation}
Finally, as usual, the $\mathcal{N}=(0,1)$ adjoint Fermi fields coming from $\mathcal{N}=(0,2)$ vector multiplets transform as in \eqref{Lambda^R_projection}. 

This field theoretic involution translates into a simple action on the generators of CY$_4$
\begin{equation}
	\sigma_0 \,:\, M_{a}\rightarrow \bar{M}_{a}\coma
\end{equation}
namely an involution that maps every generator to its conjugate. The holomorphic 4-form $\Omega^{(4,0)}$ then transforms as $\Omega^{(4,0)}\rightarrow \bar{\Omega}^{(0,4)}$, based on the discussion in Section~\ref{section_Spin(7)_from_CY4}. This, in turn, implies the invariance of the Cayley 4-form. Therefore, $\sigma_0$ combined with worldsheet parity leads to a $\Spin(7)$ orientifold. Since $\sigma_0$ applies to any CY$_4$, we refer to it as the {\it universal involution}. The resulting gauge theory is derived using the rules in Section~\ref{sec:N01theorfromorienquot}. 

In general, depending on the geometry, other involutions can also exist. In the coming sections, we will present various examples of such involutions. $\mathbb{C}^4$ is special in that, as we have previously discussed, all its anti-holomorphic involutions are equivalent to the universal one.
 
The universal involution explicitly realizes the idea 
of 
$\mathcal{N}=(0,1)$ theories as ``real slices" of  $\mathcal{N}=(0,2)$ gauge theories \cite{Gukov:2019lzi}. Moreover, in this context, the real slicing admits a beautiful geometric interpretation as the $\Spin(7)$ orientifold of a CY$_4$. We can similarly think about other involutions as different real slices of the parent theories.

\subsection{$\mathbb{C}^4/\mathbb{Z}_2$}
\label{sec:C4Z2example}

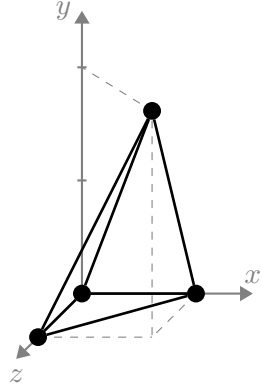
\begin{figure}[H]
    \centering
	\begin{tikzpicture}[scale=1.5]
	\draw[thick,gray,-Triangle] (0,0,0) -- node[above,pos=1] {$x$} (1.5,0,0);
	\draw[thick,gray,-Triangle] (0,0,0) -- node[left,pos=1] {$y$} (0,2.5,0);
	\draw[thick,gray] (-0.04,1,0)--(0.04,1,0);
	\draw[thick,gray] (-0.04,2,0)--(0.04,2,0);
	\draw[thin,dashed,gray] (1,0,0) -- (1,0,1) -- (1,2,1) -- (0,2,0);
	\draw[thin,dashed,gray] (0,0,1) -- (1,0,1);
	\draw[thick,gray,-Triangle] (0,0,0) -- node[below,pos=1] {$z$} (0,0,1.5);
	\node[draw=black,line width=1pt,circle,fill=black,minimum width=0.2cm,inner sep=1pt] (p1) at (0,0,0) {};
	\node[draw=black,line width=1pt,circle,fill=black,minimum width=0.2cm,inner sep=1pt] (p2) at (0,0,1) {};
	\node[draw=black,line width=1pt,circle,fill=black,minimum width=0.2cm,inner sep=1pt] (p3) at (1,0,0) {};
	\node[draw=black,line width=1pt,circle,fill=black,minimum width=0.2cm,inner sep=1pt] (p4) at (1,2,1) {};
	\draw[line width=1pt] (p1)--(p2)--(p4)--(p3)--(p1);
	\draw[line width=1pt] (p2)--(p3);
	\draw[line width=1pt] (p1)--(p4);
	\end{tikzpicture}
	\caption{Toric diagram for $\CC^4/\ZZ_2$.}
	\label{fig:C4Z2toricdiagram}
\end{figure}

Let us consider the $\CC^4/\ZZ_2$ orbifold with action $(x,y,z,w)\rightarrow (-x,-y,-z,-w)$ as the parent geometry. Its toric diagram is shown in Figure~\ref{fig:C4Z2toricdiagram}.

The corresponding 2d $\mathcal{N}=(0,2)$ theory was constructed in \cite{Franco:2015tna}. Its quiver is shown in Figure~\ref{fig:C4Z2quivN02}. 

\begin{figure}[H] 
	\centering
	\begin{subfigure}[t]{\textwidth}
	\centering
	\begin{tikzpicture}[scale=2]
	\draw[line width=1pt,redX] (-0.25,0) circle (0.25) node[fill=white,text opacity=1,fill opacity=1,draw=black,rectangle,xshift=-0.5cm,thin] {\color{redX}{$3$}} node[xshift=-0.5cm,yshift=0.75cm]{$\Lambda_{11}^1,\Lambda_{11}^2,\Lambda_{11}^3$};
	\draw[line width=1pt,redX] (2.95,0) circle (0.25) node[fill=white,text opacity=1,fill opacity=1,draw=black,rectangle,xshift=0.5cm,thin] {\color{redX}{$3$}} node[xshift=0.5cm,yshift=-0.75cm]{$\Lambda_{22}^1,\Lambda_{22}^2,\Lambda_{33}^3$};
	\node[draw=black,line width=1pt,circle,fill=yellowX,minimum width=0.75cm,inner sep=1pt] (A) at (0,0) {$1$};
	\node[draw=black,line width=1pt,circle,fill=yellowX,minimum width=0.75cm,inner sep=1pt] (B) at (2.7,0) {$2$};
	\path[Triangle-Triangle] (A) edge[line width=1pt] node[fill=white,text opacity=1,fill opacity=1,draw=black,rectangle,pos=0.25,thin] {$4$} node[fill=white,text opacity=1,fill opacity=1,draw=black,rectangle,pos=0.75,thin] {$4$} node[pos=0.30,yshift=-0.75cm] {$X_{12},Y_{12},Z_{12},W_{12}$} node[pos=0.70,yshift=0.75cm] {$X_{21},Y_{21},Z_{21},W_{21}$} (B);
	\end{tikzpicture}
	\caption{$\mathcal{N}=(0,2)$ language.}
	\label{fig:C4Z2quivN02}
	\end{subfigure}\\
	\begin{subfigure}[t]{\textwidth}
	\centering
	\begin{tikzpicture}[scale=2]
	\draw[line width=1pt,redX] (-0.4,0) circle (0.4);
	\draw[line width=1pt,redX] (-0.25,0) circle (0.25) node[fill=white,text opacity=1,fill opacity=1,draw=black,rectangle,xshift=-0.5cm,thin] {\color{redX}{$6$}} node[xshift=-0.5cm,yshift=1.1cm]{$\Lambda_{11}^1,\Lambda_{11}^2,\Lambda_{11}^3,\Lambda_{11}^{4R}$};
	\draw[line width=1pt,redX] (3.10,0) circle (0.4);
	\draw[line width=1pt,redX] (2.95,0) circle (0.25) node[fill=white,text opacity=1,fill opacity=1,draw=black,rectangle,xshift=0.5cm,thin] {\color{redX}{$6$}} node[xshift=0.5cm,yshift=-1.1cm]{$\Lambda_{22}^1,\Lambda_{22}^2,\Lambda_{33}^3,\Lambda_{22}^{4R}$};
	\node[draw=black,line width=1pt,circle,fill=yellowX,minimum width=0.75cm,inner sep=1pt] (A) at (0,0) {$1$};
	\node[draw=black,line width=1pt,circle,fill=yellowX,minimum width=0.75cm,inner sep=1pt] (B) at (2.7,0) {$2$};
	\path[Triangle-Triangle] (A) edge[line width=1pt] node[fill=white,text opacity=1,fill opacity=1,draw=black,rectangle,pos=0.25,thin] {$8$} node[fill=white,text opacity=1,fill opacity=1,draw=black,rectangle,pos=0.75,thin] {$8$} node[pos=0.30,yshift=-0.75cm] {$X_{12},Y_{12},Z_{12},W_{12}$} node[pos=0.70,yshift=0.75cm] {$X_{21},Y_{21},Z_{21},W_{21}$} (B);
	\end{tikzpicture}
	\caption{$\mathcal{N}=(0,1)$ language.}
	\label{fig:C4Z2quivN01}
	\end{subfigure}
	\caption{Quiver diagram for $\CC^4/\ZZ_2$ in $\mathcal{N}=(0,2)$ and $\mathcal{N}=(0,1)$ language.}
	\label{fig:C4Z2quiv}
\end{figure}

The $J$- and $E$-terms are:
\begin{equation}
\renewcommand{\arraystretch}{1.1}
\begin{array}{lclc}
& J                  &\text{\hspace{.5cm}} & E                    \\
\Lambda^1_{11} \,:\, & Y_{12}Z_{21}-Z_{12}Y_{21} & \quad               & W_{12}X_{21}-X_{12}W_{2 1}  \\
\Lambda^2_{11} \,:\, & Z_{12}X_{21}-X_{12}Z_{21} & \quad               & W_{12}Y_{21}-Y_{12}W_{21}   \\
\Lambda^3_{11} \,:\, & X_{12}Y_{21}-Y_{12}X_{21} & \quad               & W_{12}Z_{21}-Z_{12}W_{21}   \\
\Lambda^1_{22} \,:\, & Y_{21}Z_{12}-Z_{21}Y_{12} & \quad               & W_{21}X_{12}-X_{21}W_{12}   \\
\Lambda^2_{22} \,:\, & Z_{21}X_{12}-X_{21}Z_{12} & \quad               & W_{21}Y_{12}-Y_{21}W_{12}   \\
\Lambda^3_{22} \,:\, & X_{21}Y_{12}-Y_{21}X_{12} & \quad               & W_{21}Z_{12}-Z_{21}W_{12}   
\end{array}
\label{eq:JEterms_c4z2}
\end{equation}

Figure~\ref{fig:C4Z2quivN01} shows the quiver for this theory in $\mathcal{N}=(0,1)$ language. Denoting $W^{(0,2)}$ the superpotential obtained from \eqref{eq:JEterms_c4z2}, $W^{(0,1)}$ is given by
\begin{equation}
     W^{(0,1)} = W^{(0,2)}+
     \sum_{i,j,k=1}^2\Lambda^{4R}_{kk}( X^\dagger_{ij}X_{ij}+ Y^\dagger_{ij}Y_{ij}+ Z^\dagger_{ij}Z_{ij}+ W^\dagger_{ij}W_{ij})\fstop
\label{eq:C4Z2W01}
\end{equation}

Since the $J$- and $E$-terms in~\eqref{eq:JEterms_c4z2} are more involved, we use the HS to extract the generators of the moduli space. In Table~\ref{tab:GenerC4Z2} we present their expression in terms of chiral fields of the gauge theory.

 \begin{table}[!htp]
	\centering
	\renewcommand{\arraystretch}{1.1}
	\begin{tabular}{c|c}
		Meson    & Chiral fields  \\
		\hline
		$M_1$    & $Y_{12}Y_{21}$ \\
		$M_2$    & $Y_{12}Z_{21}=Y_{21}Z_{12}$\\
		$M_3$    & $Z_{12}Z_{21}$\\
		$M_4$    & $X_{12}Y_{21}=X_{21}Y_{12}$\\
		$M_5$    & $X_{12}Z_{21}=X_{21}Z_{12}$\\
		$M_6$    & $X_{12}X_{21}$\\
		$M_7$    & $Y_{12}W_{21}=Y_{21}W_{12}$\\
		$M_8$    & $Z_{12}W_{21}=Z_{21}W_{12}$\\
		$M_9$    & $X_{12}W_{21}=X_{21}W_{12}$\\
		$M_{10}$ & $W_{12}W_{21}$\\
	\end{tabular}
	\caption{Generators of $\CC^4/\ZZ_2$.}
	\label{tab:GenerC4Z2}
\end{table}

 The mesonic moduli space is not a complete intersection, so the PL of the HS does not terminate. We can, however, extract the relations among the generators composing the following ideal:
\begin{equation}
    \begin{split}
        \mathcal{I} = & \left\langle M_1M_3=M_2^2\coma M_1M_5=M_2M_4\coma M_3M_4=M_2M_5\coma M_1M_6=M_4^2\coma \right.\\
        & \left. M_2M_6=M_4M_5\coma M_3M_6=M_5^2\coma M_1M_8=M_2M_7\coma M_3M_7=M_2M_8\coma  \right.\\
        & \left. M_1M_9=M_4M_7\coma M_2M_9=M_4M_8\coma M_5M_7=M_2M_9\coma M_3M_9=M_5M_8\coma  \right.\\
        & \left. M_6M_7=M_4M_9\coma M_6M_8=M_5M_9\coma M_1M_{10}=M_7^2\coma M_2M_{10}=M_7M_8\coma  \right.\\
        & \left. M_3M_{10}=M_8^2\coma M_4M_{10}=M_7M_9\coma M_5M_{10}=M_8M_9\coma M_6M_{10}=M_9^2\right\rangle\fstop 
    \end{split}
\end{equation}

We now have everything necessary for identifying anti-holomorphic involutions and constructing the corresponding Spin(7) orientifolds, both from the gauge theory and from geometry.

\subsubsection{Universal Involution}

\label{sec:exampC4Z2uni} 

Let us consider the universal involution defined in Section~\ref{sec:universal involution}. It maps the two gauge groups to themselves. Chiral fields transform according to \eqref{eq:universal involution of chiral}, i.e. 
\begin{equation}
\begin{array}{cccc}
    X_{12}\rightarrow \gamma_{\Omega_1}\bar{X}_{12}\gamma_{\Omega_2}^{-1}\coma & Y_{12}\rightarrow \gamma_{\Omega_1}\bar{Y}_{12}\gamma_{\Omega_2}^{-1}\coma & Z_{12}\rightarrow \gamma_{\Omega_1}\bar{Z}_{12}\gamma_{\Omega_2}^{-1}\coma & W_{12}\rightarrow \gamma_{\Omega_1}\bar{W}_{12}\gamma_{\Omega_2}^{-1}\coma \\
    X_{21}\rightarrow \gamma_{\Omega_2}\bar{X}_{21}\gamma_{\Omega_1}^{-1}\coma & Y_{21}\rightarrow \gamma_{\Omega_2}\bar{Y}_{21}\gamma_{\Omega_1}^{-1}\coma & Z_{21}\rightarrow \gamma_{\Omega_2}\bar{Z}_{21}\gamma_{\Omega_1}^{-1}\coma & W_{21}\rightarrow \gamma_{\Omega_2}\bar{W}_{21}\gamma_{\Omega_1}^{-1}\fstop
\end{array}
\label{eq:chiralC4Z2uniinvol}
\end{equation}
The $\mathcal{N}=(0,2)$ Fermi fields transform as in \eqref{eq:universal involution of Fermi}, namely
\begin{equation}
\begin{array}{ccc}
    \Lambda^1_{11}\rightarrow \gamma_{\Omega_1}\bar{\Lambda}^1_{11}\gamma_{\Omega_1}^{-1}\coma & \Lambda^2_{11}\rightarrow \gamma_{\Omega_1}\bar{\Lambda}^2_{11}\gamma_{\Omega_1}^{-1}\coma & \Lambda^3_{11}\rightarrow \gamma_{\Omega_1}\bar{\Lambda}^3_{11}\gamma_{\Omega_1}^{-1}\coma \\
    \Lambda^1_{22}\rightarrow \gamma_{\Omega_2}\bar{\Lambda}^1_{22}\gamma_{\Omega_2}^{-1}\coma & \Lambda^2_{22}\rightarrow \gamma_{\Omega_2}\bar{\Lambda}^2_{22}\gamma_{\Omega_2}^{-1}\coma & \Lambda^3_{22}\rightarrow \gamma_{\Omega_2}\bar{\Lambda}^3_{22}\gamma_{\Omega_2}^{-1} \fstop
\end{array}
\label{eq:fermiC4Z2uniinvol}
\end{equation}
Finally, the Fermi superfields coming from the $\mathcal{N}=(0,2)$ vector multiplets transform according to \eqref{Lambda^R_projection}
\begin{equation}
\Lambda^{4R}_{11}\rightarrow \gamma_{\Omega_1}\Lambda^{4R\,\,T}_{11}\gamma_{\Omega_1}^{-1}\coma \Lambda^{4R}_{22}\rightarrow \gamma_{\Omega_2}\Lambda^{4R\,\,T}_{22}\gamma_{\Omega_2}^{-1}\fstop
\label{eq:fermiRC4Z2uniinvol}
\end{equation}
As argued in full generality in Section~\ref{sec:universal involution}, these transformations leave the superpotential $W^{(0,1)}$ in \eqref{eq:C4Z2W01} invariant.

Using Table~\ref{tab:GenerC4Z2}, we can translate this field theory involution into the geometric involution, whose action on the generators of $\mathbb{C}^4/\mathbb{Z}_2$ becomes 
\begin{equation}
	M_{a}\rightarrow \bar{M}_{a}\coma a=1,\cdots, 10 \coma
	\label{eq:C4Z2mesonuniinvol}
\end{equation}
as expected for the universal involution.

The gauge symmetry and the projections of matter fields in the orientifolded theory are controlled by $\gamma_{\Omega_1}$ and $\gamma_{\Omega_2}$. According to the discussion in Section~\ref{sec:N01theorfromorienquot}, the choices of $\gamma_{\Omega_1}$ and $\gamma_{\Omega_2}$ are not independent. In this case, they should satisfy $\gamma_{\Omega_1}=\gamma_{\Omega_2}$. To show this correlation, we consider the effect of acting with the involution twice. For example, acting on $X_{12}$ we obtain
\begin{equation}\label{eq:square of involution}
	X_{12}\rightarrow \gamma_{\Omega_1}\bar{\gamma}_{\Omega_1}X_{12}\bar{\gamma}_{\Omega_2}^{-1}\gamma_{\Omega_2}^{-1}\coma
\end{equation}
which should be equal to the identity transformation. Since $\gamma_{\Omega_i}$ is equal to $\ID_N$ or $J$, this implies that  $\gamma_{\Omega_1}=\gamma_{\Omega_2}$. Repeating this analysis for any other bifundamental field leads to the same condition. We conclude that the gauge symmetry of the orientifolded theory is 
either $\SO(N)\times \SO(N)$ or $\USp(N)\times \USp(N)$.

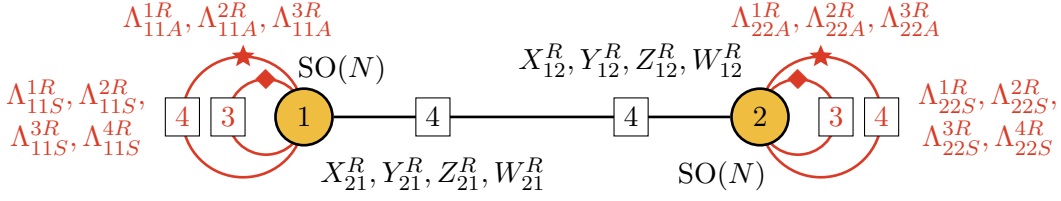
\begin{figure}[H]
	\centering
	\begin{tikzpicture}[scale=2]
	\draw[line width=1pt,redX] (-0.4,0) circle (0.4) node[fill=white,text opacity=1,fill opacity=1,draw=black,rectangle,xshift=-0.8cm,thin] {\color{redX}{$4$}} node[yshift=0.8cm,star,star points=5, star point ratio=2.25, inner sep=1pt, fill=redX, draw=redX] {} node[xshift=-2.2cm,yshift=0.0cm] {\color{redX}{\shortstack{$\Lambda^{1R}_{11S},\Lambda^{2R}_{11S}$,\\$\Lambda^{3R}_{11S},\Lambda^{4R}_{11S}$}}};
	\draw[line width=1pt,redX] (-0.25,0) circle (0.25) node[fill=white,text opacity=1,fill opacity=1,draw=black,rectangle,xshift=-0.5cm,thin] {\color{redX}{$3$}} node[yshift=0.5cm] {\color{redX}{\scriptsize{$\quadro$}}} node[xshift=-0.5cm,yshift=1.3cm] {\color{redX}{$\Lambda^{1R}_{11A},\Lambda^{2R}_{11A},\Lambda^{3R}_{11A}$}};
	\draw[line width=1pt,redX] (3.4,0) circle (0.4) node[fill=white,text opacity=1,fill opacity=1,draw=black,rectangle,xshift=0.8cm,thin] {\color{redX}{$4$}} node[yshift=0.8cm,star,star points=5, star point ratio=2.25, inner sep=1pt, fill=redX, draw=redX] {} node[xshift=2.2cm,yshift=0.0cm] {\color{redX}{\shortstack{$\Lambda^{1R}_{22S},\Lambda^{2R}_{22S}$,\\$\Lambda^{3R}_{22S},\Lambda^{4R}_{22S}$}}};
	\draw[line width=1pt,redX] (3.25,0) circle (0.25) node[fill=white,text opacity=1,fill opacity=1,draw=black,rectangle,xshift=0.5cm,thin] {\color{redX}{$3$}} node[yshift=0.5cm] {\color{redX}{\scriptsize{$\quadro$}}} node[xshift=0.5cm,yshift=1.3cm] {\color{redX}{$\Lambda^{1R}_{22A},\Lambda^{2R}_{22A},\Lambda^{3R}_{22A}$}};
	\node[draw=black,line width=1pt,circle,fill=yellowX,minimum width=0.75cm,inner sep=1pt,label={[xshift=0.5cm,yshift=-0.1cm]:$\SO(N)$}] (A) at (0,0) {$1$};
	\node[draw=black,line width=1pt,circle,fill=yellowX,minimum width=0.75cm,inner sep=1pt,label={[xshift=-0.5cm,yshift=-1.5cm]:$\SO(N)$}] (B) at (3,0) {$2$};
	\path
	(A) edge[line width=1pt] node[fill=white,text opacity=1,fill opacity=1,draw=black,rectangle,pos=0.25,thin] {$4$} node[pos=0.25,yshift=-0.75cm] {$X^R_{21},Y^R_{21},Z^R_{21},W^R_{21}$} node[fill=white,text opacity=1,fill opacity=1,draw=black,rectangle,pos=0.75,thin] {$4$} node[pos=0.75,yshift=0.75cm] {$X^R_{12},Y^R_{12},Z^R_{12},W^R_{12}$} (B);
	\end{tikzpicture}
	\caption{Quiver for the Spin(7) orientifold of $\CC^4/\ZZ_2$ using the universal involution.}
	\label{fig:C4Z2uniinvol}
	\end{figure}

For concreteness, let us focus on the $\SO(N)\times \SO(N)$ case. Figure~\ref{fig:C4Z2uniinvol} shows the corresponding quiver. There are eight real bifundamental scalars, coming from the bifundamental chiral fields in the parent.\footnote{In what follows, we will use the term bifundamental in the case of matter fields that connect pairs of nodes, even when one or both of them is either SO or USp.} Every adjoint complex Fermi in the parent is projected to one symmetric and one antisymmetric real Fermi fields, while the adjoint real Fermi fields from the $\mathcal{N}=(0,2)$ vector multiplets are projected to the symmetric representation. It is rather straightforward to write the projected superpotential but, for brevity, we omit it here and in the examples that follow. Finally, it is easy to verify the vanishing of gauge anomalies.

We would like to mention that, although the above models are built as orientifolds of the $\mathbb{C}^4/\mathbb{Z}_2$ theory, they can be equivalently regarded as $\mathbb{Z}_2$ orbifolds of the orientifolds of $\mathbb{C}^4$ in Section \ref{sec:C4example}. This viewpoint is useful to display that the models inherit the $\SO(4)^2$ global symmetry of the $\mathbb{C}^4$ orientifolds, since the $\mathbb{Z}_2$ orbifold acts in the same way on the coordinates within each 4-plet. In fact, it is easy to gather the different multiplets in $\SO(4)^2$ representations (including the gauginos in the $\mathcal{N}=(0,1)$ vector multiplet, as befits an R-symmetry). We leave the check of the $\SO(4)^2$ invariance of the interactions as an exercise for the interested reader. Similar remarks apply to other orientifolds of $\mathbb{C}^4/\mathbb{Z}_2$ in coming sections.


\subsubsection{Beyond the Universal Involution: an $\SO(N)\times \USp(N)$ Theory}
\label{sec:exampC4Z2beyonduni}

Let us now consider another involution, which also maps the two gauge groups to themselves but transforms chiral fields differently, according to
\begin{equation}
\begin{array}{cccc}
    X_{12}\rightarrow \gamma_{\Omega_1}\bar{Y}_{12}\gamma_{\Omega_2}^{-1}\coma & Y_{12}\rightarrow -\gamma_{\Omega_1}\bar{X}_{12}\gamma_{\Omega_2}^{-1}\coma & Z_{12}\rightarrow \gamma_{\Omega_1}\bar{W}_{12}\gamma_{\Omega_2}^{-1}\coma & W_{12}\rightarrow -\gamma_{\Omega_1}\bar{Z}_{12}\gamma_{\Omega_2}^{-1}\coma \\
    X_{21}\rightarrow \gamma_{\Omega_2}\bar{Y}_{21}\gamma_{\Omega_1}^{-1}\coma & Y_{21}\rightarrow -\gamma_{\Omega_2}\bar{X}_{21}\gamma_{\Omega_1}^{-1}\coma & Z_{21}\rightarrow \gamma_{\Omega_2}\bar{W}_{21}\gamma_{\Omega_1}^{-1}\coma & W_{21}\rightarrow -\gamma_{\Omega_2}\bar{Z}_{21}\gamma_{\Omega_1}^{-1}\fstop
\end{array}
\label{eq:chiralC4Z2beyuniinvol}
\end{equation}

Invariance of $W^{(0,1)}$ in \eqref{eq:C4Z2W01} implies that the Fermi fields transform as 
\begin{equation}
\begin{array}{ccc}
 \Lambda^1_{11}\rightarrow \gamma_{\Omega_1}\Lambda^{1\,\,T}_{11}\gamma_{\Omega_1}^{-1}\coma & \Lambda^2_{11}\rightarrow \gamma_{\Omega_1}\Lambda^{2\,\,T}_{11}\gamma_{\Omega_1}^{-1}\coma & \Lambda^3_{11}\rightarrow \gamma_{\Omega_1}\bar{\Lambda}^3_{11}\gamma_{\Omega_1}^{-1}\coma \\
    \Lambda^1_{22}\rightarrow \gamma_{\Omega_2}\Lambda^{1\,\,T}_{22}\gamma_{\Omega_2}^{-1}\coma & \Lambda^2_{22}\rightarrow \gamma_{\Omega_2}\Lambda^{2\,\,T}_{22}\gamma_{\Omega_2}^{-1}\coma & \Lambda^3_{22}\rightarrow \gamma_{\Omega_2}\bar{\Lambda}^3_{22}\gamma_{\Omega_2}^{-1}\coma
\end{array}
\label{eq:fermiC4Z2beyuniinvol}
\end{equation}
and
\begin{equation}
\Lambda^{4R}_{11}\rightarrow \gamma_{\Omega_1}\Lambda^{4R\,\,T}_{11}\gamma_{\Omega_1}^{-1}\coma \Lambda^{4R}_{22}\rightarrow \gamma_{\Omega_2}\Lambda^{4R\,\,T}_{22}\gamma_{\Omega_2}^{-1}\fstop
\label{eq:fermiRC4Z2beyuniinvol}
\end{equation}

Using Table~\ref{tab:GenerC4Z2}, this translates into the following geometric involution
\begin{equation}
    \begin{array}{c}
    \left(M_1,M_2,M_3,M_4,M_5,M_6,M_7,M_8,M_9,M_{10},\right)\\
    \downarrow\\
 \left(\bar{M}_6,-\bar{M}_9,\bar{M}_{10},-\bar{M}_4,\bar{M}_7,\bar{M}_1,\bar{M}_5,-\bar{M}_{8},-\bar{M}_2,\bar{M}_3\right)\fstop
 \end{array}
 \label{eq:C4Z2mesonbeyonuniinvol}
 \end{equation}

As in the previous example, the choices of $\gamma_{\Omega_1}$ and $\gamma_{\Omega_2}$ are correlated because they are connected by matter fields. From \eqref{eq:chiralC4Z2beyuniinvol}, we conclude that for each pair of chiral fields that are mapped to each other, the involution corresponds to the case $\eta=\pm\left(\begin{smallmatrix} 0&1\\-1&0
\end{smallmatrix}\right)$ in \eqref{scalar_involution}. Following to the discussion in Section~\ref{sec:N01theorfromorienquot}, in this case the gauge groups project to $\SO(N)\times \USp(N)$. We can explicitly see this constraint by considering the square of the involution on, e.g., $X_{12}$, for which we obtain
\begin{equation}\label{eq:square of involution2}
	X_{12}\rightarrow -\gamma_{\Omega_1}\bar{\gamma}_{\Omega_1}X_{12}\bar{\gamma}_{\Omega_2}^{-1}\gamma_{\Omega_2}^{-1}\coma
\end{equation}
which should be equal to the identity. This implies that $\gamma_{\Omega_1}=\ID_N$ and  $\gamma_{\Omega_2}=J$ or $\gamma_{\Omega_1}=J$ and $\gamma_{\Omega_2}=\ID_N$. The other chiral fields lead to the same condition.

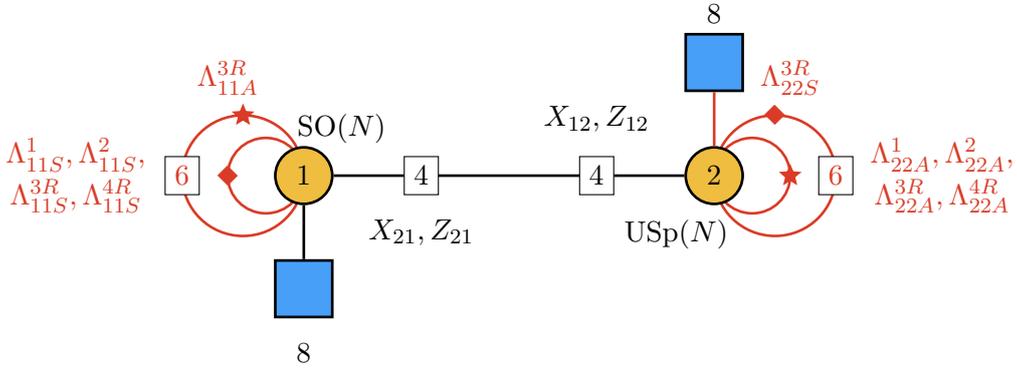
\begin{figure}[!htp]
	\centering
	\begin{tikzpicture}[scale=2]
	\draw[line width=1pt,redX] (-0.4,0) circle (0.4) node[fill=white,text opacity=1,fill opacity=1,draw=black,rectangle,xshift=-0.8cm,thin] {\color{redX}{$6$}} node[yshift=0.8cm,star,star points=5, star point ratio=2.25, inner sep=1pt, fill=redX, draw=redX] {} node[xshift=-2.2cm,yshift=0.0cm] {\color{redX}{\shortstack{$\Lambda^{1}_{11S},\Lambda^{2}_{11S}$,\\$\Lambda^{3R}_{11S},\Lambda^{4R}_{11S}$}}};
	\draw[line width=1pt,redX] (-0.25,0) circle (0.25)  node[xshift=-0.5cm] {\color{redX}{\scriptsize{$\quadro$}}} node[xshift=-0.5cm,yshift=1.3cm] {\color{redX}{$\Lambda^{3R}_{11A}$}};
	\draw[line width=1pt,redX] (3.1,0) circle (0.4) node[fill=white,text opacity=1,fill opacity=1,draw=black,rectangle,xshift=0.8cm,thin] {\color{redX}{$6$}} node[yshift=0.8cm] {\color{redX}{\scriptsize{$\quadro$}}} node[xshift=2.2cm,yshift=0.0cm] {\color{redX}{\shortstack{$\Lambda^{1}_{22A},\Lambda^{2}_{22A}$,\\$\Lambda^{3R}_{22A},\Lambda^{4R}_{22A}$}}};
	\draw[line width=1pt,redX] (2.95,0) circle (0.25) node[xshift=0.5cm,star,star points=5, star point ratio=2.25, inner sep=1pt, fill=redX, draw=redX] {} node[xshift=0.5cm,yshift=1.3cm] {\color{redX}{$\Lambda^{3R}_{22S}$}};
	\node[draw=black,line width=1pt,circle,fill=yellowX,minimum width=0.75cm,inner sep=1pt,label={[xshift=0.5cm,yshift=-0.1cm]:$\SO(N)$}] (A) at (0,0) {$1$};
	\node[draw=black,line width=1pt,circle,fill=yellowX,minimum width=0.75cm,inner sep=1pt,label={[xshift=-0.5cm,yshift=-1.5cm]:$\USp(N)$}] (B) at (2.7,0) {$2$};
 	\node[draw=black,line width=1pt,square,fill=blueX,minimum width=0.75cm,inner sep=1pt,label={[yshift=-1.5cm]:$8$}] (C) at (0,-0.75) {};
 	\node[draw=black,line width=1pt,square,fill=blueX,minimum width=0.75cm,inner sep=1pt,label={[yshift=0cm]:$8$}] (D) at (2.7,0.75) {};
	\path
	(A) edge[line width=1pt] node[fill=white,text opacity=1,fill opacity=1,draw=black,rectangle,pos=0.25,thin] {$4$} node[pos=0.25,yshift=-0.75cm] {$X_{21},Z_{21}$} node[fill=white,text opacity=1,fill opacity=1,draw=black,rectangle,pos=0.75,thin] {$4$} node[pos=0.75,yshift=0.75cm] {$X_{12},Z_{12}$} (B);
	\draw[line width=1pt] (A) -- (C);
	\draw[line width=1pt,redX] (B) -- (D);
	\end{tikzpicture}
	\caption{Quiver for the Spin(7) orientifold of $\CC^4/\ZZ_2$ using the involution in \eqref{eq:chiralC4Z2beyuniinvol}, \eqref{eq:fermiC4Z2beyuniinvol} and \eqref{eq:fermiRC4Z2beyuniinvol}. The squares indicate the number of flavors necessary to cancel gauge anomalies.}
	\label{fig:C4Z2beyuniinvol}
	\end{figure}

The resulting quiver is shown in Figure~\ref{fig:C4Z2beyuniinvol}. This theory suffers from gauge anomalies, which can be canceled by adding eight scalar flavors to the SO group and eight Fermi flavors to the USp group.

We would like to emphasize the fact that most of the orientifold theories in this paper actually do not require flavor branes to cancel their anomalies. Our expectation is that this is due to the relative simplicity of the singularities considered, at the level of their structure of collapsed cycles (for instance, their toric diagrams have no collapsed cycles), and that orientifold of more general singularities are likely to require flavor branes. This is somewhat similar to the CY$_3$ case, in which ``simple" singularities (i.e. not having interior points) generically lead to theories not requiring flavor branes, and only specific cases require them \cite{Park:1999eb}. Hence, the above example is particularly remarkable, and possibly illustrates, in a relatively simple setup, a feature which may be generic in orientifolds of more involved CY$_4$ singularities.

\section{Choice of Vector Structure}
\label{sec:vectorstructure}

\subsection{Vector Structure in Type IIB Orientifold Construction}

The $\mathbb{C}^4/\ZZ_2$ example serves to address an important point, which will apply to many others of our more general examples discussed later. As already pointed out at the end of Section~\ref{sec:HSreview}, when orientifolding by a certain geometric action, there are certain discrete choices which lead to different orientifolds for the same geometric action. One such choice is the already mentioned $\SO/\USp$ projection; in this section we discuss a second (and independent) choice, corresponding to the existence or not of vector structure in certain singularities.

This possibility was first uncovered for singularities obtained as orbifolds of flat space by even order groups, e.g. $\mathbb{C}^n/\ZZ_{2N}$, triggered by the analysis in \cite{Polchinski:1996ry} of 6d orientifold models \cite{Pradisi:1988xd,Gimon:1996rq}, in particular orientifolds of $\mathbb{C}^2/\mathbb{Z}_2$. The key observation is that in such orbifolds, the orientifold acts by mapping a sector twisted by an element $\theta^k$ to the $\theta^{-k}$-twisted sector, and hence for even order $\mathbb{Z}_{2N}$, the $\theta^N$-twisted sector is mapped to itself and there are two possible choices of sign in this action. In the open string perspective, the two possibilities correspond to choices of Chan-Paton actions satisfying
\begin{equation}
\gamma_{\theta^N}=\pm \gamma_{\Omega}\gamma_{\theta^N}^T\gamma_{\Omega}^{-1}\fstop
\end{equation}
The relation with vector structure (namely, the possibility that the gauge bundle defined by the Chan-Paton matrices admits objects in the vector representation or not) was further clarified in \cite{Berkooz:1996iz} (also \cite{Witten:1997bs}). 

Although these ideas arose in the 6d orbifold context, they are far more general. For instance, the choice of vector structure has appeared in the construction of orientifolds of toroidal orbifolds in \cite{Aldazabal:1998mr}. In such compact setups, the choice of orientifolds with vector structure sometimes requires the introduction of anti-branes~\cite{Antoniadis:1999xk, Aldazabal:1999jr}; however, this is due to untwisted RR tadpoles, and hence any choice of vector structure leads to consistent orientifolds of non-compact $\mathbb{C}^2/\mathbb{Z}_{2N}$ singularities (see e.g. the constructions in 6d in \cite{Blum:1997fw,Blum:1997mm} and in 4d in \cite{Park:1998zh,Park:1999eb}). An even more important generalization is that the existence of a discrete choice of vector structure in the orientifold action generalizes beyond orbifold singularities, and applies to a far broader set of singularities. This was tacitly included in the construction of general orientifolds of general toric Calabi-Yau 3-fold singularities in \cite{Franco:2007ii}. 

In practical terms,  the appearance of the choices of vector structure in orientifolding  arises when, for a given geometry, there are different $\ZZ_2$ symmetries on the underlying quiver gauge theory, which differ in the action on the quiver nodes: an orientifold whose action on nodes is pairwise exchange, with no nodes mapped to themselves, corresponds to an action without vector structure, whereas the presence of nodes mapped to themselves corresponds to an action with vector structure.\footnote{There are cases, e.g. orbifolds by products of cyclic groups $\mathbb{Z}_N\times \mathbb{Z}_M$ etc., in which the orientifold may act with vector structure with respect to the $\mathbb{Z}_N$ and without vector structure with respect to the $\mathbb{Z}_M$. For simplicity, we ignore these more subtle possibilities and stick to the stated convention, as a practical reference device.}

We thus expect that the choice of vector structure will arise in our present setup of $\Spin(7)$ orientifolds of Calabi-Yau 4-fold singularities. In particular, for orbifolds of $\CC^4$, this should already follow from the early analysis in \cite{Polchinski:1996ry}. From this perspective, the orientifold of $\mathbb{C}^4/\ZZ_2$ constructed in Section \ref{sec:C4Z2example} corresponds to an orientifold action with vector structure, since each of the two gauge factors of the underlying $\mathcal{N}=(0,2)$ theory are mapped to themselves. Our discussion suggests that it should be possible to construct an orientifold of the same geometry, with the same orientifold geometric action, but without vector structure. This corresponds to  the symmetry of the $\mathcal{N}=(0,2)$ theory that exchanges pairwise the two gauge factors. We will indeed build this orientifold without vector structure in the following section.

This brings about an important observation. The universal involution in Section \ref{sec:universal involution} maps each gauge factor of the $\mathcal{N}=(0,2)$ theory to itself, hence it corresponds to actions with vector structure. Therefore, in geometries admitting it, the choice of orientifold action without vector structure must correspond to orientifolds actions beyond the universal involution. Thus, the possibility of choosing the vector structure is already ensuring that the set of orientifold theories is substantially larger than the class provided by the universal involution.

\subsection{$\mathbb{C}^4/\mathbb{Z}_2$ Revisited: an Orientifold Without Vector Structure} 
\label{sec:C4Z2-revisited}

Let us revisit the $\CC^4/\ZZ_2$ theory, but this time consider an anti-holomorphic involution that maps one gauge group to the other. A possible involution of the chiral fields reads
\begin{equation}
 \begin{array}{cccc}
    X_{12}\rightarrow  \gamma_{\Omega_2}\bar{X}_{21}\gamma_{\Omega_1}^{-1}\coma & Y_{12}\rightarrow  \gamma_{\Omega_2}\bar{Y}_{21}\gamma_{\Omega_1}^{-1}\coma & Z_{12}\rightarrow  \gamma_{\Omega_2}\bar{Z}_{21}\gamma_{\Omega_1}^{-1}\coma & W_{12}\rightarrow \gamma_{\Omega_2}\bar{W}_{21}\gamma_{\Omega_1}^{-1}\coma \\
    X_{21}\rightarrow  \gamma_{\Omega_1}\bar{X}_{12}\gamma_{\Omega_2}^{-1}\coma & Y_{21}\rightarrow  \gamma_{\Omega_1}\bar{Y}_{12}\gamma_{\Omega_2}^{-1}\coma & Z_{21}\rightarrow  \gamma_{\Omega_1}\bar{Z}_{12}\gamma_{\Omega_2}^{-1}\coma & W_{21}\rightarrow \gamma_{\Omega_1}\bar{W}_{12}\gamma_{\Omega_2}^{-1}\fstop
 \end{array}
    \label{eq:chirfieldsC4Z2-revisited}
\end{equation}

Invariance of $W^{(0,1)}$ in \eqref{eq:C4Z2W01} implies that Fermi fields transform as
\begin{equation}
\begin{array}{ccc}
    \Lambda^1_{11}\rightarrow \gamma_{\Omega_2}\bar{\Lambda}^1_{22}\gamma_{\Omega_2}^{-1}\coma & \Lambda^2_{11}\rightarrow \gamma_{\Omega_2}\bar{\Lambda}^2_{22}\gamma_{\Omega_2}^{-1}\coma & \Lambda^3_{11}\rightarrow \gamma_{\Omega_2}\bar{\Lambda}^3_{22}\gamma_{\Omega_2}^{-1}\coma\\
    \Lambda^1_{22}\rightarrow \gamma_{\Omega_1}\bar{\Lambda}^1_{11}\gamma_{\Omega_1}^{-1}\coma & \Lambda^2_{22}\rightarrow \gamma_{\Omega_1}\bar{\Lambda}^2_{11}\gamma_{\Omega_1}^{-1}\coma & \Lambda^3_{22}\rightarrow \gamma_{\Omega_1}\bar{\Lambda}^3_{11}\gamma_{\Omega_1}^{-1}\coma
    \end{array}
    \label{eq:fermifieldsC4Z2-revisited}
\end{equation}
and
\begin{equation}
    \Lambda^{4R}_{11}\rightarrow \gamma_{\Omega_2}\Lambda^{4R\,\,T}_{22}\gamma_{\Omega_2}^{-1}\coma
    \Lambda^{4R}_{22}\rightarrow \gamma_{\Omega_1}\Lambda^{4R\,\,T}_{11}\gamma_{\Omega_1}^{-1}\fstop
    \label{eq:realfermifieldsC4Z2-revisited}
\end{equation}

\begin{figure}[!htp]
	\centering
	\begin{tikzpicture}[scale=2]
	\draw[line width=1pt] (0,0.43) circle (0.43) node[yshift=0.7cm,xshift=0.5cm,star,star points=5, star point ratio=2.25, inner sep=1pt, fill=black, draw] {}  node[fill=white,text opacity=1,fill opacity=1,draw=black,rectangle,yshift=0.85cm,thin] {$8$} node[yshift=0.85cm,xshift=1.85cm] {$X_S,Y_S,Z_S,W_S$};
	\draw[line width=1pt] (0,0.25) circle (0.25)  node[xshift=0.5cm] {\scriptsize{$\quadro$}} node[fill=white,text opacity=1,fill opacity=1,draw=black,rectangle,yshift=0.5cm,thin] {$8$} node[yshift=0.5cm,xshift=2.25cm] {$X_A,Y_A,Z_A,W_A$};
	\draw[line width=1pt,redX] (0,-0.25) circle (0.25)  node[fill=white,text opacity=1,fill opacity=1,draw=black,rectangle,yshift=-0.5cm,thin] {\color{redX}{$6$}} node[yshift=-0.5cm,xshift=1.75cm] {$\Lambda^1,\Lambda^2,\Lambda^3$};
	\draw[line width=1pt,redX] (0,-0.43) circle (0.43) node[yshift=-0.85cm,xshift=1.2cm] {$\Lambda^{4R}$};
	\node[draw=black,line width=1pt,circle,fill=yellowX,minimum width=0.75cm,inner sep=1pt,label={[xshift=1.25cm,yshift=-0.7cm]:$\U(N)$}] (A) at (0,0) {};
	\end{tikzpicture}
	\caption{Quiver for the Spin(7) orientifold of $\CC^4/\ZZ_2$ using the involution in \eqref{eq:chirfieldsC4Z2-revisited}, \eqref{eq:fermifieldsC4Z2-revisited} and \eqref{eq:realfermifieldsC4Z2-revisited}. The underlying geometric involution coincides with the one for the model in Figure~\ref{fig:C4Z2uniinvol}, but both theories differ in the vector structure.}
	\label{fig:C4Z2-revisited}
\end{figure}
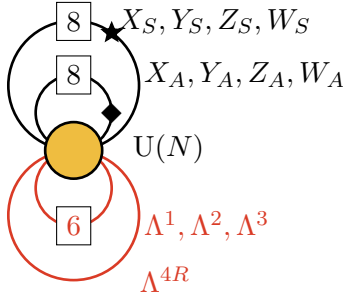

Using Table~\ref{tab:GenerC4Z2}, \eqref{eq:chirfieldsC4Z2-revisited} translates into the following geometric involution
\begin{equation}
	M_{a}\rightarrow \bar{M}_{a}\coma a=1,\cdots, 10 \coma
	\label{eq:C4Z2mesonuniinvol_without_vector_structure}
\end{equation}
which coincides with \eqref{eq:C4Z2mesonuniinvol}. This model and the one in Section~\ref{sec:exampC4Z2uni} provide concrete examples in which the same geometric action but different choices of vector structure lead to different Spin(7) orientifolds. The resulting quiver is shown in Figure~\ref{fig:C4Z2-revisited}. It is free of gauge anomalies.

 We would like to conclude this discussion with an interesting observation: in our example, the orientifold models with/without vector structure differ also in the fact that one requires flavor branes to cancel anomalies, while the other does not. In fact, this feature has also been encountered in the $4$d case of D$3$-branes at (orientifolds of) CY$_3$. For instance, in the $4$d ${\cal N}=1$ orientifolds of even order orbifolds $\mathbb{C}^2/{\mathbb{Z}_k}$  theories in \cite{Park:1999eb}, models without/with vector structure were shown to require/not require flavor D$7$-branes.\footnote{In a T-dual type IIA picture with D4-branes suspended between $k$ NS-branes, in the presence of two O6$^\prime$-planes, the two possibilities differ in having the NS-branes splitting/not splitting the O6$^\prime$-planes in halves. In the former case, the orientifold plane charge flips sign across the NS-brane and charge conservation requires the introduction of additional half-D6-branes (i.e. flavor branes) for consistency \cite{Brunner:1998jr} (see also \cite{Brunner:1997gf,Hanany:1997gh}).}
 
 \smallskip
 
\section{Beyond Orbifold Singularities} 
\label{sec:BeyondOrbif}

In this section, we construct Spin(7) orientifolds in which the parent theory is a non-orbifold toric CY$_4$.

\subsection{$D_3$}
\label{sec:D3examp}

\begin{figure}[H]
   	\centering
	\begin{tikzpicture}[scale=1.5]
	\draw[thick,gray,-Triangle] (0,0,0) -- node[above,pos=1] {$x$} (1.5,0,0);
	\draw[thick,gray,-Triangle] (0,0,0) -- node[left,pos=1] {$y$} (0,1.5,0);
	\draw[thick,gray,-Triangle] (0,0,0) -- node[below,pos=1] {$z$} (0,0,1.5);
	\node[draw=black,line width=1pt,circle,fill=black,minimum width=0.2cm,inner sep=1pt] (p1) at (0,0,0) {};
	\node[draw=black,line width=1pt,circle,fill=black,minimum width=0.2cm,inner sep=1pt] (p2) at (1,0,0) {};
	\node[draw=black,line width=1pt,circle,fill=black,minimum width=0.2cm,inner sep=1pt] (p3) at (0,0,1) {};
	\node[draw=black,line width=1pt,circle,fill=black,minimum width=0.2cm,inner sep=1pt] (p4) at (1,0,1) {};
	\node[draw=black,line width=1pt,circle,fill=black,minimum width=0.2cm,inner sep=1pt] (p5) at (0,1,0) {};
	\node[draw=black,line width=1pt,circle,fill=black,minimum width=0.2cm,inner sep=1pt] (p6) at (1,1,0) {};
	\draw[line width=1pt] (p1)--(p2)--(p4)--(p3)--(p1);
	\draw[line width=1pt] (p1)--(p5)--(p6)--(p2);
	\draw[line width=1pt] (p6)--(p4);
	\draw[line width=1pt] (p5)--(p3);
	\end{tikzpicture}
	\caption{Toric diagram for $D_3$.}
	\label{fig:D3toricdiagram}
\end{figure}
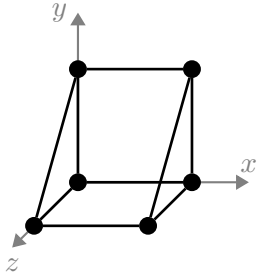

Let us consider the CY$_4$ with toric diagram shown in Figure~\ref{fig:D3toricdiagram}. This geometry is often referred to as $D_3$.

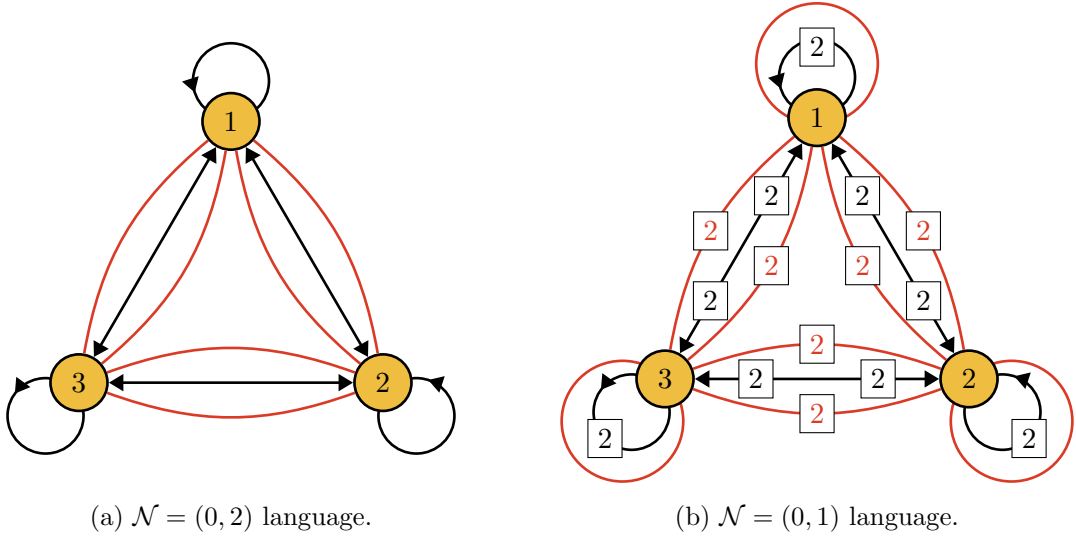
\begin{figure}[H]
	\centering
	\begin{subfigure}[t]{0.49\textwidth}
	\centering
	\begin{tikzpicture}[scale=2]
	\draw[line width=1pt,decoration={markings, mark=at position 0.40 with{\arrow{Triangle}}}, postaction={decorate}] (-0.22,-0.22) circle (0.25);
	\draw[line width=1pt,decoration={markings, mark=at position 0.20 with{\arrow{Triangle}}}, postaction={decorate}] (2.22,-0.22) circle (0.25);
	\draw[line width=1pt,decoration={markings, mark=at position 0.55 with{\arrow{Triangle}}}, postaction={decorate}] (1,2) circle (0.25);
	\node[draw=black,line width=1pt,circle,fill=yellowX,minimum width=0.75cm,inner sep=1pt] (A) at (0,0) {$3$};
	\node[draw=black,line width=1pt,circle,fill=yellowX,minimum width=0.75cm,inner sep=1pt] (B) at (2,0) {$2$};
	\node[draw=black,line width=1pt,circle,fill=yellowX,minimum width=0.75cm,inner sep=1pt] (C) at (1,1.732) {$1$};
	\path[Triangle-Triangle] (A) edge[line width=1pt] (B);
	\path[Triangle-Triangle] (B) edge[line width=1pt] (C);
	\path[Triangle-Triangle] (C) edge[line width=1pt] (A);
	\draw[line width=1pt,redX] (A) to[bend right=20]  (B);
	\draw[line width=1pt,redX] (B) to[bend right=20]  (C);
	\draw[line width=1pt,redX] (C) to[bend right=20]  (A);
	\draw[line width=1pt,redX] (A) to[bend left=20]  (B);
	\draw[line width=1pt,redX] (B) to[bend left=20]  (C);
	\draw[line width=1pt,redX] (C) to[bend left=20]  (A);
	\node at (0,-0.6) {};
	\end{tikzpicture}
	\caption{$\mathcal{N}=(0,2)$ language.}
	\label{fig:D3quivN02}
	\end{subfigure}\hfill
	\begin{subfigure}[t]{0.49\textwidth}
	\centering
	\begin{tikzpicture}[scale=2]
	\draw[line width=1pt,redX] (-0.28,-0.28) circle (0.4);
	\draw[line width=1pt,redX] (2.28,-0.28) circle (0.4);
	\draw[line width=1pt,redX] (1,2.09) circle (0.4);
	\draw[line width=1pt,decoration={markings, mark=at position 0.40 with{\arrow{Triangle}}}, postaction={decorate}] (-0.22,-0.22) circle (0.25) node[fill=white,text opacity=1,fill opacity=1,draw=black,rectangle,thin,xshift=-0.35cm,yshift=-0.35cm] {$2$};
	\draw[line width=1pt,decoration={markings, mark=at position 0.20 with{\arrow{Triangle}}}, postaction={decorate}] (2.22,-0.22) circle (0.25) node[fill=white,text opacity=1,fill opacity=1,draw=black,rectangle,thin,xshift=0.35cm,yshift=-0.35cm] {$2$};
	\draw[line width=1pt,decoration={markings, mark=at position 0.55 with{\arrow{Triangle}}}, postaction={decorate}] (1,2) circle (0.25) node[fill=white,text opacity=1,fill opacity=1,draw=black,rectangle,thin,yshift=0.4cm] {$2$};
	\node[draw=black,line width=1pt,circle,fill=yellowX,minimum width=0.75cm,inner sep=1pt] (A) at (0,0) {$3$};
	\node[draw=black,line width=1pt,circle,fill=yellowX,minimum width=0.75cm,inner sep=1pt] (B) at (2,0) {$2$};
	\node[draw=black,line width=1pt,circle,fill=yellowX,minimum width=0.75cm,inner sep=1pt] (C) at (1,1.732) {$1$};
	\path[Triangle-Triangle] (A) edge[line width=1pt] node[fill=white,text opacity=1,fill opacity=1,draw=black,rectangle,thin,pos=0.25] {$2$} node[fill=white,text opacity=1,fill opacity=1,draw=black,rectangle,thin,pos=0.75] {$2$} (B);
	\path[Triangle-Triangle] (B) edge[line width=1pt] node[fill=white,text opacity=1,fill opacity=1,draw=black,rectangle,thin,pos=0.25] {$2$} node[fill=white,text opacity=1,fill opacity=1,draw=black,rectangle,thin,pos=0.75] {$2$} (C);
	\path[Triangle-Triangle] (C) edge[line width=1pt] node[fill=white,text opacity=1,fill opacity=1,draw=black,rectangle,thin,pos=0.25] {$2$} node[fill=white,text opacity=1,fill opacity=1,draw=black,rectangle,thin,pos=0.75] {$2$} (A);
	\draw[line width=1pt,redX] (A) to[bend right=20] node[fill=white,text opacity=1,fill opacity=1,draw=black,rectangle,thin] {\color{redX}{$2$}} (B);
	\draw[line width=1pt,redX] (B) to[bend right=20] node[fill=white,text opacity=1,fill opacity=1,draw=black,rectangle,thin] {\color{redX}{$2$}} (C);
	\draw[line width=1pt,redX] (C) to[bend right=20] node[fill=white,text opacity=1,fill opacity=1,draw=black,rectangle,thin] {\color{redX}{$2$}} (A);
	\draw[line width=1pt,redX] (A) to[bend left=20] node[fill=white,text opacity=1,fill opacity=1,draw=black,rectangle,thin] {\color{redX}{$2$}} (B);
	\draw[line width=1pt,redX] (B) to[bend left=20] node[fill=white,text opacity=1,fill opacity=1,draw=black,rectangle,thin] {\color{redX}{$2$}} (C);
	\draw[line width=1pt,redX] (C) to[bend left=20] node[fill=white,text opacity=1,fill opacity=1,draw=black,rectangle,thin] {\color{redX}{$2$}} (A);
	\end{tikzpicture}
	\caption{$\mathcal{N}=(0,1)$ language.}
	\label{fig:D3quivN01}
	\end{subfigure}
	\caption{Quiver diagram for $D_3$ in $\mathcal{N}=(0,2)$ and $\mathcal{N}=(0,1)$ language.}
	\label{fig:D3quiv}
\end{figure}

The $\mathcal{N}=(0,2)$ theory on D1-branes probing $D_3$ was first derived in \cite{Franco:2015tna}. Its quiver diagram is shown in Figure~\ref{fig:D3quivN02}. 

The $J$- and $E$-terms read
\begin{equation}
\renewcommand{\arraystretch}{1.1}
\begin{array}{lclc}
& J                             &\text{\hspace{.5cm}} & E                             \\
\Lambda_{12}     \,:\, & X_{21} X_{13} X_{31} -X_{22} X_{21}  & \quad               & X_{11} X_{12} - X_{12} X_{23} X_{32} \\
\Lambda_{21}     \,:\, & X_{13} X_{31} X_{12}-X_{12} X_{22}   & \quad               & X_{23} X_{32} X_{21}-X_{21} X_{11}   \\
\Lambda_{23} \,:\, & X_{33} X_{32} - X_{32} X_{21} X_{12} & \quad               & X_{23} X_{31} X_{13} -X_{22} X_{23}  \\
\Lambda_{32}	    \,:\, & X_{23} X_{33} - X_{21} X_{12} X_{23} & \quad               & X_{32} X_{22} - X_{31} X_{13} X_{32} \\
\Lambda_{31}     \,:\, & X_{13} X_{33} - X_{12} X_{21} X_{13} & \quad               & X_{31} X_{11} - X_{32} X_{23} X_{31} \\
\Lambda_{13}     \,:\, & X_{31} X_{12} X_{21} -X_{33} X_{31}  & \quad               & X_{11} X_{13} - X_{13} X_{32} X_{23} 
\end{array}
\label{eq:D3-JEterms}
\end{equation}

Figure~\ref{fig:D3quivN01} shows the quiver for this theory in $\mathcal{N}=(0,1)$ language. The $W^{(0,1)}$ associated to \eqref{eq:D3-JEterms} is
\begin{equation}
    \begin{split}
       W^{(0,1)}=& W^{(0,2)}+\sum_{i=1}^3\Lambda_{ii}^R X^\dagger_{ii}X_{ii}+\sum_{\substack{i,j=1\\ j\neq i}}^3\Lambda_{ii}^R\left( X^\dagger_{ij}X_{ij}+ X^\dagger_{ji}X_{ji}\right)\fstop
        \end{split}
        \label{eq:D3superW01}
\end{equation}

Table~\ref{tab:GenerD3} shows the generators, which were obtained using the HS.
\begin{table}[H]
	\centering
	\renewcommand{\arraystretch}{1.1}
	\begin{tabular}{c|c}
		Meson    & Chiral fields  \\
		\hline
		$M_1$    & $X_{23}X_{32}=X_{11}$ \\
		$M_2$    & $X_{13}X_{31}=X_{22}$\\
		$M_3$    & $X_{12}X_{21}=X_{33}$\\
		$M_4$    & $X_{23}X_{31}X_{12}$\\
		$M_5$    & $X_{13}X_{32}X_{21}$
	\end{tabular}
	\caption{Generators of $D_3$.}
	\label{tab:GenerD3}
\end{table}
They satisfy the following relation
\begin{equation}
    \mathcal{I} = \left\langle M_{1}M_{2}M_{3}=M_{4}M_{5}\right\rangle\fstop
    \label{eq:D3ideal}
\end{equation}

Of course, as for all cases, we can consider the universal involution. However, in this section we will consider another involution, which gives rise to an $\SO(N)\times \U(N)$ (or $\USp(N)\times \U(N)$) gauge theory. 

\bigskip

\paragraph{$\SO(N)\times \U(N)$ Orientifold}\mbox{}

\smallskip

Let us consider an involution which, roughly speaking, acts as a reflection with respect to a vertical axis going through the middle of Figure~\ref{fig:D3quiv}. Node 1 maps to itself, while nodes 2 and 3 get identified.

Chiral fields transform according to 
\begin{equation}
    \begin{array}{ccc}
     X_{11}\rightarrow  \gamma_{\Omega_1}\bar{X}_{11}\gamma_{\Omega_1}^{-1} \coma & X_{22}\rightarrow  \gamma_{\Omega_3}\bar{X}_{33}\gamma_{\Omega_3}^{-1} \coma & X_{33}\rightarrow  \gamma_{\Omega_2}\bar{X}_{22}\gamma_{\Omega_2}^{-1} \coma \\
     X_{12}\rightarrow  \gamma_{\Omega_1}\bar{X}_{13}\gamma_{\Omega_3}^{-1} \coma & X_{21}\rightarrow  \gamma_{\Omega_3}\bar{X}_{31}\gamma_{\Omega_1}^{-1}\coma & X_{23}\rightarrow  \gamma_{\Omega_3}\bar{X}_{32}\gamma_{\Omega_2}^{-1}\coma \\ X_{32}\rightarrow  \gamma_{\Omega_2}\bar{X}_{23}\gamma_{\Omega_3}^{-1} \coma & X_{31}\rightarrow  \gamma_{\Omega_2}\bar{X}_{21}\gamma_{\Omega_1}^{-1} \coma & X_{13}\rightarrow  \gamma_{\Omega_1}\bar{X}_{12}\gamma_{\Omega_2}^{-1}\fstop   
    \end{array}
    \label{eq:D3chiralinvolution}
\end{equation}
Invariance of $W^{(0,1)}$ in \eqref{eq:D3superW01} implies that Fermi fields transform as
\begin{equation}
    \begin{array}{ccc}
    \Lambda_{12}\rightarrow \gamma_{\Omega_1}\bar{\Lambda}_{13}\gamma_{\Omega_3}^{-1}\coma & 
    \Lambda_{21}\rightarrow -\gamma_{\Omega_3}\bar{\Lambda}_{31}\gamma_{\Omega_1}^{-1}\coma & 
    \Lambda_{23}\rightarrow-\gamma_{\Omega_2}\Lambda_{23}^{T}\gamma_{\Omega_3}^{-1}\coma\\
    \Lambda_{32}\rightarrow \gamma_{\Omega_3}\Lambda_{32}^{T}\gamma_{\Omega_2}^{-1}\coma & 
    \Lambda_{31}\rightarrow -\gamma_{\Omega_2}\bar{\Lambda}_{21}\gamma_{\Omega_1}^{-1}\coma & 
    \Lambda_{13}\rightarrow \gamma_{\Omega_1}\bar{\Lambda}_{12}\gamma_{\Omega_2}^{-1} \coma
    \end{array}
    \label{eq:D3fermiinvolution}
\end{equation}
and 
\begin{equation}
\begin{array}{ccc}
\Lambda^{R}_{11}\rightarrow \gamma_{\Omega_1}\Lambda^{R\,\,T}_{11}\gamma_{\Omega_1}^{-1} \coma & 
    \Lambda^{R}_{22}\rightarrow \gamma_{\Omega_3}\Lambda^{R\,\,T}_{33}\gamma_{\Omega_3}^{-1} \coma & 
    \Lambda^{R}_{33}\rightarrow \gamma_{\Omega_2}\Lambda^{R\,\,T}_{22}\gamma_{\Omega_2}^{-1} \fstop
\end{array}
\label{eq:D3fermiRinvolution}
\end{equation}

Using Table~\ref{tab:GenerD3}, we derive the corresponding geometric involution $\sigma$ on the generators of $D_3$
\begin{equation}
    \begin{array}{c}
    \left(M_1,M_2,M_3,M_4,M_5\right)\\
    \downarrow\\
 \ \left(\bar{M}_1,\bar{M}_3,\bar{M}_2,\bar{M}_5,\bar{M}_4\right)\fstop
\end{array}
    \label{eq:D3invol}
\end{equation}
Since $D_3$ is a complete intersection, we can easily check that $\sigma$ maps the holomorphic $4$-form $\Omega^{(4,0)}$ to $\bar{\Omega}^{(0,4)}$. We define
\begin{equation}
    \Omega^{(4,0)}=\Res \frac{dM_1\wedge dM_2\wedge dM_3 \wedge dM_4 \wedge dM_5}{M_{1}M_{2}M_{3}-M_{4}M_{5}} = \frac{dM_1\wedge dM_2\wedge dM_3 \wedge dM_4}{M_4}\coma
\end{equation}
which maps to 
\begin{equation}\label{eq:antiholo 4-form of D3}
    \frac{d\bar{M}_1\wedge d\bar{M}_3\wedge d\bar{M}_2 \wedge d\bar{M}_5}{\bar{M}_5}\fstop
\end{equation}
Either choosing the residue with respect to $M_4$ to express the holomorphic $4$-form, or by using the relation in the ideal~\eqref{eq:D3ideal}, one can show that (\ref{eq:antiholo 4-form of D3}) is exactly the anti-holomorphic $4$-form $\bar{\Omega}^{(0,4)}$ of $D_3$. Based on the discussion in Section~\ref{sec:HSreview}, we conclude that $\Omega\sigma$ with $\sigma$ in (\ref{eq:D3invol}) indeed gives rise to a $\Spin(7)$ orientifold. 

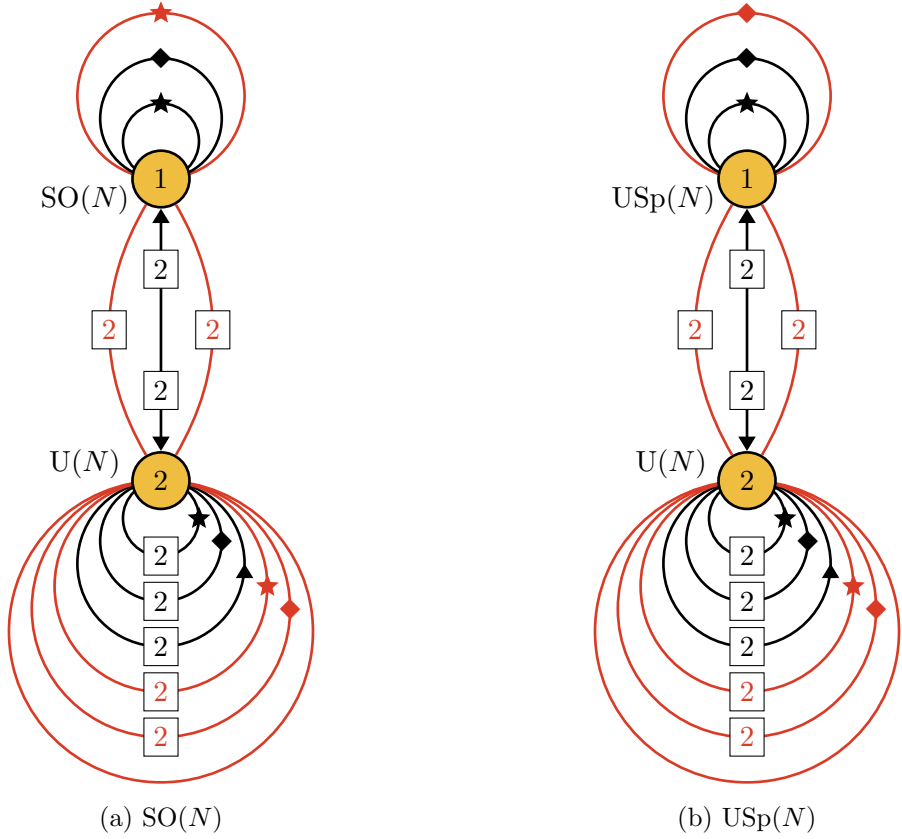
\begin{figure}[H]
    \centering
    \begin{subfigure}[t]{0.49\textwidth}
    \centering
    \begin{tikzpicture}[scale=2]
    \draw[line width=1pt,redX] (0,2.55) circle (0.55) node[yshift=1.1cm,star,star points=5, star point ratio=2.25, inner sep=1pt, fill=redX, draw=redX] {};
	\draw[line width=1pt] (0,2.4) circle (0.4) node[yshift=0.8cm] {\scriptsize{$\quadro$}};
	\draw[line width=1pt] (0,2.25) circle (0.25) node[yshift=0.5cm,star,star points=5, star point ratio=2.25, inner sep=1pt, fill=black, draw] {};
	\draw[line width=1pt] (0,-0.25) circle (0.25) node[xshift=0.5cm,star,star points=5, star point ratio=2.25, inner sep=1pt, fill=black, draw] {} node[fill=white,text opacity=1,fill opacity=1,draw=black,rectangle,thin,yshift=-0.5cm] {$2$};
	\draw[line width=1pt] (0,-0.4) circle (0.4) node[xshift=0.8cm] {\scriptsize{$\quadro$}} node[fill=white,text opacity=1,fill opacity=1,draw=black,rectangle,thin,yshift=-0.8cm] {$2$};
	\draw[line width=1pt,decoration={markings, mark=at position 0 with{\arrow{Triangle}}}, postaction={decorate}] (0,-0.55) circle (0.55) node[fill=white,text opacity=1,fill opacity=1,draw=black,rectangle,thin,yshift=-1.1cm] {$2$};
	\draw[line width=1pt,redX] (0,-0.70) circle (0.70) node[xshift=1.4cm,star,star points=5, star point ratio=2.25, inner sep=1pt, fill=redX, draw] {} node[fill=white,text opacity=1,fill opacity=1,draw=black,rectangle,thin,yshift=-1.4cm] {$2$};
	\draw[line width=1pt,redX] (0,-0.85) circle (0.85) node[xshift=1.7cm] {\scriptsize{$\quadro$}} node[fill=white,text opacity=1,fill opacity=1,draw=black,rectangle,thin,yshift=-1.7cm] {$2$};
	\draw[line width=1pt, redX] (0,-1) circle (1);
	\node[draw=black,line width=1pt,circle,fill=yellowX,minimum width=0.75cm,inner sep=1pt,label={[xshift=-1cm,yshift=-0.5cm]:$\U(N)$}] (A) at (0,0) {$2$};
	\node[draw=black,line width=1pt,circle,fill=yellowX,minimum width=0.75cm,inner sep=1pt,label={[xshift=-1cm,yshift=-1cm]:$\SO(N)$}] (B) at (0,2) {$1$};
	\path[Triangle-Triangle] 
	(A) edge[line width=1pt] node[fill=white,text opacity=1,fill opacity=1,draw=black,rectangle,thin,pos=0.25] {$2$} node[fill=white,text opacity=1,fill opacity=1,draw=black,rectangle,thin,pos=0.75] {$2$} (B);
	\draw[line width=1pt,redX] (A) to[bend right] node[fill=white,text opacity=1,fill opacity=1,draw=black,rectangle,thin] {\color{redX}{$2$}} (B);
	\draw[line width=1pt,redX] (A) to[bend left] node[fill=white,text opacity=1,fill opacity=1,draw=black,rectangle,thin] {\color{redX}{$2$}} (B);
    \end{tikzpicture}
    \caption{$\SO(N)$}
    \label{fig:o_theory_d3_SO}
    \end{subfigure}\hfill
    \begin{subfigure}[t]{0.49\textwidth}
    \centering
    \begin{tikzpicture}[scale=2]
	\draw[line width=1pt,redX] (0,2.55) circle (0.55) node[yshift=1.1cm] {\scriptsize{$\quadro$}};
	\draw[line width=1pt] (0,2.4) circle (0.4) node[yshift=0.8cm] {\scriptsize{$\quadro$}};
	\draw[line width=1pt] (0,2.25) circle (0.25) node[yshift=0.5cm,star,star points=5, star point ratio=2.25, inner sep=1pt, fill=black, draw] {};
	\draw[line width=1pt] (0,-0.25) circle (0.25) node[xshift=0.5cm,star,star points=5, star point ratio=2.25, inner sep=1pt, fill=black, draw] {} node[fill=white,text opacity=1,fill opacity=1,draw=black,rectangle,thin,yshift=-0.5cm] {$2$};
	\draw[line width=1pt] (0,-0.4) circle (0.4) node[xshift=0.8cm] {\scriptsize{$\quadro$}} node[fill=white,text opacity=1,fill opacity=1,draw=black,rectangle,thin,yshift=-0.8cm] {$2$};
	\draw[line width=1pt,decoration={markings, mark=at position 0 with{\arrow{Triangle}}}, postaction={decorate}] (0,-0.55) circle (0.55) node[fill=white,text opacity=1,fill opacity=1,draw=black,rectangle,thin,yshift=-1.1cm] {$2$};
	\draw[line width=1pt,redX] (0,-0.70) circle (0.70) node[xshift=1.4cm,star,star points=5, star point ratio=2.25, inner sep=1pt, fill=redX, draw] {} node[fill=white,text opacity=1,fill opacity=1,draw=black,rectangle,thin,yshift=-1.4cm] {$2$};
	\draw[line width=1pt,redX] (0,-0.85) circle (0.85) node[xshift=1.7cm] {\scriptsize{$\quadro$}} node[fill=white,text opacity=1,fill opacity=1,draw=black,rectangle,thin,yshift=-1.7cm] {$2$};
	\draw[line width=1pt, redX] (0,-1) circle (1);
	\node[draw=black,line width=1pt,circle,fill=yellowX,minimum width=0.75cm,inner sep=1pt,label={[xshift=-1cm,yshift=-0.5cm]:$\U(N)$}] (A) at (0,0) {$2$};
	\node[draw=black,line width=1pt,circle,fill=yellowX,minimum width=0.75cm,inner sep=1pt,label={[xshift=-1.1cm,yshift=-1cm]:$\USp(N)$}] (B) at (0,2) {$1$};
	\path[Triangle-Triangle] 
	(A) edge[line width=1pt] node[fill=white,text opacity=1,fill opacity=1,draw=black,rectangle,thin,pos=0.25] {$2$} node[fill=white,text opacity=1,fill opacity=1,draw=black,rectangle,thin,pos=0.75] {$2$} (B);
	\draw[line width=1pt,redX] (A) to[bend right] node[fill=white,text opacity=1,fill opacity=1,draw=black,rectangle,thin] {\color{redX}{$2$}} (B);
	\draw[line width=1pt,redX] (A) to[bend left] node[fill=white,text opacity=1,fill opacity=1,draw=black,rectangle,thin] {\color{redX}{$2$}} (B);
    \end{tikzpicture}
    \caption{$\USp(N)$}
    \label{fig:o_theory_d3_USp}
    \end{subfigure}
    \caption{Quivers for  the  Spin(7)  orientifolds  $D_3$ using the involution in \eqref{eq:D3chiralinvolution}, \eqref{eq:D3fermiinvolution} and \eqref{eq:D3fermiRinvolution}. The two different theories correspond to the choices $\gamma_{\Omega_1}=\ID_N$ or $J$.}
    \label{fig:o_theory_d3}
\end{figure}

Returning to the gauge theory, we obtain an $\mathcal{N}=(0,1)$ theory with gauge symmetry $\SO(N)\times\U(N)$ or $\USp(N)\times \U(N)$, depending on whether $\gamma_{\Omega_1}=\ID_N$ or $J$. The quivers for both choices are shown in Figure~\ref{fig:o_theory_d3}. Both theories are free of gauge anomalies. 

\subsection{$H_4$}

\label{sec:H4examp}

\begin{figure}[H]
    	\centering
	\begin{tikzpicture}[scale=1.5, rotate around y=-30]
	\draw[thick,gray,-Triangle] (0,0,0) -- node[above,pos=1] {$x$} (1.5,0,0);
	\draw[thick,gray,-Triangle] (0,0,0) -- node[left,pos=1] {$y$} (0,1.5,0);
	\draw[thick,gray,-Triangle] (0,0,0) -- node[below,pos=1] {$z$} (0,0,1.5);
	\draw[thin,dashed,gray] (1,0,0) -- (1,0,1) -- (1,1,1);
	\draw[thin,dashed,gray] (0,0,1) -- (1,0,1);
	\node[draw=black,line width=1pt,circle,fill=black,minimum width=0.2cm,inner sep=1pt] (O) at (0,0,0) {};
	\node[draw=black,line width=1pt,circle,fill=black,minimum width=0.2cm,inner sep=1pt] (A) at (1,0,0) {};
	\node[draw=black,line width=1pt,circle,fill=black,minimum width=0.2cm,inner sep=1pt] (B) at (0,1,0) {};
	\node[draw=black,line width=1pt,circle,fill=black,minimum width=0.2cm,inner sep=1pt] (C) at (0,0,1) {};
	\node[draw=black,line width=1pt,circle,fill=black,minimum width=0.2cm,inner sep=1pt] (D) at (1,1,1) {};
	\node[draw=black,line width=1pt,circle,fill=black,minimum width=0.2cm,inner sep=1pt] (E) at (1,1,0) {};
	\draw[line width=1pt] (O)--(A)--(E)--(B)--(O);
	\draw[line width=1pt] (O)--(C)--(B)--(D)--(A)--(C)--(D)--(E);
	\end{tikzpicture}
	\caption{Toric diagram for $H_4$.}
	\label{fig:H4toricdiagram}
\end{figure}
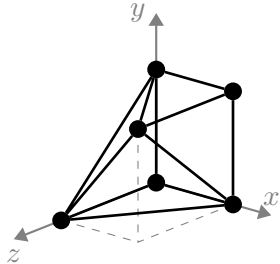

Another example that we are going to discuss is $H_4$. We show its toric diagram in Figure~\ref{fig:H4toricdiagram}. 

In particular, we will consider two $\mathcal{N}=(0,2)$ gauge theories associated with $H_4$, denoted as Phase A and Phase B. These two phases are related by $\mathcal{N}=(0,2)$ triality and were first introduced in \cite{Franco:2017cjj}. While they have different matter content and $J$- and $E$-terms, they share the same moduli space. Therefore, the generators of their moduli space and the relations among them are the same.

\subsubsection{Phase A}
\label{sec:H4PhaseA}

\begin{figure}[!htp]
	\centering
	\begin{subfigure}[t]{0.49\textwidth}
	\centering
	\begin{tikzpicture}[scale=2]
	\draw[line width=1pt,decoration={markings, mark=at position 0.75 with{\arrow{Triangle}}}, postaction={decorate}] (2.22,-0.22) circle (0.25);
	\draw[line width=1pt,redX] (-0.22,2.22) circle (0.25) node[fill=white,text opacity=1,fill opacity=1,draw=black,rectangle,xshift=-0.35cm,yshift=0.35cm,thin] {\color{redX}{$3$}};
	\node[draw=black,line width=1pt,circle,fill=yellowX,minimum width=0.75cm,inner sep=1pt] (A) at (0,0) {$4$};
	\node[draw=black,line width=1pt,circle,fill=yellowX,minimum width=0.75cm,inner sep=1pt] (B) at (2,0) {$3$};
	\node[draw=black,line width=1pt,circle,fill=yellowX,minimum width=0.75cm,inner sep=1pt] (C) at (2,2) {$2$};
	\node[draw=black,line width=1pt,circle,fill=yellowX,minimum width=0.75cm,inner sep=1pt] (D) at (0,2) {$1$};
	\path[Triangle-] (A) edge[line width=1pt] (B);
	\path[-Triangle] (B) edge[line width=1pt] (C);
	\path[Triangle-] (C) edge[line width=1pt] (D);
	\path[-Triangle] (D) edge[line width=1pt] (A);
	\draw[line width=1pt,redX] (A) to[bend right=20] (B);
	\draw[line width=1pt,redX] (B) to[bend right=20] (C);
	\draw[line width=1pt,-Triangle] (C) to[bend right=20] node[fill=white,text opacity=1,fill opacity=1,draw=black,rectangle,thin,pos=0.5] {$3$} (D);
	\draw[line width=1pt,Triangle-] (D) to[bend right=20] node[fill=white,text opacity=1,fill opacity=1,draw=black,rectangle,thin,pos=0.5] {$3$} (A);
	\draw[line width=1pt,redX] (A) to node[fill=white,text opacity=1,fill opacity=1,draw=black,rectangle,thin,pos=0.75] {\color{redX}{$2$}} (C);
	\draw[line width=1pt,redX] (D) to[bend right=10] node[fill=white,text opacity=1,fill opacity=1,draw=black,rectangle,thin,pos=0.25] {\color{redX}{$2$}} (B);
	\draw[line width=1pt,-Triangle] (D) to[bend left=10] node[fill=white,text opacity=1,fill opacity=1,draw=black,rectangle,thin,pos=0.75] {$2$} (B);
	\node at (0,-0.6) {};
	\end{tikzpicture}
	\caption{$\mathcal{N}=(0,2)$ language.}
	\label{fig:H4AquivN02}
	\end{subfigure}\hfill
	\begin{subfigure}[t]{0.49\textwidth}
	\centering
	\begin{tikzpicture}[scale=2]
	\draw[line width=1pt,decoration={markings, mark=at position 0.75 with{\arrow{Triangle}}}, postaction={decorate}] (2.22,-0.22) circle (0.25) node[fill=white,text opacity=1,fill opacity=1,draw=black,rectangle,thin,xshift=0.35cm,yshift=-0.35cm] {$2$};
	\draw[line width=1pt,redX] (-0.22,-0.22) circle (0.25);
	\draw[line width=1pt,redX] (2.22,2.22) circle (0.25);
	\draw[line width=1pt,redX] (2.28,-0.28) circle (0.4);
	\draw[line width=1pt,redX] (-0.22,2.22) circle (0.25) node[fill=white,text opacity=1,fill opacity=1,draw=black,rectangle,xshift=-0.35cm,yshift=0.35cm,thin] {\color{redX}{$4$}};
	\node[draw=black,line width=1pt,circle,fill=yellowX,minimum width=0.75cm,inner sep=1pt] (A) at (0,0) {$4$};
	\node[draw=black,line width=1pt,circle,fill=yellowX,minimum width=0.75cm,inner sep=1pt] (B) at (2,0) {$3$};
	\node[draw=black,line width=1pt,circle,fill=yellowX,minimum width=0.75cm,inner sep=1pt] (C) at (2,2) {$2$};
	\node[draw=black,line width=1pt,circle,fill=yellowX,minimum width=0.75cm,inner sep=1pt] (D) at (0,2) {$1$};
	\path[Triangle-] (A) edge[line width=1pt] node[fill=white,text opacity=1,fill opacity=1,draw=black,rectangle,thin,pos=0.5] {$2$} (B);
	\path[-Triangle] (B) edge[line width=1pt]  node[fill=white,text opacity=1,fill opacity=1,draw=black,rectangle,thin,pos=0.5] {$2$} (C);
	\path[Triangle-] (C) edge[line width=1pt] node[fill=white,text opacity=1,fill opacity=1,draw=black,rectangle,thin,pos=0.5] {$2$} (D);
	\path[-Triangle] (D) edge[line width=1pt] node[fill=white,text opacity=1,fill opacity=1,draw=black,rectangle,thin,pos=0.5] {$2$} (A);
	\draw[line width=1pt,redX] (A) to[bend right=30] node[fill=white,text opacity=1,fill opacity=1,draw=black,rectangle,thin,pos=0.5] {$2$} (B);
	\draw[line width=1pt,redX] (B) to[bend right=30]  node[fill=white,text opacity=1,fill opacity=1,draw=black,rectangle,thin,pos=0.5] {$2$} (C);
	\draw[line width=1pt,-Triangle] (C) to[bend right=30] node[fill=white,text opacity=1,fill opacity=1,draw=black,rectangle,thin,pos=0.5] {$6$} (D);
	\draw[line width=1pt,Triangle-] (D) to[bend right=30] node[fill=white,text opacity=1,fill opacity=1,draw=black,rectangle,thin,pos=0.5] {$6$} (A);
	\draw[line width=1pt,redX] (A) to node[fill=white,text opacity=1,fill opacity=1,draw=black,rectangle,thin,pos=0.75] {\color{redX}{$4$}} (C);
	\draw[line width=1pt,redX] (D) to[bend right=10] node[fill=white,text opacity=1,fill opacity=1,draw=black,rectangle,thin,pos=0.25] {\color{redX}{$4$}} (B);
	\draw[line width=1pt,-Triangle] (D) to[bend left=10] node[fill=white,text opacity=1,fill opacity=1,draw=black,rectangle,thin,pos=0.75] {$4$} (B);
	\end{tikzpicture}
	\caption{$\mathcal{N}=(0,1)$ language.}
	\label{fig:H4AquivN01}
	\end{subfigure}
	\caption{Quiver for $H_4$ in phase A.}
	\label{fig:H4_A_quiver}
\end{figure}
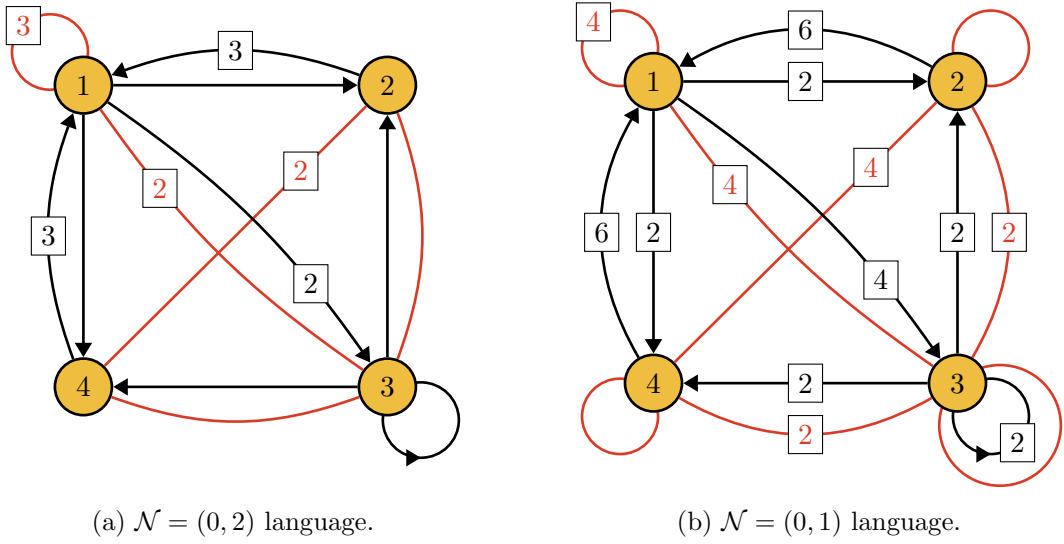

The quiver diagram for Phase A is shown in Figure~\ref{fig:H4_A_quiver}, both in $\mathcal{N}=(0,2)$ and $\mathcal{N}=(0,1)$ languages. 

The $J$- and $E$-terms are 
\begin{equation}
\renewcommand{\arraystretch}{1.1}
\begin{array}{lclc}
& J                           &\text{\hspace{.5cm}}& E                                \\
\Lambda_{11}^1 \,:\, & X_{14}X_{41} - X_{13}X_{32}Z_{21}  &                    & Y_{13}X_{34}Z_{41}-X_{12}Y_{21}         \\
\Lambda_{11}^2 \,:\, & X_{14}Y_{41} - Y_{13}X_{32}Z_{21}  &                    & X_{12}X_{21}  - X_{13}X_{34}Z_{41}      \\
\Lambda_{11}^3 \,:\, & X_{14}Z_{41} - X_{12}Z_{21}        &                    & X_{13}X_{32}Y_{21} - Y_{13}X_{34}X_{41} \\
\Lambda_{13}^1 \,:\, & X_{32}X_{21}  - X_{34}X_{41}       &                    & Y_{13}X_{33} - X_{14}Z_{41}Y_{13}       \\
\Lambda_{13}^2 \,:\, & X_{32}Y_{21}  - X_{34}Y_{41}       &                    & X_{12}Z_{21}X_{13} - X_{13}X_{33}       \\
\Lambda_{42}^1 \,:\, & X_{21}X_{14}  - Z_{21}X_{13}X_{34} &                    & Z_{41}Y_{13}X_{32} -Y_{41}X_{12}        \\
\Lambda_{42}^2 \,:\, & Y_{21}X_{14}  - Z_{21}Y_{13}X_{34} &                    & X_{41}X_{12} - Z_{41}X_{13}X_{32}       \\
\Lambda_{23} \,:\,   & X_{33}X_{32}  - X_{32}Z_{21}X_{12} &                    & Y_{21}X_{13} -X_{21}Y_{13}              \\
\Lambda_{43} \,:\,   & X_{33}X_{34}  - X_{34}Z_{41}X_{14} &                    & X_{41}Y_{13} - Y_{41}X_{13}             
\end{array}
\label{eq:H4PhaseAJEterms}
\end{equation}
The $W^{(0,1)}$ superpotential becomes
\begin{equation}
    \begin{split}
      W^{(0,1)}=&\,W^{(0,2)}
     +\Lambda_{11}^{4R}( X^\dagger_{12}X_{12}+ X^\dagger_{14}X_{14}+ X^\dagger_{21}X_{21}+ Y^\dagger_{21}Y_{21}+ Z^\dagger_{21}Z_{21}+\\
      &+ X^\dagger_{41}X_{41}+ Y^\dagger_{41}Y_{41}+ Z^\dagger_{41}Z_{41}+ X^\dagger_{13}X_{13}+ Y^\dagger_{13}Y_{13})+\\
     &+\Lambda_{22}^{R}( X^\dagger_{12}X_{12}+ X^\dagger_{32}X_{32}+ X^\dagger_{21}X_{21}+ Y^\dagger_{21}Y_{21}+ Z^\dagger_{21}Z_{21})+\\
     &+\Lambda_{33}^{R}( X^\dagger_{33}X_{33}+ X^\dagger_{32}X_{32}+ X^\dagger_{34}X_{34}+ X^\dagger_{13}X_{13}+ Y^\dagger_{13}Y_{13})+\\
     &+\Lambda_{44}^{R}( X^\dagger_{14}X_{14}+ X^\dagger_{34}X_{34}+ X^\dagger_{41}X_{41}+ Y^\dagger_{41}Y_{41}+ Z^\dagger_{41}Z_{41})\fstop
    \end{split}
    \label{eq:H4AW01sup}
\end{equation}

Table~\ref{tab:GenerH4PhaseA} shows the generators of the moduli space, which were computed using the HS, and their expression in terms of the chiral fields in phase A.
\begin{table}[H]
	\centering
	\renewcommand{\arraystretch}{1.1}
	\begin{tabular}{c|c}
		Meson & Chiral fields  \\
		\hline
		$M_1$ & $X_{33}=X_{14}Z_{41}=Z_{21}X_{12}$ \\
		$M_2$ & $Y_{21}X_{12}=Z_{41}Y_{13}X_{34}$ \\
		$M_3$ & $X_{14}Y_{41}=Z_{21}Y_{13}X_{32}$ \\
		$M_4$ & $X_{32}Y_{21}Y_{13}=X_{34}Y_{41}Y_{13}$ \\
		$M_5$ & $X_{21}X_{12}=Z_{41}X_{13}X_{34}$ \\
		$M_6$ & $X_{14}X_{41}=Z_{21}X_{13}X_{32}$ \\
		$M_7$ & $X_{32}Y_{21}X_{13}=X_{32}X_{21}Y_{13}=X_{34}Y_{41}X_{13}=X_{34}X_{41}Y_{13}$ \\
		$M_8$ & $X_{32}X_{21}X_{13}=X_{34}X_{41}X_{13}$ 
	\end{tabular}
	\caption{Generators of $H_4$ in Phase A.}
	\label{tab:GenerH4PhaseA}
\end{table}

The relations among the generators are
\begin{equation}
\begin{split}
    \mathcal{I} = &\left\langle M_1M_4=M_2M_3\coma M_1M_7=M_2M_6\coma M_1M_7=M_3M_5\coma M_2M_7=M_4M_5\coma\right.\\
    &\left.M_3M_7=M_4M_6\coma M_1M_8=M_5M_6\coma M_2M_8=M_5M_7\coma M_3M_8=M_6M_7\coma\right.\\
    &\left.M_4M_8=M_7^2\right\rangle\fstop
    \end{split}
\end{equation}

\bigskip

\paragraph{$\SO(N)\times U(N)\times \SO(N)$ Orientifold}\mbox{}

\medskip

Let us consider an anti-holomorphic involution of phase A which acts on Figure~\ref{fig:H4_A_quiver} as a reflection with respect to the diagonal connecting nodes $1$ and $3$. Then, nodes $1$ and $3$ map to themselves, while nodes $2$ and $4$ are identified.

The involution on chiral fields is
\begin{equation}
    \begin{array}{cccccccccccc}
        X_{33} & \rightarrow  &  \gamma_{\Omega_3}\bar{X}_{33}\gamma_{\Omega_3}^{-1}\coma &  
        X_{14} & \rightarrow  &  \gamma_{\Omega_1}\bar{X}_{12}\gamma_{\Omega_2}^{-1} \coma &  
        Z_{41} & \rightarrow  &  \gamma_{\Omega_2}\bar{Z}_{21}\gamma_{\Omega_1}^{-1}\coma &  
        Y_{21} & \rightarrow  & \gamma_{\Omega_4}\bar{X}_{41}\gamma_{\Omega_1}^{-1}\coma \\
        Y_{13} & \rightarrow  & \gamma_{\Omega_1}\bar{X}_{13}\gamma_{\Omega_3}^{-1}\coma &  
        X_{34} & \rightarrow  &  \gamma_{\Omega_3}\bar{X}_{32}\gamma_{\Omega_2}^{-1}\coma &  
        Y_{41} & \rightarrow  &  \gamma_{\Omega_2}\bar{X}_{21}\gamma_{\Omega_1}^{-1}\coma & 
        X_{12} & \rightarrow  &  \gamma_{\Omega_1}\bar{X}_{14}\gamma_{\Omega_4}^{-1} \coma \\
        Z_{21} & \rightarrow  &  \gamma_{\Omega_4}\bar{Z}_{41}\gamma_{\Omega_1}^{-1}\coma &  
        X_{41} & \rightarrow  & \gamma_{\Omega_2}\bar{Y}_{21}\gamma_{\Omega_1}^{-1}\coma &  
        X_{13} & \rightarrow  & \gamma_{\Omega_1}\bar{Y}_{13}\gamma_{\Omega_3}^{-1}\coma &  
        X_{32} & \rightarrow  &  \gamma_{\Omega_3}\bar{X}_{34}\gamma_{\Omega_4}^{-1}\coma\\ 
        \multicolumn{12}{c}{X_{21}\rightarrow  \gamma_{\Omega_4}\bar{Y}_{41}\gamma_{\Omega_1}^{-1}\fstop}
    \end{array}
    \label{eq:chiralH4PhaseA_o_theory}
\end{equation}

It is interesting to note that since phase A is a chiral theory, it clearly illustrates a characteristic feature of anti-holomorphic involutions: they map chiral fields to images with the same orientation, as it follows from the discussion in Section~\ref{sec:orientN01gaugeth}.

From the invariance of $W^{(0,1)}$, we obtain the transformations of the Fermi fields
\begin{equation}
    \begin{array}{ccccccccc}
    \Lambda^1_{11} & \rightarrow &  -\gamma_{\Omega_1}\Lambda^{1\,\,T}_{11}\gamma_{\Omega_1}^{-1}\coma &
    \Lambda^2_{11} & \rightarrow &  \gamma_{\Omega_1}\Lambda^{2\,\,T}_{11}\gamma_{\Omega_1}^{-1}\coma &  
    \Lambda^3_{11} & \rightarrow &  -\gamma_{\Omega_1}\bar{\Lambda}^{3}_{11}\gamma_{\Omega_1}^{-1}\coma \\
    \Lambda^1_{13} & \rightarrow &  -\gamma_{\Omega_1}\bar{\Lambda}^{2}_{13}\gamma_{\Omega_3}^{-1}\coma &
    \Lambda^2_{13} & \rightarrow &  -\gamma_{\Omega_1}\bar{\Lambda}^{1}_{13}\gamma_{\Omega_3}^{-1}\coma & 
     \Lambda^1_{42} & \rightarrow &  -\gamma_{\Omega_4}\Lambda^{1\,\,T}_{42}\gamma_{\Omega_2}^{-1}\coma \\
    \Lambda^2_{42} & \rightarrow &  \gamma_{\Omega_4}\Lambda^{2\,\,T}_{42}\gamma_{\Omega_1}^{-1}\coma & 
    \Lambda_{23}   & \rightarrow &  \gamma_{\Omega_4}\bar{\Lambda}_{43}\gamma_{\Omega_3}^{-1}\coma & 
    \Lambda_{43}    &\rightarrow&    \gamma_{\Omega_2}\bar{\Lambda}_{23}\gamma_{\Omega_3}^{-1}\coma
    \end{array}
    \label{eq:fermiH4PhaseA_o_theory}
\end{equation}
and 
\begin{equation}
    \Lambda^{4R}_{11}\rightarrow \gamma_{\Omega_1}\Lambda^{4R\,\,T}_{11}\gamma_{\Omega_1}^{-1}\coma  
    \Lambda^{R}_{22}\rightarrow \gamma_{\Omega_4}\Lambda^{R\,\,T}_{44}\gamma_{\Omega_4}^{-1}\coma
    \Lambda^{R}_{33}\rightarrow \gamma_{\Omega_3}\Lambda^{R\,\,T}_{33}\gamma_{\Omega_3}^{-1}\coma
     \Lambda^{R}_{44}\rightarrow \gamma_{\Omega_2}\Lambda^{R\,\,T}_{22}\gamma_{\Omega_2}^{-1}\fstop
    \label{eq:realfermiH4PhaseA_o_theory}
\end{equation}

Using Table~\ref{tab:GenerH4PhaseA}, we find the corresponding geometric involution $\sigma$ on the generators of $H_4$  
\begin{equation}
\begin{array}{c}
     \left(M_1,M_2,M_3,M_4,M_5,M_6,M_7,M_8\right)\\
    \downarrow\\
 \left(\bar{M}_1,\bar{M}_6,\bar{M}_5,\bar{M}_8,\bar{M}_3,\bar{M}_2,\bar{M}_7,\bar{M}_4\right)\fstop
    \end{array}
    \label{eq:H4PhaseA-invol}
\end{equation}

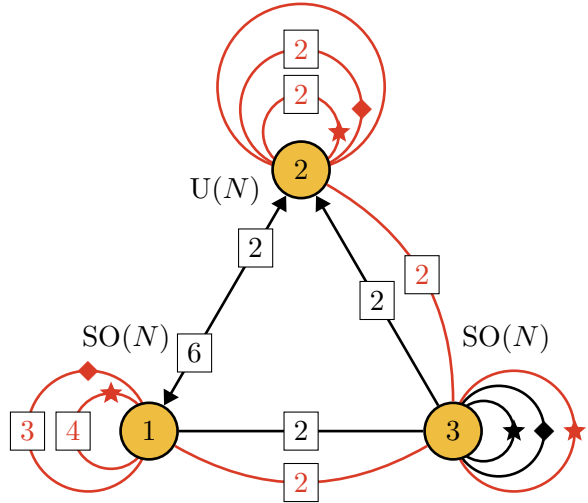
\begin{figure}[!htp]
    \centering
    \begin{tikzpicture}[scale=2]
    \draw[line width=1pt,redX] (1,1.982) circle (0.25) node[xshift=0.5cm,star,star points=5, star point ratio=2.25, inner sep=1pt, fill=redX, draw=redX] {} node[fill=white,text opacity=1,fill opacity=1,draw=black,rectangle,thin,yshift=0.5cm] {$2$};
	\draw[line width=1pt,redX] (1,2.132) circle (0.4) node[xshift=0.8cm] {\scriptsize{$\quadro$}} node[fill=white,text opacity=1,fill opacity=1,draw=black,rectangle,thin,yshift=0.8cm] {$2$};
	\draw[line width=1pt,redX] (1,2.282) circle (0.55);
	\draw[line width=1pt,redX] (-0.25,0) circle (0.25) node[yshift=0.5cm,star,star points=5, star point ratio=2.25, inner sep=1pt, fill=redX, draw=redX] {} node[fill=white,text opacity=1,fill opacity=1,draw=black,rectangle,thin,xshift=-0.5cm] {$4$};
	\draw[line width=1pt,redX] (-0.4,0) circle (0.4) node[yshift=0.8cm] {\scriptsize{$\quadro$}} node[fill=white,text opacity=1,fill opacity=1,draw=black,rectangle,thin,xshift=-0.8cm] {$3$};
	\draw[line width=1pt] (2.2,0) circle (0.2) node[xshift=0.4cm,star,star points=5, star point ratio=2.25, inner sep=1pt, fill=black, draw] {};
	\draw[line width=1pt] (2.3,0) circle (0.3) node[xshift=0.6cm] {\scriptsize{$\quadro$}};
	\draw[line width=1pt,redX] (2.4,0) circle (0.4) node[xshift=0.8cm,star,star points=5, star point ratio=2.25, inner sep=1pt, fill=redX, draw=redX] {};
	\node[draw=black,line width=1pt,circle,fill=yellowX,minimum width=0.75cm,inner sep=1pt,label={[xshift=-0.3cm,yshift=0.5cm]:$\SO(N)$}] (A) at (0,0) {$1$};
	\node[draw=black,line width=1pt,circle,fill=yellowX,minimum width=0.75cm,inner sep=1pt,label={[xshift=0.7cm,yshift=0.5cm]:$\SO(N)$}] (B) at (2,0) {$3$};
	\node[draw=black,line width=1pt,circle,fill=yellowX,minimum width=0.75cm,inner sep=1pt,label={[xshift=-1cm,yshift=-1cm]:$\U(N)$}] (C) at (1,1.732) {$2$};
	\path 
	(A) edge[line width=1pt] node[fill=white,text opacity=1,fill opacity=1,draw=black,rectangle,thin,pos=0.5] {$2$} (B);
	\path[-Triangle] (B) edge[line width=1pt] node[fill=white,text opacity=1,fill opacity=1,draw=black,rectangle,thin,pos=0.5] {$2$} (C);
	\path[Triangle-Triangle] (C) edge[line width=1pt] node[fill=white,text opacity=1,fill opacity=1,draw=black,rectangle,thin,pos=0.25] {$2$} node[fill=white,text opacity=1,fill opacity=1,draw=black,rectangle,thin,pos=0.75] {$6$} (A);
	\draw[line width=1pt,redX] (A) to[bend right] node[fill=white,text opacity=1,fill opacity=1,draw=black,rectangle,thin] {\color{redX}{$2$}} (B);
	\draw[line width=1pt,redX] (B) to[bend right] node[fill=white,text opacity=1,fill opacity=1,draw=black,rectangle,thin] {\color{redX}{$2$}} (C);
	\end{tikzpicture}
    \caption{Quiver for a  Spin(7)  orientifold of phase A of $H_4$ using the involution in \eqref{eq:chiralH4PhaseA_o_theory}, \eqref{eq:fermiH4PhaseA_o_theory} and \eqref{eq:realfermiH4PhaseA_o_theory}.}
    \label{fig:o_theory_h4_A}
\end{figure}

The orientifolded theory has gauge group $G_1(N)\times \U(N) \times  G_3(N)$. The involution of the fields connecting nodes 1 and 3 implies that in this case we must have $\gamma_{\Omega_1}=\gamma_{\Omega_3}$. Then, $G_1(N)$ and $G_3(N)$ can be either both $\SO$ or both $\USp$ gauge groups, but cannot be of different types. For example, Figure~\ref{fig:o_theory_h4_A} shows the quiver for $G_2(N)=G_3(N)=\SO(N)$. The theory is free of gauge anomalies.

\subsubsection{Phase B}

\label{sec:H4PhaseB}

\begin{figure}[H]
		\centering
		\begin{subfigure}[t]{0.49\textwidth}
			\centering
			\begin{tikzpicture}[scale=2]
			\node[draw=black,line width=1pt,circle,fill=yellowX,minimum width=0.75cm,inner sep=1pt] (A) at (0,0) {$4$};
			\node[draw=black,line width=1pt,circle,fill=yellowX,minimum width=0.75cm,inner sep=1pt] (B) at (2,0) {$3$};
			\node[draw=black,line width=1pt,circle,fill=yellowX,minimum width=0.75cm,inner sep=1pt] (C) at (2,2) {$2$};
			\node[draw=black,line width=1pt,circle,fill=yellowX,minimum width=0.75cm,inner sep=1pt] (D) at (0,2) {$1$};
			\path[-Triangle] (A) edge[line width=1pt] node[fill=white,text opacity=1,fill opacity=1,draw=black,rectangle,thin,pos=0.75] {$2$} (C);
			\path[-Triangle] (B) edge[line width=1pt] (A);
			\path[-Triangle] (D) edge[line width=1pt] node[fill=white,text opacity=1,fill opacity=1,draw=black,rectangle,thin,pos=0.75] {$2$} (B);
			\path[-Triangle] (C) edge[line width=1pt] (D);
			\draw[line width=1pt,redX] (A) to[bend right=20] node[fill=white,text opacity=1,fill opacity=1,draw=black,rectangle,thin,pos=0.5] {\color{redX}{$3$}} (B);
			\draw[line width=1pt,redX] (C) to[bend right=20] node[fill=white,text opacity=1,fill opacity=1,draw=black,rectangle,thin,pos=0.5] {\color{redX}{$3$}} (D);
			\draw[line width=1pt,Triangle-Triangle] (A) to  (D);
			\draw[line width=1pt,Triangle-Triangle] (B) to  (C);
			\node at (0,-0.42) {};
			\end{tikzpicture}
			\caption{$\mathcal{N}=(0,2)$ language.}
			\label{fig:H4BN02}
		\end{subfigure}\hfill
		\begin{subfigure}[t]{0.49\textwidth}
			\centering
			\begin{tikzpicture}[scale=2]
			\draw[line width=1pt,redX] (-0.22,-0.22) circle (0.25);
			\draw[line width=1pt,redX] (-0.22,2.22) circle (0.25);
			\draw[line width=1pt,redX] (2.22,-0.22) circle (0.25);
			\draw[line width=1pt,redX] (2.22,2.22) circle (0.25);
			\node[draw=black,line width=1pt,circle,fill=yellowX,minimum width=0.75cm,inner sep=1pt] (A) at (0,0) {$4$};
			\node[draw=black,line width=1pt,circle,fill=yellowX,minimum width=0.75cm,inner sep=1pt] (B) at (2,0) {$3$};
			\node[draw=black,line width=1pt,circle,fill=yellowX,minimum width=0.75cm,inner sep=1pt] (C) at (2,2) {$2$};
			\node[draw=black,line width=1pt,circle,fill=yellowX,minimum width=0.75cm,inner sep=1pt] (D) at (0,2) {$1$};
			\path[-Triangle] (A) edge[line width=1pt] node[fill=white,text opacity=1,fill opacity=1,draw=black,rectangle,thin,pos=0.75] {$4$} (C);
			\path[-Triangle] (B) edge[line width=1pt] node[fill=white,text opacity=1,fill opacity=1,draw=black,rectangle,thin,pos=0.5] {$2$} (A);
			\path[-Triangle] (D) edge[line width=1pt] node[fill=white,text opacity=1,fill opacity=1,draw=black,rectangle,thin,pos=0.75] {$4$} (B);
			\path[-Triangle] (C) edge[line width=1pt] node[fill=white,text opacity=1,fill opacity=1,draw=black,rectangle,thin,pos=0.5] {$2$} (D);
			\draw[line width=1pt,redX] (A) to[bend right=30] node[fill=white,text opacity=1,fill opacity=1,draw=black,rectangle,thin,pos=0.5] {\color{redX}{$6$}} (B);
			\draw[line width=1pt,redX] (C) to[bend right=30] node[fill=white,text opacity=1,fill opacity=1,draw=black,rectangle,thin,pos=0.5] {\color{redX}{$6$}} (D);
			\draw[line width=1pt,Triangle-Triangle] (A) to node[fill=white,text opacity=1,fill opacity=1,draw=black,rectangle,thin,pos=0.25] {$2$} node[fill=white,text opacity=1,fill opacity=1,draw=black,rectangle,thin,pos=0.75] {$2$} (D);
			\draw[line width=1pt,Triangle-Triangle] (B) to node[fill=white,text opacity=1,fill opacity=1,draw=black,rectangle,thin,pos=0.25] {$2$} node[fill=white,text opacity=1,fill opacity=1,draw=black,rectangle,thin,pos=0.75] {$2$} (C);
			\end{tikzpicture}
			\caption{$\mathcal{N}=(0,1)$ language.}
			\label{fig:H4BN01}
		\end{subfigure}
		\caption{Quiver for $H_4$ in phase B.}
		\label{fig:H4phaseBquiv}
	\end{figure}
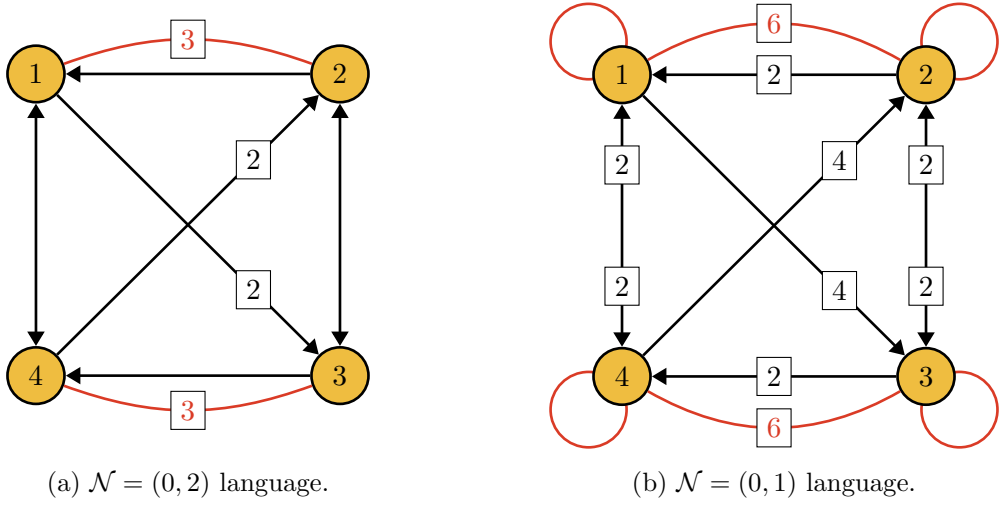

Figure~\ref{fig:H4phaseBquiv} shows the quiver for phase $B$ of $H_4$.  

The $J$- and $E$-terms are 
\begin{equation}
\renewcommand{\arraystretch}{1.1}
\begin{array}{lclc}
& J                                                                    &\text{\hspace{.5cm}}&  E                            \\
  \Lambda_{21} \,:\,     & X_{1 3}X_{34}Y_{42} - Y_{1 3}X_{34}X_{42}              &                    &  X_{21}X_{1 4}X_{41} - X_{23}X_{32}X_{21} \\
  \Lambda_{1 2}^{1} \,:\, & X_{23}X_{34}Y_{42}X_{21} - X_{21}Y_{1 3}X_{34}X_{41}  &                    &  X_{1 3}X_{32} - X_{1 4}X_{42}             \\
  \Lambda_{1 2}^{2} \,:\, & X_{21}X_{1 3}X_{34}X_{41} - X_{23}X_{34}X_{42}X_{21}  &                    &  Y_{1 3}X_{32} - X_{1 4}Y_{42}             \\
  \Lambda_{34} \,:\,     & Y_{42}X_{21}X_{1 3} - X_{42}X_{21}Y_{1 3}              &                    &  X_{34}X_{41}X_{1 4} - X_{32}X_{23}X_{34} \\
  \Lambda_{43}^{1} \,:\, & X_{34}Y_{42}X_{21}X_{1 4} - X_{32}X_{21}Y_{1 3}X_{34}  &                    &  X_{42}X_{23} - X_{41}X_{1 3}             \\                
  \Lambda_{43}^{2} \,:\, & X_{32}X_{21}X_{1 3}X_{34} - X_{34}X_{42}X_{21}X_{1 4}  &                    &  Y_{42}X_{23} - X_{41}Y_{1 3}  
\end{array}
\label{eq:H4PhaseB-JEterms}
\end{equation}

The $W^{(0,1)}$ superpotential is
\begin{equation}
    \begin{split}
        	W^{(0,1)}=&\,W^{(0,2)}+\Lambda^R_{11} (X^\dagger_{21}  X_{21}+X^\dagger_{41}  X_{41}+X^\dagger_{1 4}  X_{1 4}+X^\dagger_{1 3}  X_{1 3}+Y^\dagger_{1 3}  Y_{1 3})+\\
&+\Lambda^R_{22} (X^\dagger_{23}  X_{23}+X^\dagger_{21}  X_{21}+X^\dagger_{42}  X_{42}+X^\dagger_{32}  X_{32}+Y^\dagger_{42}  Y_{42})+\\
&+\Lambda^R_{33} (X^\dagger_{23}  X_{23}+X^\dagger_{32}  X_{32}+X^\dagger_{34}  X_{34}+X^\dagger_{1 3}  X_{1 3}+Y^\dagger_{1 3}  Y_{1 3})+\\	
&+\Lambda^R_{44} (X^\dagger_{42}  X_{42}+X^\dagger_{41}  X_{41}+X^\dagger_{34}  X_{34}+X^\dagger_{1 4}  X_{1 4}+Y^\dagger_{42}  Y_{42})\fstop
    \end{split}
    \label{eq:H4BW01sup}
\end{equation}

Table~\ref{tab:GenerH4PhaseB} shows the generators of $H_4$ in terms of the chiral fields in phase B. 
\begin{table}[H]
	\centering
	\renewcommand{\arraystretch}{1.1}
	\begin{tabular}{c|c}
		 Meson    & Chiral fields  \\
		\hline
$M_1$ & $X_{23}X_{32}=X_{41}X_{1 4}$ \\
$M_2$ & $X_{34}Y_{42}X_{23}=X_{34}X_{41}Y_{1 3}$ \\
$M_3$ & $X_{21}X_{1 4}Y_{42}=X_{21}Y_{1 3}X_{32}$ \\
$M_4$ & $X_{34}Y_{42}X_{21}Y_{1 3}$ \\
$M_5$ & $X_{34}X_{42}X_{23}=X_{34}X_{41}X_{1 3}$ \\
$M_6$ & $X_{21}X_{1 4}X_{42}=X_{21}X_{1 3}X_{32}$ \\
$M_7$ & $X_{42}X_{21}Y_{1 3}X_{34}=Y_{42}X_{21}X_{1 3}X_{34}$ \\
$M_8$ & $X_{42}X_{21}X_{1 3}X_{34}$
	\end{tabular}
	\caption{Generators of $H_4$ in Phase B.}
	\label{tab:GenerH4PhaseB}
\end{table}

They satisfy the following relations
\begin{equation}
\begin{split}
    \mathcal{I} = &\left\langle M_1M_4=M_2M_3\coma M_1M_7=M_2M_6\coma M_1M_7=M_3M_5\coma M_2M_7=M_4M_5\coma\right.\\
    &\left.M_3M_7=M_4M_6\coma M_1M_8=M_5M_6\coma M_2M_8=M_5M_7\coma M_3M_8=M_6M_7\coma\right.\\
    &\left.M_4M_8=M_7^2\right\rangle\fstop
    \end{split}
\end{equation}

This can be seen not only geometrically, but also from the gauge theory. While, as already mentioned, the generators and their relations are common to all the phases, their realizations in terms of chiral superfields in each of them are different. 

\bigskip

\paragraph{$\U(N)\times \U(N)$ Orientifold}\mbox{}

\smallskip

Let us consider an anti-holomorphic involution of phase B which acts on Figure~\ref{fig:H4phaseBquiv} as a reflection with respect to a horizontal line through the middle of the quiver. Nodes are mapped as $1 \leftrightarrow 4$ and $2 \leftrightarrow 3$.

The involution on chiral fields is
\begin{equation}
    \begin{array}{cccccccccccc}
        X_{23} & \rightarrow &   \gamma_{\Omega_3}\bar{X}_{32}\gamma_{\Omega_2}^{-1}\coma &   
        X_{41} & \rightarrow &   \gamma_{\Omega_1}\bar{X}_{14}\gamma_{\Omega_4}^{-1}\coma &   
        X_{34} & \rightarrow &   \gamma_{\Omega_2}\bar{X}_{21}\gamma_{\Omega_1}^{-1}\coma &   
        Y_{42} & \rightarrow &   \gamma_{\Omega_1}\bar{X}_{13}\gamma_{\Omega_3}^{-1}\coma \\
        Y_{13} & \rightarrow &   \gamma_{\Omega_4}\bar{X}_{42}\gamma_{\Omega_2}^{-1}\coma & 
        X_{32} & \rightarrow &   \gamma_{\Omega_2}\bar{X}_{23}\gamma_{\Omega_3}^{-1}\coma &  
        X_{14} & \rightarrow &   \gamma_{\Omega_4}\bar{X}_{41}\gamma_{\Omega_1}^{-1}\coma &  
        X_{21} & \rightarrow &   \gamma_{\Omega_3}\bar{X}_{34}\gamma_{\Omega_4}^{-1}\coma \\
        & & & 
        X_{13} & \rightarrow &   \gamma_{\Omega_4}\bar{Y}_{42}\gamma_{\Omega_2}^{-1}\coma &
        X_{42} &\rightarrow & \gamma_{\Omega_1}\bar{Y}_{13}\gamma_{\Omega_3}^{-1}\fstop
    \end{array}
    \label{eq:chiralH4PhaseBorient}
\end{equation}
Requiring the invariance of $W^{(0,1)}$, we obtain the transformations for the Fermi fields
\begin{equation}
    \begin{array}{ccccccccc}
        \Lambda_{21}     & \rightarrow &  \gamma_{\Omega_3}\bar{\Lambda}_{34} \gamma_{\Omega_4}^{-1}\coma &  
        \Lambda^1_{12}   & \rightarrow &  \gamma_{\Omega_4}\bar{\Lambda}^{2}_{43} \gamma_{\Omega_3}^{-1}\coma &  
        \Lambda^{2}_{12} & \rightarrow &  \gamma_{\Omega_4}\bar{\Lambda}^{1}_{43} \gamma_{\Omega_3}^{-1}\coma \\
        \Lambda_{34}     & \rightarrow &  \gamma_{\Omega_2}\bar{\Lambda}_{21}\gamma_{\Omega_1}^{-1}\coma &  
        \Lambda^1_{43}   & \rightarrow &  \gamma_{\Omega_1}\bar{\Lambda}^{2}_{12}\gamma_{\Omega_2}^{-1}\coma &  
        \Lambda^{2}_{43} & \rightarrow &  \gamma_{\Omega_1}\bar{\Lambda}^{1}_{12}\gamma_{\Omega_2}^{-1}\coma
    \end{array}
    \label{eq:fermiH4PhaseBorient}
\end{equation}
and 
\begin{equation}
        \Lambda^{R}_{11}\rightarrow \gamma_{\Omega_4}\Lambda^{R\,\,T}_{44}\gamma_{\Omega_4}^{-1}\coma \Lambda^{R}_{22}\rightarrow \gamma_{\Omega_3}\Lambda^{R\,\,T}_{33}\gamma_{\Omega_3}^{-1}\coma
        \Lambda^{R}_{33}\rightarrow \gamma_{\Omega_2}\Lambda^{R\,\,T}_{22}\gamma_{\Omega_2}^{-1}\coma 
        \Lambda^{R}_{44}\rightarrow \gamma_{\Omega_1}\Lambda^{R\,\,T}_{11}\gamma_{\Omega_1}^{-1}\fstop 
    \label{eq:realfermiH4PhaseBorient}
\end{equation}

Using Table~\ref{tab:GenerH4PhaseB}, we get the corresponding geometric involution $\sigma$ on the generators of $H_4$
\begin{equation}
\begin{array}{c}
     \left(M_1,M_2,M_3,M_4,M_5,M_6,M_7,M_8\right)\\
    \downarrow\\
 \left(\bar{M}_1,\bar{M}_6,\bar{M}_5,\bar{M}_8,\bar{M}_3,\bar{M}_2,\bar{M}_7,\bar{M}_4\right)\coma
    \end{array}
    \label{eq:H4PhaseB-invol}
\end{equation}
which is exactly the same anti-holomorphic involution in \eqref{eq:H4PhaseA-invol}. In Section~\ref{section_vector_structure_H4} we elaborate on the relation between both theories and the role of vector structure.

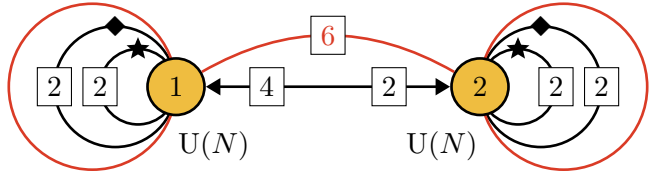
\begin{figure}[!htp]
    \centering
    \begin{tikzpicture}[scale=2]
	\draw[line width=1pt] (-2.25,0) circle (0.25) node[yshift=0.5cm,star,star points=5, star point ratio=2.25, inner sep=1pt, fill=black, draw] {} node[fill=white,text opacity=1,fill opacity=1,draw=black,rectangle,thin,xshift=-0.5cm] {$2$};
	\draw[line width=1pt] (-2.4,0) circle (0.4) node[yshift=0.8cm] {\scriptsize{$\quadro$}} node[fill=white,text opacity=1,fill opacity=1,draw=black,rectangle,thin,xshift=-0.8cm] {$2$};
	\draw[line width=1pt,redX] (-2.55,0) circle (0.55);
	\draw[line width=1pt] (0.25,0) circle (0.25) node[yshift=0.5cm,star,star points=5, star point ratio=2.25, inner sep=1pt, fill=black, draw] {} node[fill=white,text opacity=1,fill opacity=1,draw=black,rectangle,thin,xshift=0.5cm] {$2$};
	\draw[line width=1pt] (0.4,0) circle (0.4) node[yshift=0.8cm] {\scriptsize{$\quadro$}} node[fill=white,text opacity=1,fill opacity=1,draw=black,rectangle,thin,xshift=0.8cm] {$2$};
	\draw[line width=1pt,redX] (0.55,0) circle (0.55);
	\node[draw=black,line width=1pt,circle,fill=yellowX,minimum width=0.75cm,inner sep=1pt,label={[xshift=-0.5cm,yshift=-1.5cm]:$\U(N)$}] (A) at (0,0) {$2$};
	\node[draw=black,line width=1pt,circle,fill=yellowX,minimum width=0.75cm,inner sep=1pt,label={[xshift=0.5cm,yshift=-1.5cm]:$\U(N)$}] (B) at (-2,0) {$1$};
	\path[Triangle-Triangle] (A) edge[line width=1pt] node[fill=white,text opacity=1,fill opacity=1,draw=black,rectangle,thin,pos=0.25] {$2$} node[fill=white,text opacity=1,fill opacity=1,draw=black,rectangle,thin,pos=0.75] {$4$} (B);
	\draw[line width=1pt,redX] (A) to[bend right] node[fill=white,text opacity=1,fill opacity=1,draw=black,rectangle,thin] {\color{redX}{$6$}} (B);
    \end{tikzpicture}
    \caption{Quiver for a  Spin(7)  orientifold of phase B of $H_4$ using the involution in \eqref{eq:chiralH4PhaseBorient}, \eqref{eq:fermiH4PhaseBorient} and \eqref{eq:realfermiH4PhaseBorient}.}
    \label{fig:o_theory_h4_B}
\end{figure}

Figure~\ref{fig:o_theory_h4_B} shows the quiver for the orientifolded theory, which is anomaly free. 

\subsubsection{Vector Structure Explanation}

\label{section_vector_structure_H4}

On general grounds, one can expect that considering orientifolds by the same anti-holomorphic involution on geometries in different toric phases of the same geometry, should lead to equivalent $\mathcal{N}=(0,1)$ theories. Indeed, this can lead to a systematic construction of $\mathcal{N}=(0,1)$ theories related by 2d trialities, as we will discuss in a companion paper \cite{Franco:2021branetriality}.

On the other hand, this is not the case for the two orientifolds constructed in the previous section. We have seen that the $H_4$ theory admits several orientifold quotients which nevertheless correspond to the same anti-holomorphic involution, see \eqref{eq:H4PhaseA-invol} and \eqref{eq:H4PhaseB-invol}.  In this section, we show that the resulting theories are different because they correspond to orientifold quotients with or without vector structure, realized in the context of a non-orbifold singularity. 

Indeed, the structure of the orientifold action on the gauge factors follows the pattern described in Section \ref{sec:vectorstructure} for orientifolds of orbifolds of $\mathbb{C}^4$. Namely, the orientifold in Section~\ref{sec:H4PhaseA} acts on the quiver of the $H_4$ theory (in the toric phase A) by swapping the two nodes 2 and 4, while mapping nodes 1 and 3 to themselves; this corresponds to an orientifold with vector structure. On the other hand, the orientifold in Section~\ref{sec:H4PhaseB} acts on the quiver of the $H_4$ theory (in the toric phase B) by swapping 1 $\leftrightarrow$ 4 and 2 $\leftrightarrow$ 3; this corresponds to an orientifold without vector structure.

Hence, even though the two models correspond to the same underlying geometry, with an orientifold action associated to the same anti-holomorphic involution, the resulting orientifold theories are associated to genuinely different actions of the orientifold on the gauge degrees of freedom, and lead to inequivalent models.

An interesting observation is that the orientifolds with and without vector structure are obtained as orientifold quotients of the theory in two different toric phases. This effect did not arise in the context of orbifolds of $\mathbb{C}^4$, since these do not admit multiple toric phases; on the other hand, it is actually an expected phenomenon in non-orbifold singularities, as it already occurs in the context of 4d ${\cal N}=1$ theories with D3-branes at orientifold singularities. We illustrate this with the following simple example.

Consider a set of D3-branes at the tip of the non-compact CY 3-fold singularity described by the equation
\begin{equation}
    xy=z^2w^2 \fstop
\end{equation}
This corresponds to a $\IZ_2$ quotient of the conifold, of the kind introduced in \cite{Uranga:1998vf} as T-duals of 4d Hanany-Witten (HW) configurations of D4-branes suspended between NS and rotated NS-branes (aka NS$^\prime$-branes). This T-dual picture allowed to recover the same geometry from different Seiberg dual phases, as explicitly discussed in Section 3 of \cite{Feng:2001bn}. In particular, we can describe a phase A as corresponding to the type IIA configuration of D4-branes suspended in intervals separated by NS-branes ordered as NS - NS - NS$^\prime$ - NS$^\prime$ on the circle, and a phase B as corresponding to D4-branes suspended between NS-branes ordered as NS$^\prime$ - NS - NS$^\prime$ - NS on the circle. 

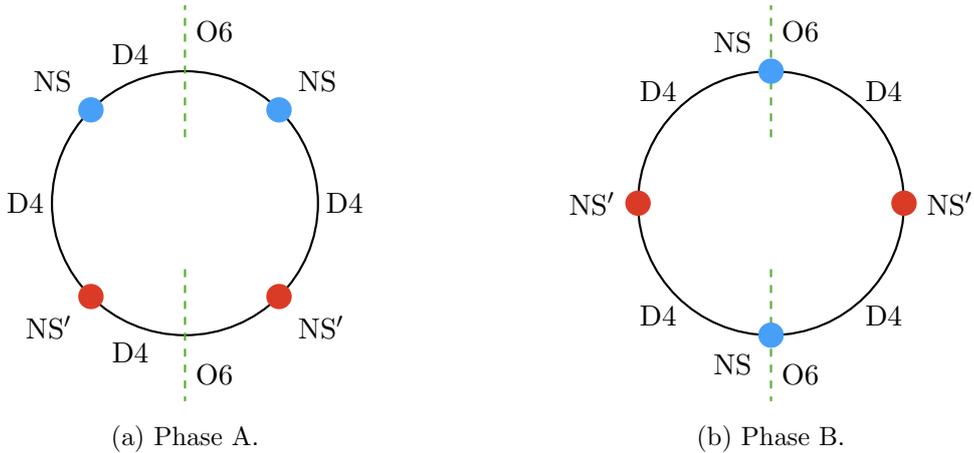
\begin{figure}[!htp]
\begin{center}
\begin{subfigure}[t]{0.49\textwidth}
\centering
\begin{tikzpicture}[scale=1.75]
\draw[thick] (0,0) circle (1);
\draw[dashed,greenX,thick] (0,0.5) -- node[black,pos=0.80,right] {O6} (0,1.5);
\draw[dashed,greenX,thick] (0,-0.5) -- node[black,pos=0.80,right] {O6} (0,-1.5);
\node[draw=blueX,line width=1pt,circle,fill=blueX,minimum width=0.3cm,inner sep=1pt,label={above right:NS}] (NS1) at (45:1) {};
\node[draw=blueX,line width=1pt,circle,fill=blueX,minimum width=0.3cm,inner sep=1pt,label={above left:NS}] (NS2) at (135:1) {};
\node[draw=redX,line width=1pt,circle,fill=redX,minimum width=0.3cm,inner sep=1pt,label={below left:NS$^\prime$}] (NSp1) at (225:1) {};
\node[draw=redX,line width=1pt,circle,fill=redX,minimum width=0.3cm,inner sep=1pt,label={below right:NS$^\prime$}] (NSp2) at (315:1) {};
\node at (0:1.2) {D4};
\node at (110:1.2) {D4};
\node at (180:1.2) {D4};
\node at (250:1.2) {D4};
\end{tikzpicture}
\caption{Phase A.}
\label{fig:NSNSpPhaseA}
\end{subfigure}\hfill
\begin{subfigure}[t]{0.49\textwidth}
\centering
\begin{tikzpicture}[scale=1.75]
\draw[thick] (0,0) circle (1);
\draw[dashed,greenX,thick] (0,0.5) -- node[black,pos=0.80,right] {O6} (0,1.5);
\draw[dashed,greenX,thick] (0,-0.5) -- node[black,pos=0.80,right] {O6} (0,-1.5);
\node[draw=blueX,line width=1pt,circle,fill=blueX,minimum width=0.3cm,inner sep=1pt,label={above left:NS}] (NS1) at (90:1) {};
\node[draw=blueX,line width=1pt,circle,fill=blueX,minimum width=0.3cm,inner sep=1pt,label={below left:NS}] (NS2) at (270:1) {};
\node[draw=redX,line width=1pt,circle,fill=redX,minimum width=0.3cm,inner sep=1pt,label={right:NS$^\prime$}] (NSp1) at (0:1) {};
\node[draw=redX,line width=1pt,circle,fill=redX,minimum width=0.3cm,inner sep=1pt,label={left:NS$^\prime$}] (NSp2) at (180:1) {};
\node at (45:1.2) {D4};
\node at (135:1.2) {D4};
\node at (225:1.2) {D4};
\node at (315:1.2) {D4};
\end{tikzpicture}
\caption{Phase B.}
\label{fig:NSNSpPhaseB}
\end{subfigure}
\caption{Configurations of D4-branes suspended between NS- and NS$^\prime$-branes in the presence of O6-planes leading to two different 4d ${\cal N}=1$ theories from orientifold quotients of the same CY$_3$ geometry differing only on the existence (Figure~\ref{fig:NSNSpPhaseA}) or not (Figure~\ref{fig:NSNSpPhaseB}) of vector structure.}
\label{fig:toric-phases-4d}
\end{center}
\end{figure}

Let us now perform an orientifold quotient in the type IIB geometry, which corresponds to, e.g., introducing O6-planes in the type IIA T-dual; this can map NS-branes to NS-branes, and NS$^\prime$-branes to NS$^\prime$-branes, and cannot swap NS- and NS$^\prime$-branes. Hence, for phase A, the only $\mathbb{Z}_2$-invariant configuration must have the orientifold swapping the two NS branes, and swapping the two NS$^\prime$-branes, see Figure \ref{fig:NSNSpPhaseA}; hence, the interval between the two NS-branes and the interval between the two NS$^\prime$-branes are both mapped to themselves under the orientifold action, while the intervals between NS- and  NS$^\prime$-branes are swapped. The result corresponds to an orientifold with vector structure.

On the other hand, for phase B, a $\mathbb{Z}_2$-invariant configuration has e.g. NS-branes mapped to themselves under the orientifold action, and the two NS$^\prime$-branes swapped, see Figure \ref{fig:NSNSpPhaseB}; hence, no interval is mapped to itself, rather the four intervals are swapped pairwise. The result corresponds to an orientifold without vector structure (there is an equivalent model obtained by having NS$^\prime$-branes on top of the orientifold plane, and the two NS-branes swapped under the orientifold action).

This illustrates the fact that the construction of orientifolds with or without vector structure, for a given geometric involution, may require their realization in different toric phases.

We have thus shown that, in order for equivalent orientifold geometric involutions to produce equivalent theories, it is necessary that they also agree on the choice of vector structure they implicitly define. This is an important ingredient in the application of orientifold quotients to $\mathcal{N}=(0,2)$ trialities to generate examples of theories displaying $\mathcal{N}=(0,1)$ triality \cite{Franco:2021branetriality}.

\section{Partial Resolution and Higgsing}
\label{sec:HiggsingPartResol}

In this section, we study partial resolutions connecting two different $\Spin(7)$ orientifolds, which translate into higgsings between the corresponding gauge theories.

\subsection{General Idea}
\label{sec:HiggsingOrient}

Consider two CY$_4$'s, CY$_4^{(1)}$ and CY$_4^{(2)}$, connected via partial resolution. Let us call the gauge theories on D1-branes probing them $\mathcal{T}_1^{(0,2)}$ and $\mathcal{T}_2^{(0,2)}$, respectively.\footnote{More precisely, we mean one of the various phases related via $\mathcal{N}=(0,2)$ triality for each CY$_4$.} Partial resolution translates into higgsing connecting the two gauge theories, in which the scalar component of one or more chiral fields gets a non-zero VEV (as usual, this is meant in the Born-Oppenheimer approximation in 2d). In the process, part of the gauge symmetry is higgsed and some matter fields may become massive. We refer to \cite{Franco:2015tna} for a more detailed discussion and explicit examples.

Let us now consider a $\Spin(7)$ orientifold $\mathcal{O}_1$ of CY$_4^{(1)}$ associated to a given anti-holomorphic involution $\sigma$. If the partial resolution considered above is symmetric under $\sigma$, it gives rise to a partial resolution of $\mathcal{O}_1$ into a $\Spin(7)$ orientifold $\mathcal{O}_2$ of CY$_4^{(2)}$. At the field theory level, the VEVs that higgs $\mathcal{T}_1^{(0,2)}\to \mathcal{T}_2^{(0,2)}$ are symmetric under the involution and project onto a higgsing between the orientifold gauge theories, $\mathcal{T}_1^{(0,1)}\to \mathcal{T}_2^{(0,1)}$. Figure~\ref{partial resolution, orientifolding and higgsing} illustrates the interplay between partial resolution, orientifolding and higgsing.

\begin{figure}[ht!]
    \centering
    \begin{subfigure}[t]{0.49\textwidth}
    \centering
       \begin{tikzpicture}[scale=2]
    \def\x{1.5};
    \node (A) at (0,0) {CY$_4^{(1)}$};
    \node (B) at (1*\x,0) {$\mathcal{O}_1$};
    \node (C) at (1*\x,-\x) {$\mathcal{O}_2$};
    \node (D) at (0,-\x) {CY$_4^{(2)}$};
    \draw[line width=1pt,-Triangle] (A)-- node[above,midway]{$\Omega\sigma$} (B);
    \draw[line width=1pt,-Triangle] (B)-- node[right,midway] {\shortstack{Partial\\resolution}} (C);
    \draw[line width=1pt,Triangle-] (C)-- node[below,midway]{$\Omega\sigma$}(D);
    \draw[line width=1pt,Triangle-] (D)-- node[left,midway] {\shortstack{Partial\\resolution}} (A);
    \end{tikzpicture}
    \caption{String background.}
    \label{fig:UniInvoParRes}
    \end{subfigure}\hfill
    \begin{subfigure}[t]{0.49\textwidth}
    \centering
    \begin{tikzpicture}[scale=2]
    \def\x{1.5};
    \node (A) at (0,0) {$\mathcal{T}_1^{(0,2)}$};
    \node (B) at (\x,0) {$\mathcal{T}_1^{(0,1)}$};
    \node (C) at (\x,-\x) {$\mathcal{T}_2^{(0,1)}$};
    \node (D) at (0,-\x) {$\mathcal{T}_2^{(0,2)}$};
    \draw[line width=1pt,-Triangle] (A)-- node[above,midway]{Projection} (B);
    \draw[line width=1pt,-Triangle] (B)-- node[right,midway] {Higgsing} (C);
    \draw[line width=1pt,Triangle-] (C)-- node[below,midway]{Projection}(D);
    \draw[line width=1pt,Triangle-] (D)-- node[left,midway] {Higgsing} (A);
    \end{tikzpicture}
    \caption{Field theory.}
    \label{fig:UniInvoHiggs}
    \end{subfigure}
    \caption{Interplay between partial resolution, orientifolding and higgsing.}
    \label{partial resolution, orientifolding and higgsing}
\end{figure}
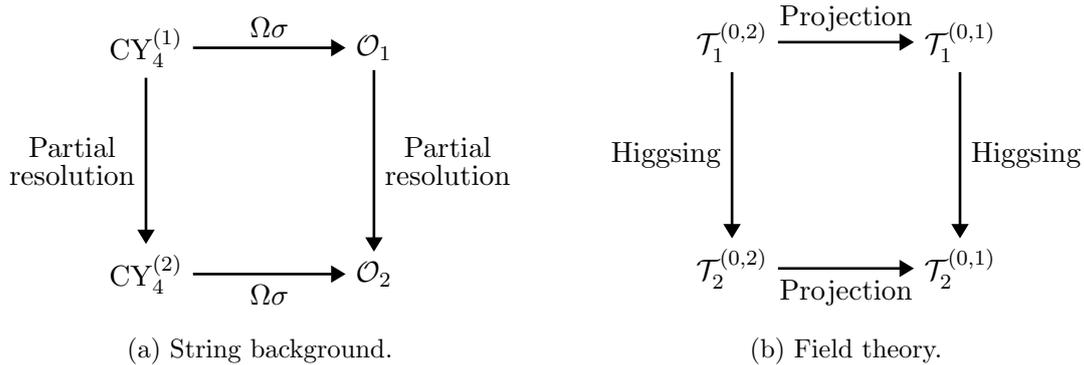

\subsection{Partial Resolution and the Universal Involution}

Interestingly, for theories obtained via the universal involution, {\it every} partial resolution between CY$_4$'s maps to a partial resolution between $\Spin(7)$ orientifolds. In this case, every field in $\mathcal{T}_1^{(0,2)}$ and $\mathcal{T}_2^{(0,2)}$ is its own orientifold image. Therefore, the condition that chiral fields and their images get VEVs simultaneously is automatically satisfied. 

Under the universal involution, VEVs and the resulting higgsing of the gauge symmetry and mass terms for some matter fields straightforwardly map from the parent to the orientifolded theory. In other words, higgsing survives the ``real slicing" of the universal involution.

\subsection{Beyond the Universal Involution: $\mathbb{C}^4/\mathbb{Z}_2\times \mathbb{Z}_2 \rightarrow \SPP\times \mathbb{C}$}
\label{sec:C4Z2Z2toSPPC}

The interplay between partial resolutions and orientifolds that we discussed above is not limited to the universal involution. 

\subsubsection*{The Parent}

Let us consider the $\mathbb{C}^4/\mathbb{Z}_2\times \mathbb{Z}_2$ orbifold, with the two $\ZZ_2$ groups generated by the actions $(1,1,0,0)$ and $(1,0,1,0)$ on $\mathbb{C}^4$, as phase rotations (in units of $\pi$). From now on, we will omit these vectors. This orbifold can be partially resolved to SPP$\times \mathbb{C}$, where SPP denotes the complex cone over the suspended pinch point. The toric diagrams and gauge theories for both geometries can be found in Appendices~\ref{app:otheory for c4z2z2} and \ref{app:otheory for sppXC}. This partial resolution and its translation into higgsing of the gauge theory has been discussed in detail in \cite{Franco:2015tna}.

The two theories are connected by turning on a VEV for $X_{13}$.\footnote{There are other choices of the chiral field getting a VEV that lead to the same resolution. They are equivalent to this choice by symmetries.} As a result, the $\U(N)_1\times \U(N)_3$ gauge groups are broken to the diagonal $\U(N)_{1/3}$. In addition, the following Fermi-chiral pairs become massive
\begin{equation}\label{eq:massive fields in (0,2)}
    \{ \Lambda_{21}, X_{32}\},~ \left\{ \Lambda_{13}, \frac{{X}_{11}-X_{33}}{2}\right\},~\{ \Lambda_{41}, X_{34}\},~\{\Lambda_{32}, X_{21}\},~\{\Lambda_{34}, X_{41}\}\fstop
\end{equation}
Integrating out the massive fields leads to the gauge theory for SPP$\times \mathbb{C}$.

\subsubsection*{The Spin(7) Orientifold}

In Appendix~\ref{app:otheory for c4z2z2}, we present a $\Spin(7)$ orientifold of $\mathbb{C}^4/\mathbb{Z}_2\times \mathbb{Z}_2$ constructed using a non-universal involution, given in~\cref{eq:C4Z2Z2chiral,eq:C4Z2Z2fermi,eq:C4Z2Z2realfermi}. The crucial point of that involution for the discussion in this section is that it maps $X_{13}$ to itself. Following the discussion in Section~\ref{sec:HiggsingOrient}, the resolution/higgsing of the parent is therefore projected onto one for the $\Spin(7)$ orientifold.

In the $\Spin(7)$ orientifold of $\mathbb{C}^4/\mathbb{Z}_2 \times \mathbb{Z}_2$, the higgsing associated to this partial resolution proceeds by giving a VEV to $X_{13}^R$. This breaks the $\SO(N)_1\times \SO(N)_3$ gauge symmetry into the diagonal $\SO(N)_{1/3}$. In addition, the combination of real Fermi fields $\frac{\Lambda_{11S}^R+\Lambda_{33S}^R}{2}$, coming from the $\mathcal{N}=(0,2)$ vector multiplets of gauge groups 1 and 3, become massive. Finally, The following fields also become massive
\begin{equation}\label{eq:massive fields in (0,1)}
   \Lambda_{21}, ~X_{32},~\Lambda_{13}^R,~ \frac{{X}_{11S}^R-X_{33S}^R}{2},~\frac{{X}_{11A}^R-X_{33A}^R}{2},~\Lambda_{32}, X_{21}\fstop
\end{equation}
 Integrating them out, each of the surviving bifundamentals of $\SO(N)_1\times \SO(N)_3$ becomes a symmetric and an antisymmetric of $\SO(N)_{1/3}$. The resulting theory is exactly the one associated for the $\Spin(7)$ orientifold of SPP$\times \mathbb{C}$ in Figure~\ref{fig:o_theory_SPP}, generated by the anti-holomorphic involution in \cref{eq:SPPchiral,eq:SPPfermi,eq:SPPrealfermi}.

\section{Conclusions}

\label{section_conclusions}

In this paper, we initiated the geometric engineering of 2d $\mathcal{N}=(0,1)$ gauge theories by means of D1-branes probing (orientifolds of) Spin(7) cones. In particular, we introduced Spin(7) orientifolds, which are constructed by starting from a CY$_4$ cone and quotienting it by a combination of an anti-holomorphic involution leading to a Spin(7) cone and worldsheet parity. 

We illustrated this construction with various examples, including theories coming from both orbifold and non-orbifold parent singularities, discussed the r\^ole of the choice of vector structure in the orientifold quotient, and studied partial resolutions. 

Spin(7) orientifolds explicitly realize the perspective on 2d $\mathcal{N}=(0,1)$  theories as real slices of $\mathcal{N}=(0,2)$ ones. Remarkably, this projection is mapped to Joyce’s construction of Spin(7) manifolds as quotients of CY$_4$’s by anti-holomorphic involutions.

We envision multiple directions for future research. To name a few: 
\begin{itemize}

\item In general, the map between Spin(7) orientifolds and 2d $\mathcal{N}=(0,1)$ gauge theories is not one-to-one but one-to-many. We will investigate this issue in \cite{Franco:2021branetriality}, showing that this non-uniqueness provides a geometric understanding of $\mathcal{N}=(0,1)$ triality.

\item Another interesting direction is to construct the gauge theories on D1-branes probing Spin(7) manifolds obtained from CY$_4$’s via Joyce’s construction, without the additional quotient by worldsheet parity leading to Spin(7) orientifolds. A significant part of the results of this paper would also be useful for such setups. We plan to study this problem in a future work. 

\item In Section~\ref{sec:HiggsingPartResol}, we considered resolutions of Spin(7) orientifolds. It would be interesting to investigate deformations and their gauge theory counterpart.  Understanding deformations for CY$_4$’s and their translation to the associated 2d $\mathcal{N}=(0,2)$ gauge theories would be a useful preliminary step, which is interesting in its own right. 

\end{itemize}

We hope that the novel perspective on 2d $\mathcal{N}=(0,1)$ introduced in this paper will provide a useful tool for understanding their dynamics.

\acknowledgments

We would like to thank Massimo Porrati and Sergei Gukov for enjoyable discussions. The research of S. F. was supported by the U.S. National Science Foundation grants PHY-1820721 and PHY-2112729. A. M. received funding from ``la Caixa" Foundation (ID 100010434) with fellowship code LCF/BQ/IN18/11660045 and from the European Union’s Horizon 2020 research and innovation programme under the Marie Sk\l odowska-Curie grant agreement No. 713673 until September 2021. The work of A. M. is supported in part by Deutsche Forschungsgemeinschaft under Germany's Excellence Strategy EXC 2121 Quantum Universe 390833306. The work of A. U. is supported by the Spanish Research Agency (Agencia Espanola de Investigaci\'on) through the grants IFT Centro de Excelencia Severo Ochoa SEV-2016-0597, and the grant GC2018-095976-B-C21 from MCIU/AEI/FEDER, UE. S. F. and X. Y. would like to thank the Simons Center for Geometry and Physics for hospitality during part of this work. X. Y. is also grateful to the Caltech Particle Theory Group, and specially to Sergei Gukov and Nathan Benjamin, for hosting him during the final stages of this paper.

\newpage 
\appendix

\section{$\CC^4/\ZZ_2\times \ZZ_2$ and $\SPP\times \CC$}
\label{app:C4Z2Z2SPPC}

In this appendix we present two additional examples of Spin(7) manifolds, which are considered in Section~\ref{sec:C4Z2Z2toSPPC} to discuss partial resolutions.

\subsection{$\CC^4/\ZZ_2\times \ZZ_2$}
\label{app:otheory for c4z2z2}

\begin{figure}[H]
    \centering
	\begin{tikzpicture}[scale=1.5]
	\draw[thick,gray,-Triangle] (0,0,0) -- node[above,pos=1] {$x$} (2.5,0,0);
	\draw[thick,gray,-Triangle] (0,0,0) -- node[left,pos=1] {$y$} (0,1.5,0);
	\draw[thick,gray,-Triangle] (0,0,0) -- node[below,pos=1] {$z$} (0,0,2.5);
	\node[draw=black,line width=1pt,circle,fill=black,minimum width=0.2cm,inner sep=1pt] (O) at (0,0,0) {};
	\node[draw=black,line width=1pt,circle,fill=black,minimum width=0.2cm,inner sep=1pt] (X1) at (1,0,0) {};
	\node[draw=black,line width=1pt,circle,fill=black,minimum width=0.2cm,inner sep=1pt] (X2) at (2,0,0) {};
	\node[draw=black,line width=1pt,circle,fill=black,minimum width=0.2cm,inner sep=1pt] (Z1) at (0,0,1) {};
	\node[draw=black,line width=1pt,circle,fill=black,minimum width=0.2cm,inner sep=1pt] (Z2) at (0,0,2) {};
	\node[draw=black,line width=1pt,circle,fill=black,minimum width=0.2cm,inner sep=1pt] (XZ) at (1,0,1) {};
	\node[draw=black,line width=1pt,circle,fill=black,minimum width=0.2cm,inner sep=1pt] (Y) at (0,1,0) {};
	\draw[line width=1pt] (O)--(X2)--(Z2)--(O);
	\draw[line width=1pt] (Y)--(O);
	\draw[line width=1pt] (Y)--(X1);
	\draw[line width=1pt] (Y)--(X2);
	\draw[line width=1pt] (Y)--(Z1);
	\draw[line width=1pt] (Y)--(Z2);
	\draw[line width=1pt] (Y)--(XZ);
	\end{tikzpicture}
	\caption{Toric diagram for $\mathbb{C}^4/\mathbb{Z}_2\times \mathbb{Z}_2$.}
	\label{fig:C4Z2Z2toricdiagram}
\end{figure}
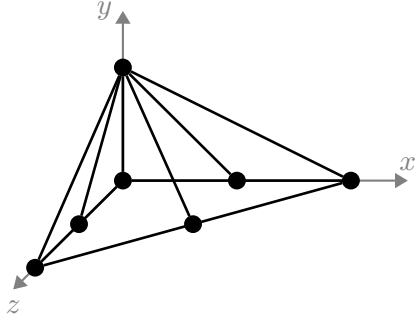

Figure~\ref{fig:C4Z2Z2toricdiagram} shows the toric diagram for the $\CC^4/\ZZ_2\times \ZZ_2$ orbifold.

\begin{figure}[H]
		\centering
 		\begin{subfigure}[t]{0.49\textwidth}
 			\centering
			\begin{tikzpicture}[scale=2]
			\draw[line width=1pt,decoration={markings, mark=at position 0.5 with{\arrow{Triangle}}}, postaction={decorate}] (-0.22,-0.22) circle (0.25);
			\draw[line width=1pt,decoration={markings, mark=at position 0.7 with{\arrow{Triangle}}}, postaction={decorate}] (-0.22,2.22) circle (0.25);
			\draw[line width=1pt,decoration={markings, mark=at position 0.75 with{\arrow{Triangle}}}, postaction={decorate}] (2.22,-0.22) circle (0.25);
			\draw[line width=1pt,decoration={markings, mark=at position 0.5 with{\arrow{Triangle}}}, postaction={decorate}] (2.22,2.22) circle (0.25);
			\node[draw=black,line width=1pt,circle,fill=yellowX,minimum width=0.75cm,inner sep=1pt] (A) at (0,0) {$4$};
			\node[draw=black,line width=1pt,circle,fill=yellowX,minimum width=0.75cm,inner sep=1pt] (B) at (2,0) {$3$};
			\node[draw=black,line width=1pt,circle,fill=yellowX,minimum width=0.75cm,inner sep=1pt] (C) at (2,2) {$2$};
			\node[draw=black,line width=1pt,circle,fill=yellowX,minimum width=0.75cm,inner sep=1pt] (D) at (0,2) {$1$};
			\path[Triangle-Triangle] (A) edge[line width=1pt] (B);
			\path[Triangle-Triangle] (B) edge[line width=1pt] (C);
			\path[Triangle-Triangle] (C) edge[line width=1pt] (D);
			\path[Triangle-Triangle] (D) edge[line width=1pt] (A);
			\path[Triangle-Triangle] (A) edge[line width=1pt] (C);
			\path[Triangle-Triangle] (B) edge[line width=1pt] (D);
			\draw[line width=1pt,redX] (A) to[bend right=20] (B);
			\draw[line width=1pt,redX] (A) to[bend left=20] (B);
			\draw[line width=1pt,redX] (B) to[bend right=20] (C);
			\draw[line width=1pt,redX] (B) to[bend left=20] (C);
			\draw[line width=1pt,redX] (C) to[bend right=20] (D);
			\draw[line width=1pt,redX] (C) to[bend left=20] (D);
			\draw[line width=1pt,redX] (D) to[bend right=20] (A);
			\draw[line width=1pt,redX] (D) to[bend left=20] (A);
			\draw[line width=1pt,redX] (A) to[bend right=20] (C);
			\draw[line width=1pt,redX] (A) to[bend left=20] (C);
			\draw[line width=1pt,redX] (B) to[bend right=20] (D);
			\draw[line width=1pt,redX] (B) to[bend left=20] (D);
			\node at (0,-0.65) {};
			\end{tikzpicture}
			\caption{$\mathcal{N}=(0,2)$ language.}
			\label{fig:C4Z2Z2N02}
		\end{subfigure}\hfill
		\begin{subfigure}[t]{0.49\textwidth}
			\centering
			\begin{tikzpicture}[scale=2]
			\draw[line width=1pt,decoration={markings, mark=at position 0.5 with{\arrow{Triangle}}}, postaction={decorate}] (-0.22,-0.22) circle (0.25) node[fill=white,text opacity=1,fill opacity=1,draw=black,rectangle,thin,xshift=-0.35cm,yshift=-0.35cm] {$2$};
			\draw[line width=1pt,decoration={markings, mark=at position 0.70 with{\arrow{Triangle}}}, postaction={decorate}] (-0.22,2.22) circle (0.25) node[fill=white,text opacity=1,fill opacity=1,draw=black,rectangle,thin,xshift=-0.35cm,yshift=0.35cm] {$2$};
			\draw[line width=1pt,decoration={markings, mark=at position 0.75 with{\arrow{Triangle}}}, postaction={decorate}] (2.22,-0.22) circle (0.25) node[fill=white,text opacity=1,fill opacity=1,draw=black,rectangle,thin,xshift=0.35cm,yshift=-0.35cm] {$2$};
			\draw[line width=1pt,decoration={markings, mark=at position 0.5 with{\arrow{Triangle}}}, postaction={decorate}] (2.22,2.22) circle (0.25) node[fill=white,text opacity=1,fill opacity=1,draw=black,rectangle,thin,xshift=0.35cm,yshift=0.35cm] {$2$};
			\draw[line width=1pt,redX] (-0.28,-0.28) circle (0.4);
			\draw[line width=1pt,redX] (-0.28,2.28) circle (0.4);
			\draw[line width=1pt,redX] (2.28,-0.28) circle (0.4);
			\draw[line width=1pt,redX] (2.28,2.28) circle (0.4);
			\node[draw=black,line width=1pt,circle,fill=yellowX,minimum width=0.75cm,inner sep=1pt] (A) at (0,0) {$4$};
			\node[draw=black,line width=1pt,circle,fill=yellowX,minimum width=0.75cm,inner sep=1pt] (B) at (2,0) {$3$};
			\node[draw=black,line width=1pt,circle,fill=yellowX,minimum width=0.75cm,inner sep=1pt] (C) at (2,2) {$2$};
			\node[draw=black,line width=1pt,circle,fill=yellowX,minimum width=0.75cm,inner sep=1pt] (D) at (0,2) {$1$};
			\path[Triangle-Triangle] (A) edge[line width=1pt] node[fill=white,text opacity=1,fill opacity=1,draw=black,rectangle,thin,pos=0.25] {$2$} node[fill=white,text opacity=1,fill opacity=1,draw=black,rectangle,thin,pos=0.75] {$2$} (B);
			\path[Triangle-Triangle] (B) edge[line width=1pt] node[fill=white,text opacity=1,fill opacity=1,draw=black,rectangle,thin,pos=0.25] {$2$} node[fill=white,text opacity=1,fill opacity=1,draw=black,rectangle,thin,pos=0.75] {$2$}(C);
			\path[Triangle-Triangle] (C) edge[line width=1pt] node[fill=white,text opacity=1,fill opacity=1,draw=black,rectangle,thin,pos=0.25] {$2$} node[fill=white,text opacity=1,fill opacity=1,draw=black,rectangle,thin,pos=0.75] {$2$}(D);
			\path[Triangle-Triangle] (D) edge[line width=1pt] node[fill=white,text opacity=1,fill opacity=1,draw=black,rectangle,thin,pos=0.25] {$2$} node[fill=white,text opacity=1,fill opacity=1,draw=black,rectangle,thin,pos=0.75] {$2$}(A);
			\path[Triangle-Triangle] (A) edge[line width=1pt] node[fill=white,text opacity=1,fill opacity=1,draw=black,rectangle,thin,pos=0.15] {$2$} node[fill=white,text opacity=1,fill opacity=1,draw=black,rectangle,thin,pos=0.85] {$2$}(C);
			\path[Triangle-Triangle] (B) edge[line width=1pt] node[fill=white,text opacity=1,fill opacity=1,draw=black,rectangle,thin,pos=0.15] {$2$} node[fill=white,text opacity=1,fill opacity=1,draw=black,rectangle,thin,pos=0.85] {$2$}(D);
			\draw[line width=1pt,redX] (A) to[bend right=20] node[fill=white,text opacity=1,fill opacity=1,draw=black,rectangle,thin,pos=0.5] {\color{redX}{$2$}} (B);
			\draw[line width=1pt,redX] (A) to[bend left=20] node[fill=white,text opacity=1,fill opacity=1,draw=black,rectangle,thin,pos=0.5] {\color{redX}{$2$}} (B);
			\draw[line width=1pt,redX] (B) to[bend right=20] node[fill=white,text opacity=1,fill opacity=1,draw=black,rectangle,thin,pos=0.5] {\color{redX}{$2$}} (C);
			\draw[line width=1pt,redX] (B) to[bend left=20] node[fill=white,text opacity=1,fill opacity=1,draw=black,rectangle,thin,pos=0.5] {\color{redX}{$2$}} (C);
			\draw[line width=1pt,redX] (C) to[bend right=20] node[fill=white,text opacity=1,fill opacity=1,draw=black,rectangle,thin,pos=0.5] {\color{redX}{$2$}} (D);
			\draw[line width=1pt,redX] (C) to[bend left=20] node[fill=white,text opacity=1,fill opacity=1,draw=black,rectangle,thin,pos=0.5] {\color{redX}{$2$}} (D);
			\draw[line width=1pt,redX] (D) to[bend right=20] node[fill=white,text opacity=1,fill opacity=1,draw=black,rectangle,thin,pos=0.5] {\color{redX}{$2$}} (A);
			\draw[line width=1pt,redX] (D) to[bend left=20] node[fill=white,text opacity=1,fill opacity=1,draw=black,rectangle,thin,pos=0.5] {\color{redX}{$2$}} (A);
			\draw[line width=1pt,redX] (A) to[bend right=20] node[fill=white,text opacity=1,fill opacity=1,draw=black,rectangle,thin,pos=0.78] {\color{redX}{$2$}} (C);
			\draw[line width=1pt,redX] (A) to[bend left=20] node[fill=white,text opacity=1,fill opacity=1,draw=black,rectangle,thin,pos=0.22] {\color{redX}{$2$}} (C);
			\draw[line width=1pt,redX] (B) to[bend right=20] node[fill=white,text opacity=1,fill opacity=1,draw=black,rectangle,thin,pos=0.75] {\color{redX}{$2$}} (D);
			\draw[line width=1pt,redX] (B) to[bend left=20] node[fill=white,text opacity=1,fill opacity=1,draw=black,rectangle,thin,pos=0.25] {\color{redX}{$2$}} (D);
			\end{tikzpicture}
			\caption{$\mathcal{N}=(0,1)$ language.}
			\label{fig:C4Z2Z2N01}
		\end{subfigure}
		\caption{Quiver diagram for $\mathbb{C}^4/\mathbb{Z}_2\times \mathbb{Z}_2$ in $\mathcal{N}=(0,2)$ and $\mathcal{N}=(0,1)$ language.}
		\label{fig:C4Z2Z2quiv}
	\end{figure}
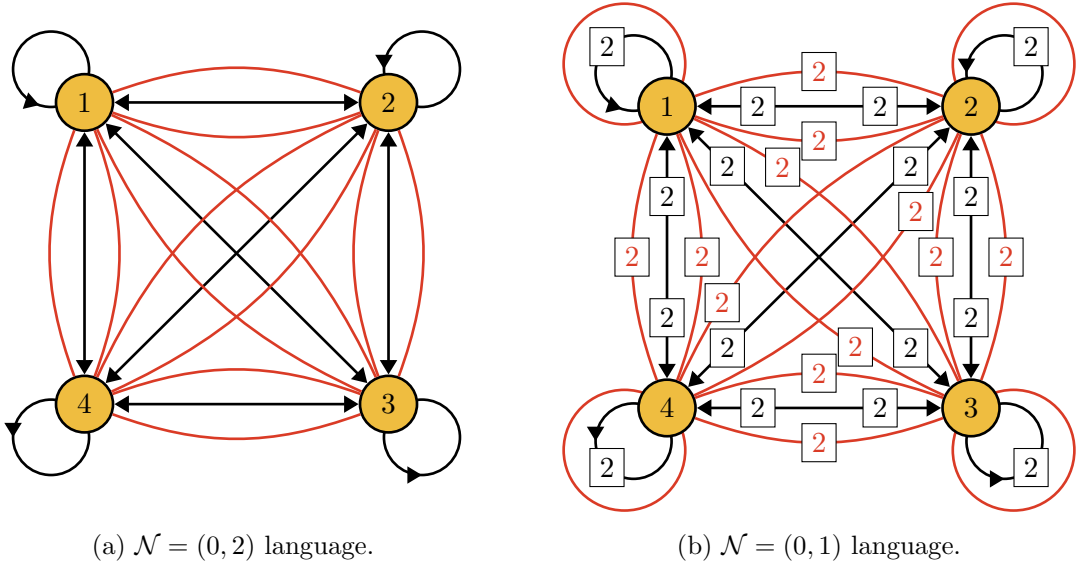

The gauge theory for D1-branes probing this orbifold was first constructed in \cite{Franco:2015tna}. Its quiver diagram is shown in Figure~\ref{fig:C4Z2Z2N02}.

The $J$- and $E$-terms are
\begin{equation}
    \begin{array}{lclc}
& J                                              &\text{\hspace{.5cm}}& E                         \\
\Lambda_{12} \,:\, & X_{23}X_{31} - X_{24}X_{41} &                    &X_{11}X_{12} - X_{12}X_{22}\\ 
\Lambda_{21} \,:\, & X_{14}X_{42} - X_{13}X_{32} &                    &X_{22}X_{21} - X_{21}X_{11}\\ 
\Lambda_{13} \,:\, & X_{34}X_{41} - X_{32}X_{21} &                    &X_{11}X_{13} - X_{13}X_{33}\\ 
\Lambda_{31} \,:\, & X_{12}X_{23} - X_{14}X_{43} &                    &X_{33}X_{31} - X_{31}X_{11}\\ 
\Lambda_{14} \,:\, & X_{42}X_{21} - X_{43}X_{31} &                    &X_{11}X_{14} - X_{14}X_{44}\\ 
\Lambda_{41} \,:\, & X_{13}X_{34} - X_{12}X_{24} &                    &X_{44}X_{41} - X_{41}X_{11}\\ 
\Lambda_{23} \,:\, & X_{31}X_{12} - X_{34}X_{42} &                    &X_{22}X_{23} - X_{23}X_{33}\\ 
\Lambda_{32} \,:\, & X_{24}X_{43} - X_{21}X_{13} &                    &X_{33}X_{32} - X_{32}X_{22}\\ 
\Lambda_{24} \,:\, & X_{43}X_{32} - X_{41}X_{12} &                    &X_{22}X_{24} - X_{24}X_{44}\\ 
\Lambda_{42} \,:\, & X_{21}X_{14} - X_{23}X_{34} &                    &X_{44}X_{42} - X_{42}X_{22}\\ 
\Lambda_{34} \,:\, & X_{41}X_{13} - X_{42}X_{23} &                    &X_{33}X_{34} - X_{34}X_{44}\\
\Lambda_{43} \,:\, & X_{32}X_{24} - X_{31}X_{14} &                    &X_{44}X_{43} - X_{43}X_{33}
\end{array}
\label{J_E_C4/Z2xZ2}
\end{equation}

Figure~\ref{fig:C4Z2Z2N01} shows the quiver for this theory in $\mathcal{N}= (0,1)$ language. The $W^{(0,1)}$ associated to \eqref{J_E_C4/Z2xZ2} is
\begin{equation}
    \begin{split}
        W^{(0,1)} = & \, W^{(0,2)}+ \sum_{\substack{i,j=1\\i\neq j}}^4\Lambda^{R}_{ii}\left(X^\dagger_{ij}X_{ij}+X^\dagger_{ji}X_{ji}\right)+\sum_{i=1}^4\Lambda^{R}_{ii}X^\dagger_{ii}X_{ii}\fstop
    \end{split}
\end{equation}

Table~\ref{tab:GenerC4Z2Z2} shows the generators of the moduli space and their expression in terms of chiral fields.

 \begin{table}[!htp]
	\centering
	\renewcommand{\arraystretch}{1.1}
	\begin{tabular}{c|c}
		Meson    & Chiral fields  \\
		\hline
		$M_1$    & $X_{12}X_{21}=X_{43}X_{34}$ \\
		$M_2$    & $X_{13}X_{31}=X_{24}Z_{42}$\\
		$M_3$    & $X_{14}X_{41}=X_{23}X_{32}$\\
		$M_4$    & $X_{14}X_{42}X_{21}=X_{23}X_{31}X_{12}=X_{32}X_{24}X_{43}=X_{41}X_{13}X_{34}=$\\ 
		         & $=X_{14}X_{43}X_{31}=X_{23}X_{34}X_{42}=X_{32}X_{21}X_{13}=X_{41}X_{12}X_{24}$\\
		$M_5$    & $X_{11}=X_{22}=X_{33}=X_{44}$\\
	\end{tabular}
	\caption{Generators of $\CC^4/\ZZ_2\times \ZZ_2$.}
	\label{tab:GenerC4Z2Z2}
\end{table}

They are subject to the following relation
\begin{equation}
    \mathcal{I}=\left\langle M_1M_2M_3=M_4^2\right\rangle\fstop
\end{equation}

\bigskip

\paragraph{$\SO(N)\times \U(N)\times \SO(N)$ Orientifold}\mbox{}

\medskip

Let us consider an anti-holomorphic involution which acts on Figure~\ref{fig:C4Z2Z2quiv} as a reflection with respect to the diagonal connecting nodes 1 and 3. Then, nodes 1 and 3 map to themselves, while nodes 2 and 4 are identified.

The involution on chiral fields is
 \begin{equation}\label{eq:C4Z2Z2chiral}
 \begin{array}{cccccccccccc}
 	 X_{12} & \rightarrow &   \gamma_{\Omega_1}\bar{X}_{14}\gamma_{\Omega_4}^{-1}\coma & 
 	 X_{21} & \rightarrow &  \gamma_{\Omega_4}\bar{X}_{41}\gamma_{\Omega_1}^{-1} \coma  & 
 	 X_{13} & \rightarrow &   \gamma_{\Omega_1}\bar{X}_{13}\gamma_{\Omega_3}^{-1} \coma & 
 	 X_{31} & \rightarrow &   \gamma_{\Omega_3}\bar{X}_{31}\gamma_{\Omega_1}^{-1} \coma \\
 	 X_{14} & \rightarrow &   \gamma_{\Omega_1}\bar{X}_{12}\gamma_{\Omega_2}^{-1}\coma & 
 	 X_{41} & \rightarrow &  \gamma_{\Omega_2}\bar{X}_{21}\gamma_{\Omega_1}^{-1} \coma  & 
 	 X_{43} & \rightarrow &   \gamma_{\Omega_2}\bar{X}_{23}\gamma_{\Omega_3}^{-1} \coma & 
 	 X_{34} & \rightarrow &   \gamma_{\Omega_3}\bar{X}_{32}\gamma_{\Omega_2}^{-1} \coma \\
 	 X_{24} & \rightarrow &   \gamma_{\Omega_4}\bar{X}_{42}\gamma_{\Omega_2}^{-1}\coma & 
 	 X_{42} & \rightarrow &  \gamma_{\Omega_2}\bar{X}_{24}\gamma_{\Omega_4}^{-1} \coma  & 
 	 X_{23} & \rightarrow &   \gamma_{\Omega_4}\bar{X}_{43}\gamma_{\Omega_3}^{-1} \coma & 
 	 X_{32} & \rightarrow &   \gamma_{\Omega_3}\bar{X}_{34}\gamma_{\Omega_4}^{-1} \coma \\
 	 X_{11} & \rightarrow &   -\gamma_{\Omega_1}\bar{X}_{11}\gamma_{\Omega_1}^{-1}\coma & 
 	 X_{22} & \rightarrow &  -\gamma_{\Omega_4}\bar{X}_{44}\gamma_{\Omega_4}^{-1} \coma  & 
 	 X_{33} & \rightarrow &  -\gamma_{\Omega_3}\bar{X}_{33}\gamma_{\Omega_3}^{-1} \coma & 
 	 X_{44} & \rightarrow &   -\gamma_{\Omega_2}\bar{X}_{22}\gamma_{\Omega_2}^{-1} \fstop
 \end{array}
 \end{equation}

From the invariance of $W^{(0,1)}$, we obtain the transformations of the Fermi fields
 \begin{equation}\label{eq:C4Z2Z2fermi}
     \begin{array}{cccccccccccc}
 	 \Lambda_{12} & \rightarrow &   -\gamma_{\Omega_1}\bar{\Lambda}_{14}\gamma_{\Omega_4}^{-1}\coma & 
 	 \Lambda_{21} & \rightarrow &  -\gamma_{\Omega_4}\bar{\Lambda}_{41}\gamma_{\Omega_1}^{-1} \coma  & 
 	 \Lambda_{13} & \rightarrow &   -\gamma_{\Omega_1}\bar{\Lambda}_{13}\gamma_{\Omega_3}^{-1} \coma & 
 	 \Lambda_{31} & \rightarrow &   -\gamma_{\Omega_3}\bar{\Lambda}_{31}\gamma_{\Omega_1}^{-1} \coma \\
 	 \Lambda_{14} & \rightarrow &   -\gamma_{\Omega_1}\bar{\Lambda}_{12}\gamma_{\Omega_2}^{-1}\coma & 
 	 \Lambda_{41} & \rightarrow &  -\gamma_{\Omega_2}\bar{\Lambda}_{21}\gamma_{\Omega_1}^{-1} \coma  & 
 	 \Lambda_{43} & \rightarrow &   -\gamma_{\Omega_2}\bar{\Lambda}_{23}\gamma_{\Omega_3}^{-1} \coma & 
 	 \Lambda_{34} & \rightarrow &   -\gamma_{\Omega_3}\bar{\Lambda}_{32}\gamma_{\Omega_2}^{-1} \coma \\
 	 \Lambda_{24} & \rightarrow &   -\gamma_{\Omega_4}\bar{\Lambda}_{42}\gamma_{\Omega_2}^{-1}\coma & 
 	 \Lambda_{42} & \rightarrow &  -\gamma_{\Omega_2}\bar{\Lambda}_{24}\gamma_{\Omega_4}^{-1} \coma  & 
 	 \Lambda_{23} & \rightarrow &   -\gamma_{\Omega_4}\bar{\Lambda}_{43}\gamma_{\Omega_3}^{-1} \coma & 
 	 \Lambda_{32} & \rightarrow &   -\gamma_{\Omega_3}\bar{\Lambda}_{34}\gamma_{\Omega_4}^{-1} \coma 
 \end{array}
 \end{equation}
 and
\begin{equation}\label{eq:C4Z2Z2realfermi}
 \begin{array}{cccccccccccc}
 	 \Lambda^{R}_{11} & \rightarrow &   \gamma_{\Omega_3}\Lambda^{R\,\,T}_{11}\gamma_{\Omega_3}^{-1}\coma & 
 	 \Lambda^{R}_{22} & \rightarrow &  \gamma_{\Omega_2}\Lambda^{R\,\,T}_{44}\gamma_{\Omega_2}^{-1} \coma  & 
 	 \Lambda^{R}_{33} & \rightarrow &   \gamma_{\Omega_1}\Lambda^{R\,\,T}_{33}\gamma_{\Omega_1}^{-1} \coma & 
 	 \Lambda^{R}_{44} & \rightarrow &   \gamma_{\Omega_4}\Lambda^{R\,\,T}_{22}\gamma_{\Omega_4}^{-1} \fstop
 \end{array}
 \end{equation}

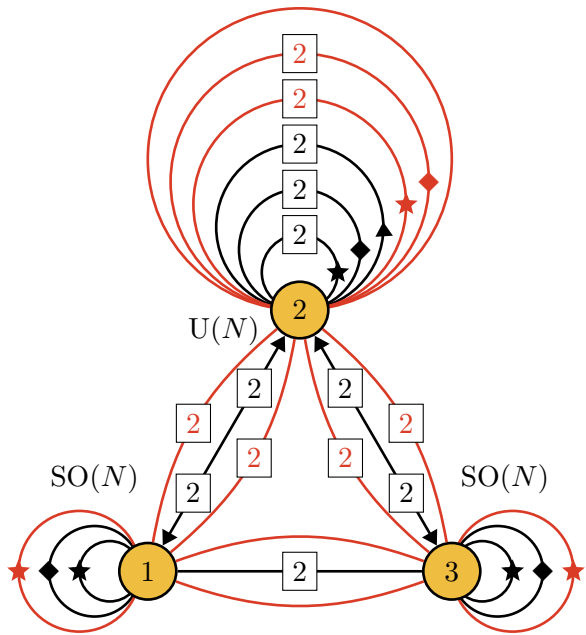
\begin{figure}[H]
    \centering
    \begin{tikzpicture}[scale=2]
    \draw[line width=1pt] (1,1.982) circle (0.25) node[xshift=0.5cm,star,star points=5, star point ratio=2.25, inner sep=1pt, fill=black, draw=black] {} node[fill=white,text opacity=1,fill opacity=1,draw=black,rectangle,thin,yshift=0.5cm] {$2$};
	\draw[line width=1pt] (1,2.132) circle (0.4) node[xshift=0.8cm] {\scriptsize{$\quadro$}} node[fill=white,text opacity=1,fill opacity=1,draw=black,rectangle,thin,yshift=0.8cm] {$2$};
	\draw[line width=1pt,decoration={markings, mark=at position 0.01 with{\arrow{Triangle}}}, postaction={decorate}] (1,2.282) circle (0.55) node[fill=white,text opacity=1,fill opacity=1,draw=black,rectangle,thin,yshift=1.1cm] {$2$};
	\draw[line width=1pt,redX] (1,2.432) circle (0.70) node[xshift=1.4cm,star,star points=5, star point ratio=2.25, inner sep=1pt, fill=redX, draw=redX] {} node[fill=white,text opacity=1,fill opacity=1,draw=black,rectangle,thin,yshift=1.4cm] {$2$};
	\draw[line width=1pt,redX] (1,2.582) circle (0.85) node[xshift=1.7cm] {\scriptsize{$\quadro$}} node[fill=white,text opacity=1,fill opacity=1,draw=black,rectangle,thin,yshift=1.7cm] {$2$};
	\draw[line width=1pt,redX] (1,2.732) circle (1);
	\draw[line width=1pt] (-0.25,0) circle (0.2) node[xshift=-0.4cm,star,star points=5, star point ratio=2.25, inner sep=1pt, fill=black, draw] {};
	\draw[line width=1pt] (-0.35,0) circle (0.3) node[xshift=-0.6cm] {\scriptsize{$\quadro$}};
	\draw[line width=1pt,redX] (-0.45,0) circle (0.4) node[xshift=-0.8cm,star,star points=5, star point ratio=2.25, inner sep=1pt, fill=redX, draw=redX] {};
	\draw[line width=1pt] (2.2,0) circle (0.2) node[xshift=0.4cm,star,star points=5, star point ratio=2.25, inner sep=1pt, fill=black, draw] {};
	\draw[line width=1pt] (2.3,0) circle (0.3) node[xshift=0.6cm] {\scriptsize{$\quadro$}};
	\draw[line width=1pt,redX] (2.4,0) circle (0.4) node[xshift=0.8cm,star,star points=5, star point ratio=2.25, inner sep=1pt, fill=redX, draw=redX] {};
	\node[draw=black,line width=1pt,circle,fill=yellowX,minimum width=0.75cm,inner sep=1pt,label={[xshift=-0.7cm,yshift=0.5cm]:$\SO(N)$}] (A) at (0,0) {$1$};
	\node[draw=black,line width=1pt,circle,fill=yellowX,minimum width=0.75cm,inner sep=1pt,label={[xshift=0.7cm,yshift=0.5cm]:$\SO(N)$}] (B) at (2,0) {$3$};
	\node[draw=black,line width=1pt,circle,fill=yellowX,minimum width=0.75cm,inner sep=1pt,label={[xshift=-1cm,yshift=-1cm]:$\U(N)$}] (C) at (1,1.732) {$2$};
	\path
	(A) edge[line width=1pt] node[fill=white,text opacity=1,fill opacity=1,draw=black,rectangle,thin,pos=0.5] {$2$} (B);
	\path[Triangle-Triangle] (B) edge[line width=1pt] node[fill=white,text opacity=1,fill opacity=1,draw=black,rectangle,thin,pos=0.25] {$2$} node[fill=white,text opacity=1,fill opacity=1,draw=black,rectangle,thin,pos=0.75] {$2$} (C);
	\path[Triangle-Triangle] (C) edge[line width=1pt] node[fill=white,text opacity=1,fill opacity=1,draw=black,rectangle,thin,pos=0.25] {$2$} node[fill=white,text opacity=1,fill opacity=1,draw=black,rectangle,thin,pos=0.75] {$2$} (A);
	\draw[line width=1pt,redX] (A) to[bend right=20]  (B);
	\draw[line width=1pt,redX] (A) to[bend left=20]  (B);
	\draw[line width=1pt,redX] (B) to[bend right=20] node[fill=white,text opacity=1,fill opacity=1,draw=black,rectangle,thin] {\color{redX}{$2$}} (C);
	\draw[line width=1pt,redX] (B) to[bend left=20] node[fill=white,text opacity=1,fill opacity=1,draw=black,rectangle,thin] {\color{redX}{$2$}} (C);
	\draw[line width=1pt,redX] (A) to[bend right=20] node[fill=white,text opacity=1,fill opacity=1,draw=black,rectangle,thin] {\color{redX}{$2$}} (C);
	\draw[line width=1pt,redX] (A) to[bend left=20] node[fill=white,text opacity=1,fill opacity=1,draw=black,rectangle,thin] {\color{redX}{$2$}} (C);
    \end{tikzpicture}
    \caption{Quiver for the  Spin(7) orientifold of $\mathbb{C}^4/\mathbb{Z}_2\times \mathbb{Z}_2$ using the involution in \eqref{eq:C4Z2Z2chiral}, \eqref{eq:C4Z2Z2fermi} and \eqref{eq:C4Z2Z2realfermi}.}
    \label{fig:o_theory_C4Z2Z2}
\end{figure}

 Using Table~\ref{tab:GenerC4Z2Z2}, the transformation in the field theory translates into the geometric involution $\sigma$ acting on the generators as
 \begin{equation}
 \begin{array}{c}
     \left(M_1,M_2,M_3,M_4,M_5\right)\\
   \downarrow\\
   \left(\bar{M}_3,\bar{M}_2,\bar{M}_1,\bar{M}_4,-\bar{M}_5\right)\coma
 \end{array}
 \end{equation}
 which, as expected, is different from and inequivalent to the universal involution $M_a\rightarrow \bar{M}_a$. The quiver for the resulting theory is shown in Figure~\ref{fig:o_theory_C4Z2Z2}, which is also free of gauge anomalies. 
 
\subsection{SPP$\times\CC$}

\label{app:otheory for sppXC}

\begin{figure}[H]
    \centering
	\begin{tikzpicture}[scale=1.5, rotate around y=-10]
	\draw[thick,gray,-Triangle] (0,0,0) -- node[above,pos=1] {$x$} (1.5,0,0);
	\draw[thick,gray,-Triangle] (0,0,0) -- node[left,pos=1] {$y$} (0,1.5,0);
	\draw[thick,gray,-Triangle] (0,0,0) -- node[below,pos=1] {$z$} (0,0,2.5);
	\node[draw=black,line width=1pt,circle,fill=black,minimum width=0.2cm,inner sep=1pt] (O) at (0,0,0) {};
	\node[draw=black,line width=1pt,circle,fill=black,minimum width=0.2cm,inner sep=1pt] (X1) at (1,0,0) {};
	\node[draw=black,line width=1pt,circle,fill=black,minimum width=0.2cm,inner sep=1pt] (Z1) at (0,0,1) {};
	\node[draw=black,line width=1pt,circle,fill=black,minimum width=0.2cm,inner sep=1pt] (Z2) at (0,0,2) {};
	\node[draw=black,line width=1pt,circle,fill=black,minimum width=0.2cm,inner sep=1pt] (XZ) at (1,0,1) {};
	\node[draw=black,line width=1pt,circle,fill=black,minimum width=0.2cm,inner sep=1pt] (Y) at (0,1,0) {};
	\draw[line width=1pt] (O)--(X1)--(XZ)--(Z2)--(O);
	\draw[line width=1pt] (Y)--(O);
	\draw[line width=1pt] (Y)--(X1);
	\draw[line width=1pt] (Y)--(Z1);
	\draw[line width=1pt] (Y)--(Z2);
	\draw[line width=1pt] (Y)--(XZ);
	\end{tikzpicture}
	\caption{Toric diagram for SPP$\times\CC$.}
	\label{fig:SPPtoricdiagram}
\end{figure}
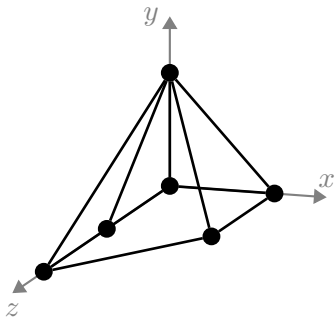

Figure~\ref{fig:SPPtoricdiagram} shows the toric diagram for SPP$\times\CC$.

The gauge theory for D1-branes probing this CY$_4$ was introduced in \cite{Franco:2015tna}. Its quiver diagram is shown in Figure~\ref{fig:SPPquivN02}.

\begin{figure}[H]
	\centering
	\begin{subfigure}[t]{0.49\textwidth}
	\centering
	\begin{tikzpicture}[scale=2]
	\draw[line width=1pt,decoration={markings, mark=at position 0.40 with{\arrow{Triangle}}}, postaction={decorate}] (-0.22,-0.22) circle (0.25);
	\draw[line width=1pt,decoration={markings, mark=at position 0.20 with{\arrow{Triangle}}}, postaction={decorate}] (2.22,-0.22) circle (0.25);
	\draw[line width=1pt,decoration={markings, mark=at position 0.55 with{\arrow{Triangle}}}, postaction={decorate}] (1,2) circle (0.25) node[fill=white,text opacity=1,fill opacity=1,draw=black,rectangle,thin,yshift=0.4cm] {$2$};
	\draw[line width=1pt,redX] (1,2.09) circle (0.4);
	\node[draw=black,line width=1pt,circle,fill=yellowX,minimum width=0.75cm,inner sep=1pt] (A) at (0,0) {$3$};
	\node[draw=black,line width=1pt,circle,fill=yellowX,minimum width=0.75cm,inner sep=1pt] (B) at (2,0) {$2$};
	\node[draw=black,line width=1pt,circle,fill=yellowX,minimum width=0.75cm,inner sep=1pt] (C) at (1,1.732) {$1$};
	\path[Triangle-Triangle] (A) edge[line width=1pt] (B);
	\path[Triangle-Triangle] (B) edge[line width=1pt] (C);
	\path[Triangle-Triangle] (C) edge[line width=1pt] (A);
	\draw[line width=1pt,redX] (A) to[bend right=20]  (B);
	\draw[line width=1pt,redX] (B) to[bend right=20]  (C);
	\draw[line width=1pt,redX] (C) to[bend right=20]  (A);
	\draw[line width=1pt,redX] (A) to[bend left=20]  (B);
	\draw[line width=1pt,redX] (B) to[bend left=20]  (C);
	\draw[line width=1pt,redX] (C) to[bend left=20]  (A);
	\node at (0,-0.6) {};
	\end{tikzpicture}
	\caption{$\mathcal{N}=(0,2)$ language.}
	\label{fig:SPPquivN02}
	\end{subfigure}\hfill
	\begin{subfigure}[t]{0.49\textwidth}
	\centering
	\begin{tikzpicture}[scale=2]
	\draw[line width=1pt,redX] (-0.28,-0.28) circle (0.4);
	\draw[line width=1pt,redX] (2.28,-0.28) circle (0.4);
	\draw[line width=1pt,redX] (1,2.09) circle (0.4) node[fill=white,text opacity=1,fill opacity=1,draw=black,rectangle,thin,yshift=0.8cm] {$3$};
	\draw[line width=1pt,decoration={markings, mark=at position 0.40 with{\arrow{Triangle}}}, postaction={decorate}] (-0.22,-0.22) circle (0.25) node[fill=white,text opacity=1,fill opacity=1,draw=black,rectangle,thin,xshift=-0.35cm,yshift=-0.35cm] {$2$};
	\draw[line width=1pt,decoration={markings, mark=at position 0.20 with{\arrow{Triangle}}}, postaction={decorate}] (2.22,-0.22) circle (0.25) node[fill=white,text opacity=1,fill opacity=1,draw=black,rectangle,thin,xshift=0.35cm,yshift=-0.35cm] {$2$};
	\draw[line width=1pt,decoration={markings, mark=at position 0.55 with{\arrow{Triangle}}}, postaction={decorate}] (1,2) circle (0.25) node[fill=white,text opacity=1,fill opacity=1,draw=black,rectangle,thin,yshift=0.4cm] {$4$};
	\node[draw=black,line width=1pt,circle,fill=yellowX,minimum width=0.75cm,inner sep=1pt] (A) at (0,0) {$3$};
	\node[draw=black,line width=1pt,circle,fill=yellowX,minimum width=0.75cm,inner sep=1pt] (B) at (2,0) {$2$};
	\node[draw=black,line width=1pt,circle,fill=yellowX,minimum width=0.75cm,inner sep=1pt] (C) at (1,1.732) {$1$};
	\path[Triangle-Triangle] (A) edge[line width=1pt] node[fill=white,text opacity=1,fill opacity=1,draw=black,rectangle,thin,pos=0.25] {$2$} node[fill=white,text opacity=1,fill opacity=1,draw=black,rectangle,thin,pos=0.75] {$2$} (B);
	\path[Triangle-Triangle] (B) edge[line width=1pt] node[fill=white,text opacity=1,fill opacity=1,draw=black,rectangle,thin,pos=0.25] {$2$} node[fill=white,text opacity=1,fill opacity=1,draw=black,rectangle,thin,pos=0.75] {$2$} (C);
	\path[Triangle-Triangle] (C) edge[line width=1pt] node[fill=white,text opacity=1,fill opacity=1,draw=black,rectangle,thin,pos=0.25] {$2$} node[fill=white,text opacity=1,fill opacity=1,draw=black,rectangle,thin,pos=0.75] {$2$} (A);
	\draw[line width=1pt,redX] (A) to[bend right=20] node[fill=white,text opacity=1,fill opacity=1,draw=black,rectangle,thin] {\color{redX}{$2$}} (B);
	\draw[line width=1pt,redX] (B) to[bend right=20] node[fill=white,text opacity=1,fill opacity=1,draw=black,rectangle,thin] {\color{redX}{$2$}} (C);
	\draw[line width=1pt,redX] (C) to[bend right=20] node[fill=white,text opacity=1,fill opacity=1,draw=black,rectangle,thin] {\color{redX}{$2$}} (A);
	\draw[line width=1pt,redX] (A) to[bend left=20] node[fill=white,text opacity=1,fill opacity=1,draw=black,rectangle,thin] {\color{redX}{$2$}} (B);
	\draw[line width=1pt,redX] (B) to[bend left=20] node[fill=white,text opacity=1,fill opacity=1,draw=black,rectangle,thin] {\color{redX}{$2$}} (C);
	\draw[line width=1pt,redX] (C) to[bend left=20] node[fill=white,text opacity=1,fill opacity=1,draw=black,rectangle,thin] {\color{redX}{$2$}} (A);
	\end{tikzpicture}
	\caption{$\mathcal{N}=(0,1)$ language.}
	\label{fig:SPPquivN01}
	\end{subfigure}
	\caption{Quiver diagram for SPP$\times\CC$ in $\mathcal{N}=(0,2)$ and $\mathcal{N}=(0,1)$ language.}
	\label{fig:SPPquiv}
\end{figure}

The $J$- and $E$-terms are
\begin{equation}
    \begin{array}{lclc}
                   & J                                       &   \text{\hspace{.5cm}}&   E                     \\
\Lambda_{11} \,:\, & X_{13}X_{31} - X_{12}X_{21}             &                    &   \Phi_{11}X_{11} - X_{11}\Phi_{11} \\
\Lambda_{21} \,:\, & X_{12}X_{23}X_{32} - X_{11}X_{12}       &                    &   \Phi_{22}X_{21} - X_{21}\Phi_{11} \\
\Lambda_{12} \,:\, & X_{21}X_{11} - X_{23}X_{32}X_{21}       &                    &    X_{12}\Phi_{22}-\Phi_{11}X_{12} \\
\Lambda_{31} \,:\, & X_{13}X_{32}X_{23}- X_{11}X_{13}        &                    &    X_{31}\Phi_{11}-\Phi_{33}X_{31} \\
\Lambda_{13} \,:\, & X_{31}X_{11} - X_{32}X_{23}X_{31}       &                    &   \Phi_{11}X_{13} - X_{13}\Phi_{33} \\
\Lambda_{32} \,:\, & X_{21}X_{12}X_{23} -X_{23}X_{31}X_{13}  &                    &   \Phi_{33}X_{32} - X_{32}\Phi_{22} \\
\Lambda_{23} \,:\, & X_{32}X_{21}X_{12} - X_{31}X_{13}X_{32} &                    &   \Phi_{22}X_{23} - X_{23}\Phi_{33} 
\end{array}
\label{eq:SPPxC-JETerms}
\end{equation}

Figure~\ref{fig:SPPquivN01} shows the quiver for this theory in $\mathcal{N}= (0,1)$ language. The $W^{(0,1)}$ associated to \eqref{eq:SPPxC-JETerms} is
\begin{equation}
    W^{(0,1)} = W^{(0,2)}+ \Lambda^R_{11}X^\dagger_{11}X_{11}+ \sum_{\substack{i,j=1\\i\neq j}}^3\Lambda^R_{ii}\left(X^\dagger_{ij}X_{ij}+X^\dagger_{ji}X_{ji}\right)+\sum_{i=1}^3\Lambda^R_{ii}\Phi^\dagger_{ii}\Phi_{ii}\fstop
\end{equation}

\begin{table}[!htp]
   \centering
	\renewcommand{\arraystretch}{1.1}
	\begin{tabular}{c|c}
		Meson    & Chiral fields  \\
		\hline
	$M_1$ & $X_{13}X_{32}X_{21}$ \\	
    $M_2$ & $X_{11}=X_{23}X_{32}$ \\
    $M_3$ & $X_{31}X_{12}X_{23}$ \\
    $M_4$ & $X_{12}X_{21}=X_{13}X_{31}$ \\
    $M_5$ & $\Phi_{11}=\Phi_{22}=\Phi_{33}$ 
    \end{tabular}
    \caption{Generators of SPP$\times\CC$.}
    \label{tab:GenerSPP}
\end{table}

Table~\ref{tab:GenerSPP} shows the generators of SPP$\times\CC$ and their expression in terms of chiral fields. They satisfy the following relation
\begin{equation}
    \mathcal{I}=\left\langle M_1M_3=M_2M_4^2\right\rangle\fstop
\end{equation}

\bigskip

\paragraph{$\SO(N)\times \U(N)$ Orientifold}\mbox{}

\medskip

Let us consider an anti-holomorphic involution which acts on Figure~\ref{fig:SPPquiv} as a reflection with respect to a vertical line going through node 1. Then, node 1 maps to itself, while nodes 2 and 3 are identified.

The involution on chiral fields is
 \begin{equation}\label{eq:SPPchiral}
 \begin{array}{cccccccccccc}
 	 X_{12} & \rightarrow &   \gamma_{\Omega_1}\bar{X}_{13}\gamma_{\Omega_3}^{-1}\coma & 
 	 X_{21} & \rightarrow &  \gamma_{\Omega_3}\bar{X}_{31}\gamma_{\Omega_1}^{-1} \coma  & 
 	 X_{13} & \rightarrow &   \gamma_{\Omega_1}\bar{X}_{12}\gamma_{\Omega_2}^{-1} \coma & 
 	 X_{31} & \rightarrow &   \gamma_{\Omega_2}\bar{X}_{21}\gamma_{\Omega_1}^{-1} \coma \\
 	 X_{23} & \rightarrow &   \gamma_{\Omega_3}\bar{X}_{32}\gamma_{\Omega_2}^{-1} \coma & 
 	 X_{32} & \rightarrow &   \gamma_{\Omega_2}\bar{X}_{23}\gamma_{\Omega_3}^{-1} \coma &
 	 X_{11} & \rightarrow &   \gamma_{\Omega_1}\bar{X}_{11}\gamma_{\Omega_1}^{-1}\coma & 
 	 \Phi_{11} & \rightarrow &  -\gamma_{\Omega_1}\bar{\Phi}_{11}\gamma_{\Omega_1}^{-1} \coma \\ 
 & & &
 \Phi_{22} & \rightarrow &  -\gamma_{\Omega_3}\bar{\Phi}_{33}\gamma_{\Omega_3}^{-1} \coma & 
 	 \Phi_{33} & \rightarrow &   -\gamma_{\Omega_2}\bar{\Phi}_{22}\gamma_{\Omega_2}^{-1} \fstop
& & &
 \end{array}
 \end{equation}

From the invariance of $W^{(0,1)}$, we obtain the transformations of the Fermi fields
 \begin{equation}\label{eq:SPPfermi}
     \begin{array}{cccccccccccc}
 	 \Lambda_{11} & \rightarrow &   -\gamma_{\Omega_1}\bar{\Lambda}_{11}\gamma_{\Omega_1}^{-1}\coma & 
 	 \Lambda_{21} & \rightarrow &  \gamma_{\Omega_3}\bar{\Lambda}_{31}\gamma_{\Omega_1}^{-1} \coma  & 
 	 \Lambda_{12} & \rightarrow &   \gamma_{\Omega_1}\bar{\Lambda}_{13}\gamma_{\Omega_3}^{-1} \coma & 
 	 \Lambda_{31} & \rightarrow &   \gamma_{\Omega_2}\bar{\Lambda}_{21}\gamma_{\Omega_1}^{-1} \coma \\
 	\multicolumn{12}{c}{\begin{array}{ccccccccc}
 	   \Lambda_{13} & \rightarrow &   \gamma_{\Omega_1}\bar{\Lambda}_{12}\gamma_{\Omega_2}^{-1}\coma & 
 	 \Lambda_{32} & \rightarrow &  -\gamma_{\Omega_2}\bar{\Lambda}_{23}\gamma_{\Omega_3}^{-1} \coma  & 
 	 \Lambda_{23} & \rightarrow &   -\gamma_{\Omega_3}\bar{\Lambda}_{32}\gamma_{\Omega_2}^{-1} \coma
 \end{array}
 }
 \end{array}
 \end{equation}
 and
\begin{equation}\label{eq:SPPrealfermi}
 \begin{array}{ccccccccc}
 	 \Lambda^{R}_{11} & \rightarrow &   \gamma_{\Omega_1}\Lambda^{R\,\,T}_{11}\gamma_{\Omega_1}^{-1}\coma & 
 	 \Lambda^{R}_{22} & \rightarrow &  \gamma_{\Omega_3}\Lambda^{R\,\,T}_{33}\gamma_{\Omega_3}^{-1} \coma  & 
 	 \Lambda^{R}_{33} & \rightarrow &   \gamma_{\Omega_2}\Lambda^{R\,\,T}_{22}\gamma_{\Omega_2}^{-1} \fstop
 \end{array}
 \end{equation}
 
 Using Table~\ref{tab:GenerSPP}, the transformation in the field theory translates into the geometric involution $\sigma$ acting on the generators as
 \begin{equation}
 \begin{array}{c}
     \left(M_1,M_2,M_3,M_4,M_5\right)\\
   \downarrow\\
   \left(\bar{M}_3,\bar{M}_2,\bar{M}_1,\bar{M}_4,-\bar{M}_5\right)\coma
 \end{array}
 \end{equation}
which is different from and inequivalent to the universal involution $M_a\rightarrow \bar{M}_a$. The quiver for the resulting theory is shown in Figure~\ref{fig:o_theory_C4Z2Z2}, which is also free of gauge anomalies.

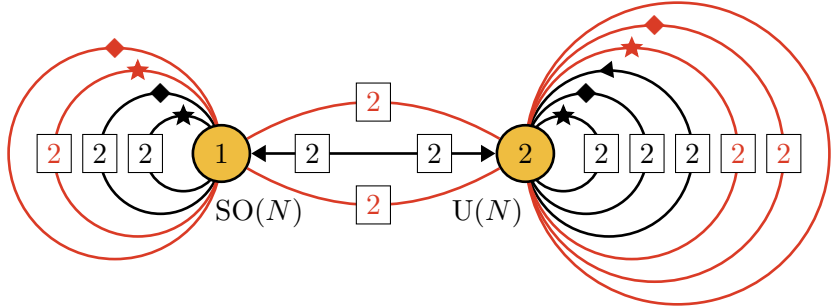
\begin{figure}[H]
    \centering
    \begin{tikzpicture}[scale=2]
	\draw[line width=1pt] (-2.25,0) circle (0.25) node[yshift=0.5cm,star,star points=5, star point ratio=2.25, inner sep=1pt, fill=black, draw] {} node[fill=white,text opacity=1,fill opacity=1,draw=black,rectangle,thin,xshift=-0.5cm] {$2$};
	\draw[line width=1pt] (-2.4,0) circle (0.4) node[yshift=0.8cm] {\scriptsize{$\quadro$}} node[fill=white,text opacity=1,fill opacity=1,draw=black,rectangle,thin,xshift=-0.8cm] {$2$};
	\draw[line width=1pt,redX] (-2.55,0) circle (0.55) node[yshift=1.1cm,star,star points=5, star point ratio=2.25, inner sep=1pt, fill=redX, draw=redX] {} node[fill=white,text opacity=1,fill opacity=1,draw=black,rectangle,thin,xshift=-1.1cm] {$2$};
	\draw[line width=1pt,redX] (-2.7,0) circle (0.7) node[yshift=1.4cm] {\scriptsize{$\quadro$}};
	\draw[line width=1pt] (0.25,0) circle (0.25) node[yshift=0.5cm,star,star points=5, star point ratio=2.25, inner sep=1pt, fill=black, draw] {} node[fill=white,text opacity=1,fill opacity=1,draw=black,rectangle,thin,xshift=0.5cm] {$2$};
	\draw[line width=1pt] (0.4,0) circle (0.4) node[yshift=0.8cm] {\scriptsize{$\quadro$}} node[fill=white,text opacity=1,fill opacity=1,draw=black,rectangle,thin,xshift=0.8cm] {$2$};
	\draw[line width=1pt,decoration={markings, mark=at position 0.27 with{\arrow{Triangle}}}, postaction={decorate}] (0.55,0) circle (0.55)  node[fill=white,text opacity=1,fill opacity=1,draw=black,rectangle,thin,xshift=1.1cm] {$2$};
	\draw[line width=1pt,redX] (0.7,0) circle (0.7) node[yshift=1.4cm,star,star points=5, star point ratio=2.25, inner sep=1pt, fill=redX, draw=redX] {} node[fill=white,text opacity=1,fill opacity=1,draw=black,rectangle,thin,xshift=1.4cm] {$2$};
	\draw[line width=1pt,redX] (0.85,0) circle (0.85) node[yshift=1.7cm] {\scriptsize{$\quadro$}} node[fill=white,text opacity=1,fill opacity=1,draw=black,rectangle,thin,xshift=1.7cm] {$2$};
	\draw[line width=1pt,redX] (1,0) circle (1);
	\node[draw=black,line width=1pt,circle,fill=yellowX,minimum width=0.75cm,inner sep=1pt,label={[xshift=-0.5cm,yshift=-1.5cm]:$\U(N)$}] (A) at (0,0) {$2$};
	\node[draw=black,line width=1pt,circle,fill=yellowX,minimum width=0.75cm,inner sep=1pt,label={[xshift=0.5cm,yshift=-1.5cm]:$\SO(N)$}] (B) at (-2,0) {$1$};
	\path[Triangle-Triangle] (A) edge[line width=1pt] node[fill=white,text opacity=1,fill opacity=1,draw=black,rectangle,thin,pos=0.25] {$2$} node[fill=white,text opacity=1,fill opacity=1,draw=black,rectangle,thin,pos=0.75] {$2$} (B);
	\draw[line width=1pt,redX] (A) to[bend right] node[fill=white,text opacity=1,fill opacity=1,draw=black,rectangle,thin] {\color{redX}{$2$}} (B);
	\draw[line width=1pt,redX] (A) to[bend left] node[fill=white,text opacity=1,fill opacity=1,draw=black,rectangle,thin] {\color{redX}{$2$}} (B);
    \end{tikzpicture}
    \caption{Quiver for a  Spin(7)  orientifold of $\SPP\times \CC$ using the involution in \eqref{eq:SPPchiral}, \eqref{eq:SPPfermi} and \eqref{eq:SPPrealfermi}.}
    \label{fig:o_theory_SPP}
\end{figure}

\bibliographystyle{JHEP}
\bibliography{ref}

\end{document}